\documentclass[%
aps,
 pra,%
 amsmath,amssymb,showpacs,
 reprint,%
 floatfix,
twocolumn
]{revtex4-1}

\usepackage{graphicx}
\usepackage{dcolumn}
\usepackage{bm}

\allowdisplaybreaks[2]

\usepackage{color}

\newcommand{\PR }{Phys. Rev. }
\newcommand{\PRL }{Phys. Rev. Lett. }
\newcommand{\PRA }{\PR A }

\newcommand{\NJP }{New J. Phys. }
\newcommand{\Eqref}[1]{\mbox{Eq.~(\ref{#1})}}
\newcommand{\figref}[1]{\mbox{Fig.~\ref{#1}}}
\newcommand{\secref}[1]{\mbox{Sec.~\ref{#1}}}
\newcommand{\appref}[1]{\mbox{App.~\ref{#1}}}
\newcommand{\vv}[1]{\mathbf{#1}}         
\newcommand{\mathperiod}{\,.} 
\newcommand{\mathcomma}{\,,}  
\newcommand{\ee}{\mathrm{e}}             
\newcommand{\ii}{\mathrm{i}}             
\newcommand{\abs}[1]{\left|{#1}\right|}
\newcommand{\abssmall}[1]{|{#1}|}
\newcommand{\ket}[1]{\left|{#1}\right\rangle}
\newcommand{\ketsmall}[1]{|{#1}\rangle}
\newcommand{\bra}[1]{\left\langle{#1}\right|}
\newcommand{\brasmall}[1]{\langle{#1}|}
\newcommand{\expvalsmall}[1]{\langle{#1}\rangle}
\newcommand{\ie}{\mbox{i.\,e.}\nolinebreak[4]}
\newcommand{\eg}{\mbox{e.\,g.}\nolinebreak[4]}
\newcommand{\cf}{{cf.}~}
\newcommand{\anfz}[1]{``{#1}''}

\begin{document}


\title{Probing few-excitation eigenstates of interacting atoms on a lattice\\by observing their collective light emission in the far field}

\author{Paolo~Longo}
\email{paolo.longo@mpi-hd.mpg.de}
\author{J\"org~Evers}%
\affiliation{Max Planck Institute for Nuclear Physics, Saupfercheckweg 1, 69117 Heidelberg, Germany}

\date{\today}

\begin{abstract}
{
The collective emission from a one-dimensional chain of interacting two-level 
atoms coupled to a common electromagnetic reservoir is investigated.
We derive the system's dissipative few-excitation eigenstates, and analyze 
its static properties, including the collective dipole moments and 
branching ratios between different eigenstates. 
Next, we study the dynamics, and characterize the light emitted
or scattered by such a system via different far-field observables. 
Throughout the analysis, we consider spontaneous emission from an excited state 
as well as two different pump field setups, and contrast the two extreme cases of 
non-interacting and strongly interacting atoms. 
For the latter case, the two-excitation submanifold contains a two-body bound state, 
and we find that the two cases lead to different far-field signatures.
Finally we exploit these signatures to characterize the wavefunctions
of the collective eigenstates. 
For this, we identify a direct relation between the collective branching ratio 
and the momentum distribution of the collective eigenstates' wavefunction. 
This provides a method to proof the existence of certain collective eigenstates and to access their
wave function without the need to individually address and/or manipulate single atoms.
}
\end{abstract}



\maketitle

\section{Introduction}
In recent years, tailored lattice systems of (artificial) atoms 
have turned into a thriving field of experimental and theoretical
research across many different subdisciplines of quantum optics.
Still, the typical questions and problems addressed in the context of atomic 
lattice systems interacting with light often relate to basic and generic properties.
Physical realizations of artificial atomic lattice systems cover 
a wide range of technologies with which light--matter interactions can be studied.
To name just a few, these include
cold atoms in optical lattices \cite{zoller05,winkler06,fukuhara13}, 
fiber-based settings \cite{rauschenbeutel10}, 
atom--cavity networks and on-chip photonics \cite{angelakis08,koch13,plenio08,benson13}, 
or cold polar molecules \cite{whaley14}.

One particular line of research involves static and dynamic properties of the system in the 
few-excitation subspace~\cite{valiente08,valiente09,evertz12,longo13}. 
In particular few-excitation eigenstates in one dimension 
are also connected to the topic of more \anfz{exotic} eigenstates such as two-body bound states 
on a lattice \cite{winkler06,fukuhara13,valiente08,valiente09,longo13,longo14}. 
These represent a nice example for how a \anfz{historical} prediction originally put forward by 
Bethe \cite{bethe32} is investigated today by means of cutting-edge experimental techniques \cite{winkler06,fukuhara13}.

However, oftentimes, related experiments are very demanding in that they require {\it in situ} tuning of 
parameters and/or rely on single-site manipulation. 
This motivates alternative approaches such as
the coupling to a probing light field~\cite{Weitenberg11,porras08,lesanovsky10,Mekhov12,lulingarxiv,longo14}.
Essentially, the key question here is about how much information on the eigenstates
of an atomic lattice system can be inferred from the scattered light in the far field 
when the system is probed optically.

\begin{figure}[tb]
 \centering
 \includegraphics[width=0.48\textwidth]{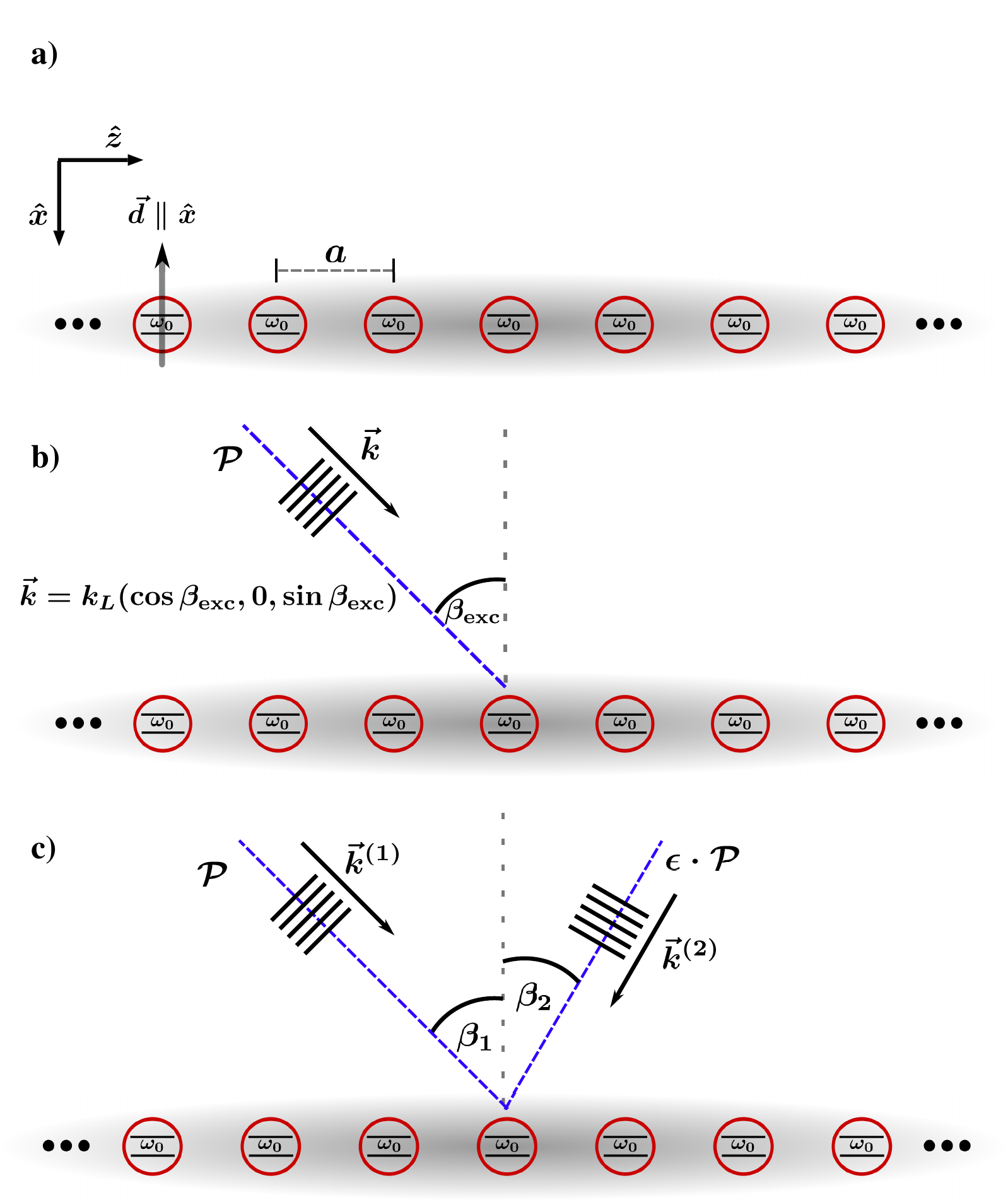}
 \caption{\label{fig:overview}
	  (color online) General structure of the analysis. (a) shows the basic setup. 
	  The two-level atoms (bare transition energy $\omega_0$) are arranged on a one-dimensional
	  lattice (lattice constant $a$) along the $z$ direction.
	  The atoms' dipole moments are uniformly aligned and lie in the $x$-$z$ plane
	  (in this sketch, we have exemplarily chosen an angle between the dipole moments 
	  and the atomic chain of $\theta=\pi/2$). 
	  The whole chain is coupled to a common reservoir indicated by the shading. 
	  In (b), the first considered pump-field configuration is shown, with an incoherent 
	  pump rate $\abssmall{\mathcal{P}}^2$ applied at an excitation angle $\beta_{\mathrm{exc}}$. 
	  The pump's electric field polarization vector lies in the $x$-$z$ plane and is 
	  perpendicular to the driving field's wave vector $\vv{k}$.
	  To separate the scattered field from the driving field, a detection out-of-plane
	  (\protect\ie, in the $y$-$z$ plane) could be performed.
	  (c) Two-pump setup with excitation angles $\beta_1$ and $\beta_2$. 
	  The ratio of the driving fields' pump rates is $\epsilon^2$.}
\end{figure}

A recent addition to the zoo of lattice systems arises from the emerging field of 
x-ray quantum optics~\cite{xrayreview}. 
Nuclear transitions driven, e.g., by synchrotron light~\cite{xraybook}, are a particularly promising 
implementation, fueled by a number of recent theoretical~\cite{theo1,theo2,theo3,theo3a,theo4,theo5,theo6} 
and experimental~\cite{exp1,exp2,exp3,exp4,exp5,exp6,exp7,exp8,nano,preprint} works. 
Typically, the nuclei are embedded in a solid state target, which may exhibit a crystalline structure. 
Facilitated by the M\"o{\ss}bauer effect~\cite{mossbauer,mossbauer2}, 
suitable nuclei such as ${}^{57}$Fe have the additional benefit of offering recoil-less absorption 
or emission of light.
Nanostructuring of the target, e.g., into thin-film waveguide structures~\cite{exp4,exp6,exp7,theo6} 
or nano-wires~\cite{nano}, allows one to explore lower-dimensional geometries. 
Furthermore, in state-of-the-art experiments, source limitations naturally restrict the nuclei 
to the low-excitation subspace. 
Finally, since the nuclei are probed via scattered light observed in the far field, again the 
question arises, how much this light reveals about the system. 

Motivated by this question, here, we study the far-field signatures that emerge in the collective emission 
from dipole--dipole coupled two-level atoms on a one-dimensional lattice (see \figref{fig:overview}). 
To this end, we develop an analytical framework reminiscent of spin physics, which allows us to discuss 
the relevant physical mechanisms in a broader context. 
We start by deriving the dissipative few-excitation eigenstates of the coupled atoms subject to a 
common reservoir. 
Next, we determine the collective dipole moments and related branching ratios for transitions between 
the eigenstates of the atomic chain. 
As a key result of this analysis, we identify a connection between the momentum distribution of the 
collective atomic states' wavefunctions and these branching ratios.
This analysis further enables us to determine the relevant level space for the spontaneous emission from 
an excited eigenstate, as well as under the influence of weak incoherent driving fields. 
We then study the dynamics within this relevant level space by means of a master equation. 
Relating the electric far field operator to the collective atomic eigenbasis, we can evaluate far-field 
observables such as the emitted intensity, the emission spectrum, and an intensity correlation function. 
This approach is of interest since it does not require single-atom addressability or manipulation techniques, 
and it is compatible with recent experiments in nuclear quantum optics.

In our analysis, we particularly contrast the two extreme cases of non-interacting and strongly 
interacting atoms. 
For the latter case, the two-excitation submanifold comprises not only scattering states but also 
two-body bound states of atomic excitation. 
We further find that the two cases lead to different dynamics, involving characteristic branching ratios 
between the eigenstates. 
These differences also manifest themselves in all optical far-field observables. 
As an application, we discuss different driving field setups to characterize the collective eigenstates' 
wavefunctions via the scattered light.  
Finally, as an outlook, we calculate second-order correlation functions of the light in the far field. 

This paper is organized as follows.
In \secref{sec:fundamentals}, we introduce our model and discuss it's static properties. 
Furthermore, we determine the electric far-field operator in the collective atomic eigenbasis and 
define our observables.
Based on a Lindblad master equation, we analyze in \secref{sec:spontemdyn} the angle-dependent far-field 
pattern emerging from the spontaneous emission of the system's eigenstates. 
In \secref{sec:steadystatesigs}, we discuss the far-field properties in the presence of external driving fields. 
We conclude the paper in \secref{sec:conclusion}. 
Appendices~\ref{sec:appphdof}--\ref{sec:appetasums} provide technical details on the
calculations performed.

\section{Fundamentals}
\label{sec:fundamentals}

\subsection{The Model}
We start with the Hamiltonian ($\hbar \equiv 1$) of $M \equiv N+1 \gg 1$ ($N$ is even) two-level atoms 
that are periodically arranged on a one-dimensional lattice (lattice constant $a$),
are linearly coupled to a photon bath, and are subject to a pair interaction 
$V_{nm}=V_{\abs{n-m}}$ (where $V_0=0$):
\begin{eqnarray}
 \label{eq:origH}
 \nonumber
 H = & & \sum_{n=-\frac{N}{2}}^{\frac{N}{2}} \omega_n \sigma^{+}_n \sigma^{-}_{n} 
    + \sum_k \epsilon_k a^{\dagger}_k a^{\phantom \dagger}_k 
    \\ 
    &+& \sum_{n=-\frac{N}{2}}^{\frac{N}{2}} \sum_k \left( g^{\phantom *}_{nk} \sigma^{+}_n a^{\phantom \dagger}_k  
				      +  g^{*}_{nk} \sigma^{-}_n a^{\dagger}_k \right) \\ \nonumber
    &+& \frac{1}{2} \sum_{n=-\frac{N}{2}}^{\frac{N}{2}} \sum_{m=-\frac{N}{2}}^{\frac{N}{2}} V_{\abs{n-m}} \sigma^{+}_n \sigma^{-}_n \sigma^{+}_m \sigma^{-}_m
    \mathperiod
\end{eqnarray}
The atomic indices refer to lattice sites,
$\omega_n = \omega_{0}~\forall~n$ denotes the atomic transition energies,
and $g_{nk}$ signifies the coupling of atom $n$ to mode $k$ of a photon bath 
(whose dispersion relation is $\epsilon_k$).
The photonic creation and annihilation operators $a^{\dagger}_k$ and $a^{\phantom \dagger}_k$
are bosonic operators and the index $k$ runs over all modes of the photon reservoir.
The atomic raising (lowering) operator of atom $n$ is denoted by
$\sigma^{+}_n$ ($\sigma^{-}_n$), satisfying 
the commutation relation 
$\left[ \sigma^{+}_n , \sigma^{-}_m \right] = \delta_{nm} \sigma^z_n$, 
where $\sigma^{+}_n = \ketsmall{\uparrow}_n \brasmall{\downarrow}_n$, 
$\sigma^{-}_n = \left( \sigma^{+}_n \right)^{\dagger}$, and
$\sigma^z_n = \ketsmall{\uparrow}_n \brasmall{\uparrow}_n - \ketsmall{\downarrow}_n \brasmall{\downarrow}_n$.
Here, $\ketsmall{\uparrow}_n$ ($\ketsmall{\downarrow}_n$) denotes the excited (ground)
state of atom $n$.

Upon eliminating the photonic degrees of freedom similarly to 
the formalism presented in Ref.~\cite{li12} (see \appref{sec:appphdof} for details),
we arrive at the
effective, non-Hermitian Hamiltonian 
\begin{eqnarray}
 \nonumber
  \label{eq:Heff} 
  H_{\mathrm{eff}}
      &=& \sum_{n=-\frac{N}{2}}^{\frac{N}{2}} \omega_0 ~\sigma^+_n \sigma^-_n 
       - \frac{\ii}{2} \sum_{n=-\frac{N}{2}}^{\frac{N}{2}} \sum_{m=-\frac{N}{2}}^{\frac{N}{2}} \Gamma_{\abs{n-m}} \sigma^+_n \sigma^-_m \\
      && ~+~ \frac{1}{2} \sum_{n=-\frac{N}{2}}^{\frac{N}{2}} \sum_{m=-\frac{N}{2}}^{\frac{N}{2}} V_{\abs{n-m}} \sigma^+_n \sigma^-_n \sigma^+_m \sigma^-_m
 \mathperiod
\end{eqnarray}
The individual terms in this Hamiltonian can be understood as follows. 
Being coupled to an electromagnetic reservoir, an individual atom experiences 
an energy shift (Lamb shift) of $\mathrm{Im}(\Gamma_0)/2$ and it is subject
to spontaneous decay at a rate of $\mathrm{Re}(\Gamma_0) \equiv \gamma_0$.
Photons can be exchanged between nearby atoms via the common electromagnetic
reservoir by virtue of dipole--dipole coupling.
These processes are incorporated into the terms with $\mathrm{Im}(\Gamma_{\abs{n-m} \geq 1})$,
while the dissipative part (\ie, $\mathrm{Re}(\Gamma_{\abs{n-m} \geq 1})$) 
leads to a modification of the spontaneous emission rate.
Furthermore, two atoms at a relative distance of $\abs{n-m} a$ experience an
interaction-induced energy shift $V_{\abs{n-m}}$ if they are both in the excited state.
Hamiltonian~(\ref{eq:Heff}) can also be seen as describing interacting and dissipative 
spin-$1/2$ excitations on a one-dimensional lattice.

In this paper, we aim at investigating the regime of an extended sample,
where the bare atomic emission wavelength 
$\lambda_{\mathrm{at}} = 2\pi/k_{\mathrm{at}} = 2\pi c / \omega_{0}$
($c$ is the speed of light)
is smaller than the lattice constant,
\ie, $\lambda_{\mathrm{at}} / a < 1$,
contrasting the established \anfz{small-volume limit} originally 
investigated by Dicke~\cite{dicke54,haroche82}. 
In the following, we restrict the dipole--dipole
coupling and the atom--atom interactions to nearest
neighbors.
The only parameters left in this regime are 
$\Gamma_0$, $\Gamma_1$, and $U \equiv V_1$, 
resulting in the tight-binding formulation
\begin{eqnarray}
 \label{eq:Hefftb} 
  H^{\mathrm{tb}}_{\mathrm{eff}}  
      &=& \sum_{n=-\frac{N}{2}}^{\frac{N}{2}} \left( \omega_0 - \frac{\ii \Gamma_0}{2} \right)~ \sigma^+_n \sigma^-_n \\
      \nonumber
      && ~-~ \frac{\ii \Gamma_1}{2} \sum_{n=-\frac{N}{2}}^{\frac{N}{2}-1} \left( \sigma^+_{n+1} \sigma^-_n + \mathrm{h.c.} \right) 
      \\ \nonumber
      && ~+~ U \sum_{n=-\frac{N}{2}}^{\frac{N}{2}-1} \sigma^+_{n+1} \sigma^-_{n+1} \sigma^+_n \sigma^-_n
 \mathperiod
\end{eqnarray}
For atoms coupled to free space, 
the complex rates that enter Hamiltonians~(\ref{eq:Heff}) and~(\ref{eq:Hefftb}) 
are given by~\cite{li12}
\begin{eqnarray}
 \label{eq:fullrates}
 \frac{\Gamma_0}{\gamma_0} &=& 1 - \frac{2 \ii}{\pi} \mathcomma \\
 \frac{\Gamma_{x \neq 0}}{\gamma_0} &=& A_x \sin^2\theta + B_x \cdot \frac{3 \cos^2 \theta - 1}{2} \mathcomma \\
 A_x &=& - \frac{3 \ii \ee^{\ii \xi_x}}{2 \xi_x} \mathcomma \\
 \nonumber
 B_x &=& \frac{3}{\xi^3_x} \cdot 
	    \big{[} \sin \xi_x - \xi_x \cos \xi_x \\ 
      && \hspace{1.2cm} ~-~ \ii \left( \cos \xi_x + \xi_x \sin \xi_x \right) \big{]} \mathcomma \\
 \label{eq:fullrateslast}
 \xi_x &=& k_{\mathrm{at}} a \abs{x} = \omega_{0} a \abs{x} / c 
 \mathcomma
\end{eqnarray}
where $x \equiv n-m$ 
and $\theta$ denotes the angle between the atomic chain ($z$ axis) and
the atomic dipole moment $\vv{d}$ 
(see also \figref{fig:overview}a)).

Exemplarily, orders of magnitudes in the field of Rydberg atoms trapped in an optical lattice
are \cite{weidemueller13} 
$\lambda_\mathrm{at} \sim 500~\mathrm{nm}$, ($\omega_{0} / 2 \pi \sim 500~\mathrm{THz}$), 
$\gamma_0 \sim \mathrm{MHz}$, and $U \sim 50~\mathrm{GHz}$ 
(for $a \sim 1~\mu\mathrm{m}$), 
representing separated scales $\omega_{0} \gg U \gg \gamma_0$.
Alternatively, in the realm of x-ray quantum optics 
\cite{exp1,exp2,exp3,exp4,exp5,exp6,exp7,exp8,nano}, 
for instance the ${}^{57}$Fe M\"ossbauer transition 
is characterized by $\omega_{0}/2\pi \sim   10^{18}$~Hz,  
$\gamma_0 \sim $~MHz,  $a \sim \mathrm{pm}$, and $U=0$.
For these systems, crystalline solid state targets naturally provide ordered arrays of x-ray emitters.

\subsection{Eigenstates and Eigenvalues}
\label{subsec:eigenstates}
In this Section,
we investigate the non-Hermitian eigenvalue problem 
\begin{equation}
 \left( H_{\mathrm{eff}} - E \right) \ketsmall{\Psi}=0
 \mathperiod
\end{equation}
The resulting eigenvalues $E$ are complex and the real part plays the role of the eigenstate's
excitation energy, whereas $-2 \mathrm{Im}(E)$ can be interpreted as the decay rate 
of the eigenstate's occupation number \cite{li12}.
Also note that for a non-Hermitian Hamiltonian, the left eigenstates are in general not just the
Hermitian conjugate of the right eigenstates (as it would be for a Hermitian Hamiltonian).
Generally, the notion of biorthogonality needs to be taken into account 
(for instance, see Ref.~\cite{biorth}).
However,
these details will not be important in the course of this paper since we are led by 
the following train of thought.

First, we use an effective non-Hermitian Hamiltonian to obtain the energies and
decay rates in the system's eigenbasis.
Later on, when switching to a Lindblad formulation (\secref{subsec:lindbladeq}), the coherent
time evolution is given with respect to the \anfz{closed} system described by the Hermitian part of the
Hamiltonian which is realized formally by setting the real parts of the complex rates to zero,
\ie, $\mathrm{Re}(\Gamma_x)=0$.
Note that setting the real part of the complex decay rates to zero results in a Hermitian Hamiltonian,
yielding real energy eigenvalues (which are the system's transition energies).
Expressed differently, the coherent part of the Lindblad equation is diagonal with respect 
to the \anfz{non-dissipative} basis (which we also utilize for the calculation of matrix elements).
Likewise, the dissipators in the incoherent part of the Lindblad equation contain the
eigenstates' total decay rates as obtained from the imaginary part of the complex eigenenergies
of the original non-Hermitian Hamiltonian~\cite{noteequiv}.

\subsubsection{Single-Excitation Eigenstates and Eigenvalues}
\begin{figure*}[thp]
 \centering
 \includegraphics[width=	\textwidth]{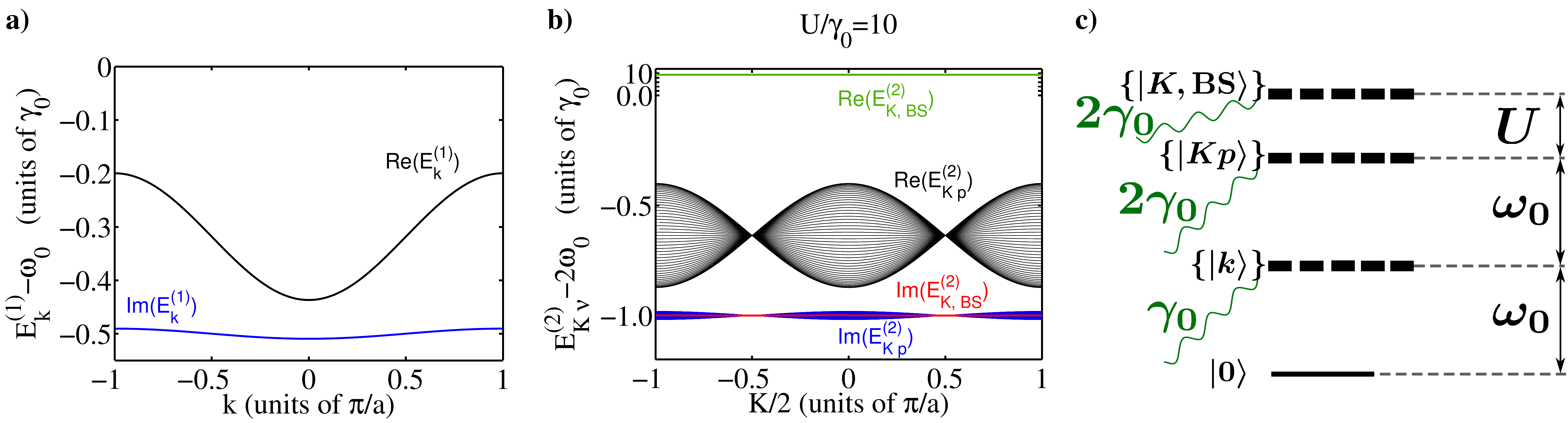}
 \caption{\label{fig:disprels}
	  (color online).
	  Complex dispersion relations for single-excitation states $\ketsmall{k}$
	  (a))
	  and two-excitation states $\ketsmall{K \nu}$ (b)).
	  Note that the energies of the bound states are detached from the quasicontinuum
	  of scattering states.
	  The parameters chosen here are $\lambda_{\mathrm{at}}/a=0.5$ and $\theta=\pi/2$, 
	  resulting in $\Gamma_0/\gamma_0=1-0.637\ii$ and $\Gamma_1/\gamma_0=0.009-0.119\ii$.
	  For $\omega_0 \gg U \gg \gamma_0$, we can resort to a simplified level scheme
	  of quasidegenerate bands (c)).
	  }
\end{figure*}
A solution to the single-excitation eigenproblem (see \appref{subsec:appeigeq1} for details) 
can be formulated in terms of a Bloch wave 
\begin{equation}
 \ket{k}=\sum_n \frac{\ee^{\ii k a n}}{\sqrt{M}} \sigma^+_n \ket{0}
 \mathcomma
\end{equation}
where $ka = -\pi + 2 \pi \ell / M$ ($\ell=0,1,\dots,M-1$)
is a wavenumber from the first Brillouin zone
and $\ketsmall{0}$ signifies the vacuum state.
The complex eigenvalues for the general Hamiltonian~(\ref{eq:Heff}) are essentially
given as the Fourier lattice transform of the complex rates, \ie,
\begin{eqnarray}
 \label{eq:eigen1ex1dlatt}
 \nonumber
 E^{(1)}_k &=& \omega_{0} - \frac{\ii}{2} \sum_{x=-\frac{N}{2}}^{\frac{N}{2}} \Gamma_{\abs{x}} \ee^{-\ii k a x} \\
     &=& \omega_{0} - \frac{\ii}{2} \Gamma_0 - \ii \sum_{x=1}^{\frac{N}{2}} \Gamma_{x} \cos(k a x) \mathperiod
\end{eqnarray}
Rewritten in terms of the collective Lamb shift and decay rate we have
\begin{eqnarray}
 \label{eq:1exrealimagpart}
 \mathrm{Re}(E^{(1)}_k) &=& \omega_{0} + \frac{1}{2} \mathrm{Im}(\Gamma_0) + \sum_{x=1}^{\frac{N}{2}} \mathrm{Im}(\Gamma_x) \cos(k a x) \mathcomma \\
 \label{eq:1exrealimagpartB}
 \Gamma_k &\equiv& -2 \mathrm{Im}(E^{(1)}_k) = \gamma_0 + 2 \sum_{x=1}^{\frac{N}{2}} \mathrm{Re}(\Gamma_x) \cos(k a x) \mathperiod
\end{eqnarray}
As a side note we would like to point out that the superradiant Dicke state \cite{dicke54,haroche82} 
in the small-volume limit $a \rightarrow 0$ 
is also accounted for in the above expressions.
For $a \rightarrow 0$, we have $\Gamma_x \rightarrow \Gamma_0$ (see Eqs.~(\ref{eq:fullrates})--(\ref{eq:fullrateslast}))
so that the state with $k=0$ (symmetric Dicke state) exhibits a decay rate as well as a Lamb shift 
proportional to the number of emitters $M$ in the volume.
Hence, $\Gamma_{k=0} \overset{a \rightarrow 0}{\rightarrow} M \gamma_0$
and $\mathrm{Re}(E^{(1)}_k=0) \overset{a \rightarrow 0}{\rightarrow} \omega_{0} - M \gamma_0 / \pi$.

However, for the remainder of this paper, we focus on the regime of an extended sample and utilize the 
tight binding Hamiltonian~(\ref{eq:Hefftb}).
In that case, the above formulae reduce to
\begin{eqnarray}
 \label{eq:eigen1extblatt_B}
 \mathrm{Re}(E^{(1)}_k) &=& \omega_{0} + \frac{1}{2} \mathrm{Im}(\Gamma_0) +  \mathrm{Im}(\Gamma_1) \cos(ka) \simeq \omega_{0} \mathcomma \\
 \label{eq:eigen1extblatt_C}
 \Gamma_k &=& \gamma_0 + 2 \mathrm{Re}(\Gamma_1) \cos(ka) \simeq \gamma_0 \mathcomma
\end{eqnarray}
where the approximated expressions are valid for $\gamma_0 / \omega_{0} \ll 1$, 
describing sharp optical resonances.
The single-excitation dispersion relation for the tight-binding case is depicted 
in \figref{fig:disprels}a).

\subsubsection{Two-Excitation Eigenstates and Eigenvalues}
\label{subsubsec:twoexcst}
The solution to the two-excitation problem 
(see \appref{subsec:appeigeq2} for details) can be written in terms of
a product of the center-of-mass motion 
(described by a plane wave with a center-of-mass wavenumber $K$) and a
relative wavefunction $\Psi^{(K \nu)}_{\abs{n_1-n_2}}$, \ie,
\begin{equation}
 \label{eq:twobodyansatz}
 \ket{K \nu} = \frac{1}{2\sqrt{M}} \sum_{n_1 n_2} 
	      \ee^{\ii \frac{Ka}{2} \left(n_1+n_2\right)} 
	      \cdot \Psi^{(K \nu)}_{\abs{n_1-n_2}} 
	      ~\sigma^+_{n_1} \sigma^+_{n_2} \ket{0}
	      \mathperiod
\end{equation}
Here, $K/2$ is from the first Brillouin zone
and $\nu$ is a quantum number still to be determined.
For $n_1=n_2$, the wavefunction needs to vanish ($\Psi^{(K \nu)}_0 = 0$) 
since a single atom cannot be doubly excited, expressing the fact that the excitations 
of a 1D spin-$1/2$ chain are hard-core bosons.

Note that this eigenproblem is similar to the problem of two excitations in the extended Bose-Hubbard model
\cite{valiente09,valiente08}
and also occurs in the study of biexcitons in arrays of coupled chromophores \cite{agranovichbook,spano91,spano95}.
Originally put forward by Bethe \cite{bethe32}
(and, for instance, also addressed in 
Refs.~\cite{bethe32,winkler06,fukuhara13,valiente09,valiente08,longo13}), 
is the remarkable fact that a complete basis of the two-excitation submanifold comprises 
scattering states \emph{and} bound states.
We will now discuss these two classes of solutions.

{\it Scattering States.---}
For each center-of-mass wavenumber $K$, we have scattering states characterized by their relative
wavenumber $p$ (from the first Brillouin zone). 
The relative wave function is of the form ($x \neq 0$)
\begin{equation}
 \label{eq:ansatzscattstat}
 \Psi^{(K p)}_x = \frac{1}{\sqrt{M}} \left( \ee^{\ii pa \abs{x}} 
      + \ee^{\ii \delta_{Kp}} \ee^{-\ii pa \abs{x}} \right) 
      \mathcomma
\end{equation}
where $\delta_{K p}$ is the scattering phase shift due to the atom--atom interaction.
The corresponding complex energy eigenvalues can be written as
\begin{eqnarray}
 \label{eq:eigen2extblatt_B}
 \nonumber
 \mathrm{Re}(E^{(2)}_{Kp}) &=& 2 \omega_{0} + \mathrm{Im}(\Gamma_0) + 2 \mathrm{Im}(\Gamma_1) \cos\left( \frac{Ka}{2} \right)\cos(pa)  
 \\
 &\simeq& 2 \omega_{0}
 \mathcomma \\
 \label{eq:eigen2extblatt_C}
 \nonumber
 \Gamma^{Kp}_{\mathrm{tot}} &\equiv& -2 \mathrm{Im}(E^{(2)}_{Kp}) 
 \\
 &=&
 2 \gamma_0 + 4 \mathrm{Re}(\Gamma_1) \cos\left( \frac{Ka}{2} \right)\cos(pa) 
 \\ \nonumber
 &\simeq& 2 \gamma_0
 \mathperiod
\end{eqnarray}
This two-excitation dispersion relation is depicted in \figref{fig:disprels}b).
As in Eqs.~(\ref{eq:eigen1extblatt_B}) and~(\ref{eq:eigen1extblatt_C}),
the approximated expressions are valid for $\gamma_0 / \omega_{0} \ll 1$.
Note that the eigenvalues~(\ref{eq:eigen2extblatt_B}) and~(\ref{eq:eigen2extblatt_C})
do not depend on the scattering potential~$U$.
In fact, the (complex) eigenenergy is 
just the energy of the \anfz{free} particles as it is always the case in scattering theory.
This becomes most apparent when expressing the center-of-mass and the relative wavenumbers in terms of 
single-particle wavenumbers $k_1$ and $k_2$, \ie,
$E^{(2)}_{Kp} = E^{(2)}_{k1+k2, (k_1-k_2)/2} = E^{(1)}_{k_1} + E^{(1)}_{k_2}$.
In particular, when there is no interaction ($U = 0$) the full many-body
solution can be written as a direct product of single-excitation states and the many-body eigenenergy
is the sum of the single-excitation energies.

In the introduction of \secref{subsec:eigenstates}, we mentioned that
the eigenstate's wavefunction coefficients 
belonging to the non-dissipative, Hermitian system (realized by setting 
$\mathrm{Re}(\Gamma_0)=\mathrm{Re}(\Gamma_1)=0$) are important
for a later calculation of matrix elements.
The corresponding scattering phase shift for the non-dissipative system is~\cite{fnphaseshift}
\begin{equation}
 \label{eq:phaseshiftclosedsys}
 \ee^{\ii \delta_{Kp}} = 
      - \frac{ \mathrm{Im}(\Gamma_1) \cos\left( \frac{Ka}{2} \right) - U \ee^{\ii pa}}
	    {\mathrm{Im}(\Gamma_1) \cos\left( \frac{Ka}{2} \right) - U \ee^{- \ii pa}} 
	    \mathperiod
\end{equation}
For the remainder of this paper, we will concentrate on the two cases
of non-interacting atoms ($U=0$) and strong atom--atom interactions (\anfz{large $U$ limit}).
For $U=0$ the phase shift is 
\begin{equation}
 \label{eq:expfacU0}
 \ee^{\ii \delta_{Kp}} = -1~~~~~(U=0)
 \mathperiod
\end{equation}
This is a result of the hard-core constraint $\Psi^{(K \nu)}_0 = 0$ 
and can be understood as an infinite repulsion at zero relative coordinate.
Conversely, in the limit of strong atom--atom interactions, 
where $U \gg \gamma_0 > \mathrm{Im}(\Gamma_1)$,
the phase shift is 
\begin{equation}
 \label{eq:expfacUinf}
 \ee^{\ii \delta_{Kp}} = - \ee^{2 \ii pa}~~~~~(U \gg \gamma_0)
 \mathperiod
\end{equation}
Further note that, for both $U=0$ and large $U$, the $p=0$-wavefunction vanishes.
This also includes the so-called symmetric Dicke state with $K=p=0$ 
for which the relative phases between the different wavefunction coefficients is 
constant.
Unlike for the original Dicke model,
two excitations in our tight-binding model interfere destructively for $p=0$ 
due to the mentioned hard-core constraint 
(see $\ee^{\ii \delta_{Kp}} = -1$ in Eqs.~(\ref{eq:expfacU0}), (\ref{eq:expfacUinf}), 
and~(\ref{eq:ansatzscattstat})).

{\it Bound States.---}
Additionally, the system may also feature bound states \cite{valiente08,valiente09,longo13},
whose relative wavefunction is of the form ($x \neq 0$)
\begin{equation}
 \label{eq:ansatzboundstat}
 \Psi^{(K, \mathrm{BS})}_x = \alpha^{\abs{x}-1}_K
 \mathcomma
\end{equation}
where
$\alpha_K = - \ii \Gamma_1 \cos\left( \frac{Ka}{2} \right)/U$
and the complex energy eigenvalues are
\begin{eqnarray}
 \label{eq:eigen2extblattBS_B}
 \mathrm{Re}({E}^{(2)}_{K, \mathrm{BS}}) &=& 
 2\omega_{0} + \mathrm{Im}(\Gamma_0) +  U \\ \nonumber
 && ~- \cos^2\left( \frac{Ka}{2}\right) \cdot \frac{ \left[ \left( \mathrm{Re}(\Gamma_1) \right)^2  -\left( \mathrm{Im}(\Gamma_1) \right)^2  \right] }{U} 
 \\ \nonumber
 &\simeq& 2\omega_{0} +  U \mathcomma \\
 \label{eq:eigen2extblattBS_C}
 \nonumber
 \Gamma^{K, \mathrm{BS}}_{\mathrm{tot}} &\equiv& -2 \mathrm{Im}({E}^{(2)}_{K, \mathrm{BS}}) 
 \\
 &=& 2\gamma_0 + 4 \cos^2\left( \frac{Ka}{2} \right) \cdot \frac{\mathrm{Re}(\Gamma_1) \mathrm{Im}(\Gamma_1)  }{U}  \\
 \nonumber
 &\simeq& 2 \gamma_0
 \mathperiod
\end{eqnarray}
Again,
the approximated expressions are valid for $\omega_{0} \gg U \gg \gamma_0$, 
\ie, for sharp optical resonances and strong atom--atom interactions.
The dispersion relation for the two-body bound states is compared with the scattering states' 
properties in \figref{fig:disprels}b).
Figure~\ref{fig:disprels}c) summarizes the few-excitation Hilbert space in terms of a simplified level scheme 
consisting of quasi-degenerate bands (which is valid for $\omega_0 \gg \gamma_0$).

Note that---in contrast to the scattering states---the energy of the bound states also 
depends on the dissipative property $\mathrm{Re}(\Gamma_1)$.
This can be understood as a dissipation-induced energy shift due to the 
dressing of the bound state through the reservoir.
However, this effect is negligible for strong atom--atom interactions $U \gg \gamma_0$.

If we, as before, imagine the non-dissipative system (where $\mathrm{Re}(\Gamma_0)=\mathrm{Re}(\Gamma_1)=0$)
and require the relative wavefunction to be spatially confined, the condition $\abs{\alpha_K}<1$
translates into 
\begin{equation}
\label{eq:critforexist}
\abs{ {\mathrm{Im}(\Gamma_1)} \cos\left( \frac{Ka}{2} \right)   } < \abs{U}
\mathperiod
\end{equation}
This criterion for the existence of a two-body bound state with a center-of-mass wavenumber $K$
is always fulfilled for $U \gg \gamma_0$.
In this regime of strong atom--atom interactions, 
the two-body bound state is tightly confined with respect to the relative coordinate, 
\ie, 
\begin{equation}
 \Psi^{(K, \mathrm{BS})}_x = \delta_{\abssmall{x},1}~~~~~(U \gg \gamma_0)
 \mathcomma
\end{equation}
describing a composite two-excitation object moving along
the lattice. 
Because only neighboring sites are occupied here, the
minimal spatial separation between two excitations is given by the lattice 
constant $a$.

The nearest neighbor interaction~$U$ can be understood as the discrete variant 
of a $\delta$-like potential.
Therefore, for each center-of-mass wavenumber $K$, there is at most one 
bound state in the spectrum.
There are no bound states in the case of non-interacting atoms ($U=0$).

When comparing the complex eigenenergies of scattering states and bound states,
we find that two excitations approximately all decay at a rate of 
$\Gamma^{K \nu}_{\mathrm{tot}} \simeq 2 \gamma_0$ (for both
$\nu=p$ and $\nu = \mathrm{BS}$).
However, the bound states' energies are detached from the quasicontinuum of scattering states
due to an interaction induced shift which is on the order of $U$ (see also \figref{fig:disprels}).

\subsection{Collective Dipole Moments and Momentum Distribution}
\begin{figure*}[thp]
 \centering
  \includegraphics[width=\textwidth]{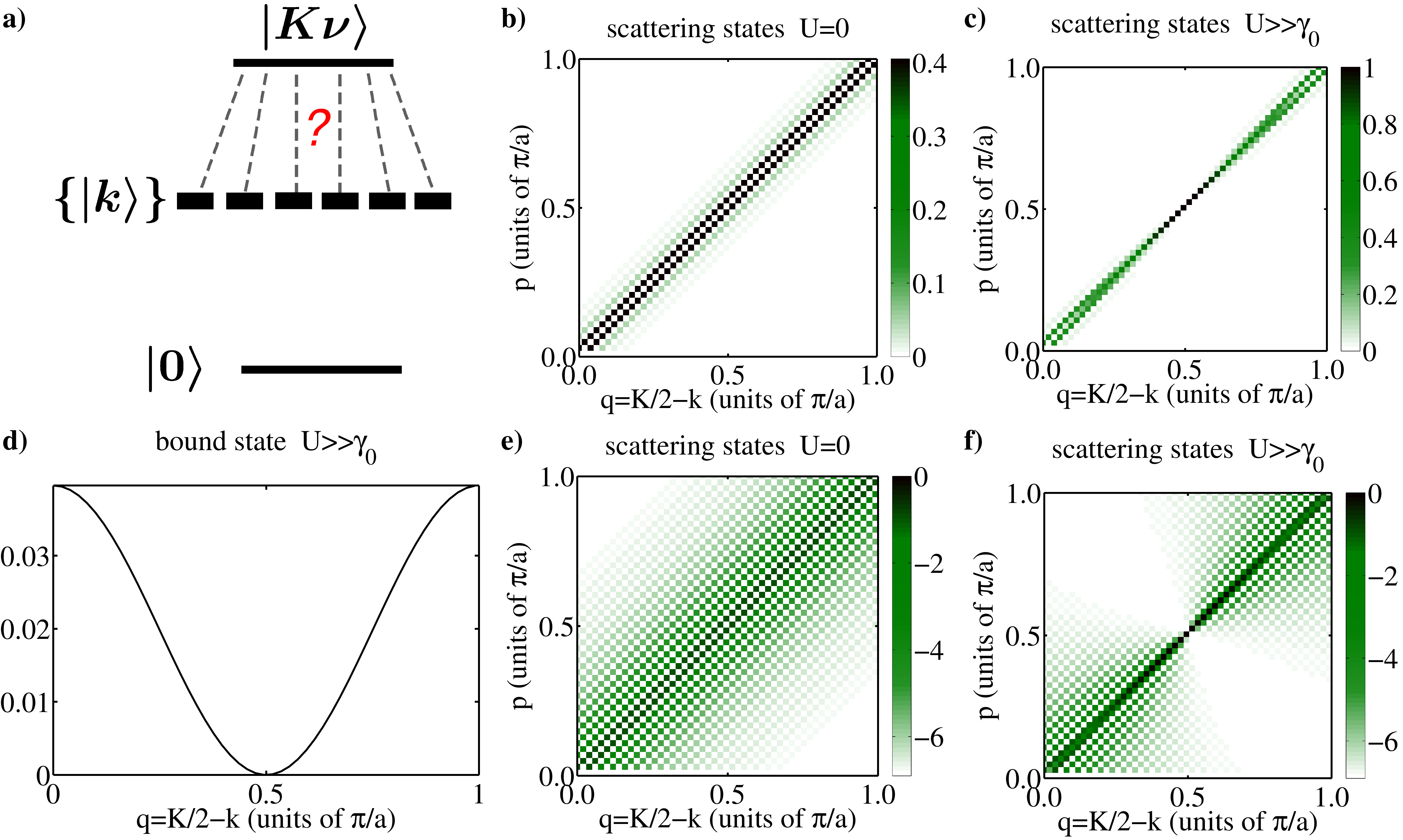}
  \caption{\label{fig:momdists}
	    (color online).
	    a)
	    How does a two-excitation state $\ketsmall{K \nu}$ decay into the
	    sub-manifold of single-excitation states $\{ \ketsmall{k} \}$? 
	    This can be answered with the help of the matrix elements of 
	    the collective dipole moment operator, which are intimately
	    linked to the relative wavefunction's momentum distribution
	    $\abssmall{\bar{\eta}^{(K \nu)}_{q}}^2$.
	    b)
	    Momentum distribution $\abssmall{\bar{\eta}^{(K p)}_{q}}^2$ 
	    of the relative wavefunction for scattering states ($M=101$)
	    for $U=0$ and 
	    c)
	    $U \gg \gamma_0$.
	    d)
	    Momentum distribution $\abssmall{\bar{\eta}^{(K, \mathrm{BS})}_{q}}^2$ of a bound state  
	    for $U \gg \gamma_0$ according to \Eqref{eq:etaUlargeBS}.
	    e), f)
	    Logarithmic plot of the scattering state's momentum distributions
	    b) and c)
	    (for better visualization, we have actually plotted
	    $\ln \left( \abssmall{\bar{\eta}^{(K p)}_{q}}^2 + c \right)$
	    with $c=10^{-3}$).}
\end{figure*}

We now turn to the problem of how in detail a two-excitation eigenstate
decays (\cf \figref{fig:momdists}a)) and show that the relevant key quantity here is 
intimately linked to the state's momentum distribution.

\subsubsection{Dipole Moments and Branching Ratio}
As the effective Hamiltonian~(\ref{eq:Heff}) lacks the detailed information
of the various possible decay paths (degrees of freedom have been integrated out),
we need to consider the collective operator
\begin{eqnarray}
 \label{eq:collop}
 \nonumber
 D^- &=& \sum_{\mu} \sum_{y=-\frac{N}{2}}^{\frac{N}{2}} D^-_{y \mu} \mathcomma \\
 D^-_{y \mu}  
	      &=& g^{*}_{y \mu} \sigma^-_y a^{\dagger}_{\mu} \\ \nonumber
	      &=& g^{*}_{\mu} \frac{ \ee^{- \ii \mu a y } }{\sqrt{M}} 
			\sigma^-_y a^{\dagger}_{\mu} \mathcomma
\end{eqnarray}
which induces one-photon transitions.
Here, $g_{\mu}$ signifies the atom--photon coupling strength 
(which can be considered being wave number independent across
the spectral window that is relevant here), and $\mu$ denotes the wavenumber 
that is transferred to the photon field.
This collective operator transfers a collective $n$-excitation atomic state
to the sub-manifold of $n-1$ excitations, which is accompanied by the emission
of a photon.

To identify the possible decay paths, we 
calculate the transition matrix elements
$\sum_f \brasmall{f} D^- \ketsmall{i}$
with respect to the non-dissipative eigenstates 
(as explained in the introduction of \secref{subsec:eigenstates}).
As an initial state, we take the direct product of the photon reservoir being in the vacuum state
and the atomic system being in a two-excitation eigenstate, \ie,
$\ketsmall{i} = \ketsmall{K \nu} \otimes \ketsmall{\{0\}}$.
Consequently, the target states
$\ketsmall{f} = \ketsmall{k} \otimes \ketsmall{1_{\mu}}$
consist of an atomic single-excitation state 
and a single photon (in mode $\mu$).
Explicit calculation reveals that 
\begin{eqnarray}
 \nonumber
 \sum_f \brasmall{f} D^- \ketsmall{i} & = & 
 \sum_{\mu} \sum_{y} g^{*}_{\mu} \frac{ \ee^{- \ii \mu a y } }{\sqrt{M}}
 \brasmall{k} \sigma^-_y \ketsmall{K \nu}
 \\
 &=&
 g^*_{K-k} \bar{\eta}^{(K \nu)}_{\frac{K}{2}-k} 
 \mathcomma
\end{eqnarray}
where the quantity $\bar{\eta}^{(K \nu)}_{\frac{K}{2}-k}$
(for details, we refer to \secref{sec:eta})
can be interpreted as a collective dipole moment.
This quantity (which will be discussed later in this Section)
can be used to define the branching ratio $b^{(K \nu)}_k$ 
for the decay $\ketsmall{K \nu} \rightarrow \ketsmall{k}$ via
\begin{eqnarray}
 \label{eq:defbranchrat}
 \nonumber
 b^{(K \nu)}_k 
      &\equiv& 
      \frac{\abs{g_{K-k}}^2 \abs{\bar{\eta}^{(K \nu)}_{\frac{K}{2}-k}}^2 }{\displaystyle \sum_{q=-\frac{\pi}{a}}^{\frac{\pi}{a}} \abs{g_{K-q}}^2 \abs{\bar{\eta}^{(K \nu)}_{\frac{K}{2}-q}}^2} \\
      &\simeq& \frac{\abs{\bar{\eta}^{(K \nu)}_{\frac{K}{2}-k}}^2 }{\displaystyle \sum_{q=\frac{K}{2}-\frac{\pi}{a}}^{\frac{K}{2}+\frac{\pi}{a}} \abs{\bar{\eta}^{(K \nu)}_q}^2}
      = \frac{1}{2} \cdot \abs{\bar{\eta}^{(K \nu)}_{\frac{K}{2}-k}}^2
      \mathperiod
\end{eqnarray}
In the second last step, we have exploited that the coupling strength is practically 
constant over the spectral range that corresponds to the wavenumbers involved.
Also note that 
$\sum_{q} \abssmall{\bar{\eta}^{(K \nu)}_q}^2 = 2$ holds (see \secref{sec:eta}),
which results from the branching ratio's sum over all decay channels $\sum_k b^{(K \nu)}_k=1$.
The partial rate for the decay $\ketsmall{K \nu} \rightarrow \ketsmall{k}$
is therefore 
\begin{equation}
 \Gamma^{K \nu}_k \equiv b^{K \nu}_k \cdot \Gamma^{K \nu}_{\mathrm{tot}} 
    = \frac{1}{2} \abs{\bar{\eta}^{(K \nu)}_{\frac{K}{2}-k}}^2 \cdot \Gamma^{K \nu}_{\mathrm{tot}}
 \mathperiod
\end{equation}
Hence, the collective dipole moment determines the branching ratio.

\subsubsection{Momentum Distribution, Direct and Background Fluorescence}
\label{subsubsec:dirbgetc}
The quantity $\abssmall{\bar{\eta}^{(K \nu)}_q}^2$
has yet another precise physical meaning.
The Fourier transform of the two-body wave function
(see \Eqref{eq:twobodyansatz0} for details),
\ie, 
\begin{eqnarray}
 \nonumber
 && 
 \sum_{n_1=-\frac{N}{2}}^{\frac{N}{2}} 
 \sum_{n_2=-\frac{N}{2}}^{\frac{N}{2}}
 \frac{\ee^{-\ii k_1 a n_1}}{\sqrt{M}} \frac{\ee^{-\ii k_2 a n_2}}{\sqrt{M}} 
 \Phi_{n_1 n_2}
 =  \\ \nonumber
 &&
 \frac{1}{2M^{\frac{3}{2}}}
 \sum_{n_1=-\frac{N}{2}}^{\frac{N}{2}} 
 \sum_{n_2=-\frac{N}{2}}^{\frac{N}{2}}
 \ee^{\ii \left( \frac{K}{2}-k_1 \right)a n_1}
 \ee^{\ii \left( \frac{K}{2}-k_2 \right)a n_2}
 \Psi^{(K \nu)}_{n_1-n_2}
 =
 \\ 
 &&
 ~~~ \frac{1}{2} \bar{\eta}^{(K\nu)}_{\frac{1}{2}\left( k_1-k_2 \right)} ~ \delta_{ \left[ K-k_1-k_2 \right]_{\frac{2\pi}{a}} , 0}
 \mathcomma
\end{eqnarray}
demonstrates that $\abssmall{\bar{\eta}^{(K \nu)}_q}^2$ can indeed be interpreted as the 
momentum distribution of the relative wave function with respect to a relative momentum $q$.
The Kronecker-$\delta$ expresses the conservation of the center-of-mass wavenumber and the
notation $\left[ a \right]_{b}$ stands for \anfz{$a$ modulo $b$}.
As a result,
the information about the individual decay paths of a two excitation eigenstate,
\ie, collective dipole moments and branching ratios,
are fully encoded in the momentum distribution of the eigenstate's relative wavefunction.

The explicit expressions are given as ($q \geq 0$ and $p>0$, see \appref{sec:eta} for details)
\begin{eqnarray}
 \label{eq:etascattredgen}
 \bar{\eta}^{(K p)}_{q} &=&
 \begin{cases}
  \frac{1+\ee^{\ii \delta_{Kp}}}{2} 
  & 
  p = q \\
  \frac{2}{M} \ee^{\ii \frac{\delta_{Kp}}{2}}
  \frac{\sin\left[ \frac{1}{2} \left( \delta_{Kp} - \left(p-q\right)a \right) \right]}{\sin\left[\frac{1}{2}\left(p-q\right)a\right]}
  & 
 \left[\frac{\left(p-q\right)aM}{2\pi}\right]_{2} \neq 0
 \end{cases}
 \mathcomma
 \\
 \label{eq:etaBSred}
 \bar{\eta}^{(K, \mathrm{BS})}_{q} &=& \frac{2}{\sqrt{M}} \cdot \frac{\cos(qa) - \alpha_K}{1-2\alpha_K \cos(qa) + \alpha_K^2}
 \mathperiod
\end{eqnarray}
For the case of non-interacting atoms ($U=0$), we have
\begin{eqnarray}
 \label{eq:etaU0}
 \abs{\bar{\eta}^{(K, p>0)}_{q}}^2 &=&
 \begin{cases}
  0
  & 
  p = q \\
  \frac{4}{M^2} \frac{1}{\tan^2\left[\frac{1}{2}\left(p-q\right)a\right]} 
  & 
  \left[\frac{\left(p-q\right)aM}{2\pi}\right]_{2} \neq 0
 \end{cases}
 \mathcomma
\end{eqnarray}
whereas in the regime of strong atom--atom interactions ($U \gg \gamma_0$) 
we are left with
\begin{eqnarray}
 \label{eq:etaUlargscatt}
 \abs{\bar{\eta}^{(K, p>0)}_{q}}^2 &=&
 \begin{cases}
  \sin^2\left(pa\right)
  & 
  p = q \\
  \frac{4}{M^2} \frac{1+ \cos\left[\left(p+q\right)a\right]}{1- \cos\left[\left(p-q\right)a\right]} 
  & 
  \left[\frac{\left(p-q\right)aM}{2\pi}\right]_{2} \neq 0
 \end{cases}
 \mathcomma \\
 \label{eq:etaUlargeBS}
 \abs{\bar{\eta}^{(K, \mathrm{BS})}_{q}}^2 &=& \frac{4}{M} \cos^2\left(qa\right)
 \mathperiod
\end{eqnarray}
Note that the symmetry properties 
$\abssmall{\bar{\eta}^{(K p)}_{q}}^2 = \abssmall{\bar{\eta}^{(K p)}_{-q}}^2$
and
$\abssmall{\bar{\eta}^{(K p)}_{q}}^2 = \abssmall{\bar{\eta}^{(K, -p)}_{q}}^2$ hold.

In \figref{fig:momdists}b)--d), we plot the relative wavefunction's momentum 
distributions~(\ref{eq:etaU0})--(\ref{eq:etaUlargeBS}) on a linear scale.
These figures reveal many details of the two-excitation states' properties and their
respective decay channels.
We start the discussion with a scattering state $\ketsmall{K p}$ for the case 
of non-interacting atoms ($U=0$).
Given the quantum numbers $K$ and $p$, we can extract 
the values of $q = K/2-k$ for which the momentum distribution is non-zero
from \figref{fig:momdists}b).
This determines the wavenumbers $\{k\}$ of the intermediate single-excitation 
states that contribute to the overall decay 
$\ketsmall{Kp} \rightarrow \{ \ketsmall{k} \} \rightarrow \ketsmall{0}$.
Since the main diagonal in \figref{fig:momdists}b) is strictly zero, we 
can conclude that the decay channel with $\pm q=p$ is fully suppressed
for $U=0$.
Upon expressing the center-of-mass and relative wavenumbers in terms of 
single-excitation wavenumbers $k_1$ and $k_2$ 
(\ie, $K=k_1+k_1$, $p=(k_1-k_2)/2$), it becomes apparent that 
this \anfz{direct} channel with $\pm q=p$ (in the following also termed 
\anfz{direct fluorescence}) stands for decay processes where the intermediate
single-excitation states are $\ketsmall{k_1}$ and $\ketsmall{k_2}$.
In other words, the direct fluorescence refers to decay processes for which 
the single-excitation wavenumbers are individually conserved.
This suppression which occurs for $U=0$ is a consequence of the interaction-induced
phase shift $\ee^{\ii \delta_{Kp}}=-1$ between the two excitations 
(see Eqs.~(\ref{eq:phaseshiftclosedsys}) and~(\ref{eq:etascattredgen})).
Even though the atoms are non-interacting, the two excitations experience an
infinitely strong on-site repulsion due to the hard-core constraint of spin-$1/2$
excitations ($\Psi^{(K \nu)}_0=0$, see also \secref{subsubsec:twoexcst}),
ultimately leading to the suppression of the direct channel.
In addition to that,
we have a \anfz{background} channel (also termed \anfz{background} 
fluorescence) for $\pm q \neq p$, yielding non-zero entries on the 
secondary diagonals in \figref{fig:momdists}b) (see also \figref{fig:momdists}e) 
for a logarithmic scale).
The background fluorescence plays the dominant role for $U=0$ and represents
decay channels which do not conserve the individual single-excitation wavenumbers
(the intermediate single-excitation states exhibit wavenumbers $k \neq k_1, k_2$).

Conversely, the details of a two-excitation scattering state's decay in
the regime $U \gg \gamma_0$ depend on the relative wavenumber $p$.
From the relative wavefunction's momentum distribution in \figref{fig:momdists}c)
(see also \figref{fig:momdists}f) for a logarithmic scale),
we infer that both direct and background fluorescence channels can occur.
Specifically, at $p=\pi/2a$ the background channel is fully suppressed and only 
the direct fluorescence contributes.
However, for relative wavenumbers away from $p=\pi/2a$, a combination of
direct and background terms can be observed. 
Ultimately, close to $p=0$ or $p=\pi/a$, the background fluorescence dominates
and \figref{fig:momdists}c) resembles \figref{fig:momdists}b) (note the different scales).
On the whole, the relative wavefunction's momentum distribution for a scattering state
in the regime $U \gg \gamma_0$ exhibits features as a function of the relative wavenumber $p$,
whereas the regime of non-interacting atoms ($U=0$) yields a featureless distribution 
(\cf \figref{fig:momdists}b) and~e)).

Finally, we discuss the respective momentum distribution of a bound 
state $\ketsmall{K, \mathrm{BS}}$ for $U \gg \gamma_0$.
For a bound state, the concepts of \anfz{direct} and \anfz{background fluorescence} 
do not exist since
the concept of a relative wavenumber $p$ as a quantum number does not apply
(a bound state is fully characterized by its center-of-mass wavenumber $K$).
In \figref{fig:momdists}d) we see that the momentum distribution is a broad function
in momentum space (\ie, non-zero over a wide range of $q$).
Hence, there are many intermediate single-excitation states to which a two-body bound state can decay.
This property clearly contrasts a bound state from scattering states and will eventually 
lead to the characteristic and distinct features in the emission patterns which we discuss
later.

\subsection{Far-Field Observables}
\label{subsec:ffobs}
In this paper, we focus on the far-field signatures that emerge from the light that is 
scattered by the atoms.
The Glauber decomposition of the electric field operator for the scattered far field 
can be written as \cite{glauber,mandelwolf}
\begin{equation}
 \label{eq:eopfarfield}
 \hat{\vv{E}}^{(-)}(\vv{r},t) =  \xi \vv{w}(\vv{r}) \sum_n \sigma^{+}_{n}(t-t_n) \mathcomma
\end{equation}
where
$\xi  = \omega_{0}^2/4 \pi \epsilon_0 c^2$ ($\epsilon_0$ is the vacuum permittivity),
\begin{equation}
 \vv{w}(\vv{r}) =  \frac{\vv{d}}{r} - \left(\vv{d} \cdot \vv{r}\right) \frac{\vv{r}}{r^3} 
 ~~~~~(r=\abssmall{\vv{r}})
\end{equation}
signifies the far-field pattern of a single dipole,
and $\hat{\vv{E}}^{(+)} = \left( \hat{\vv{E}}^{(-)} \right)^{\dagger}$.
The retarded times in the argument of the atomic operators in \Eqref{eq:eopfarfield} are
\begin{eqnarray}
 t_n &=& \frac{1}{c} \abssmall{\vv{r}-\vv{r}_n} = t_n(\beta_{\mathrm{det}}) \\
     &\simeq& \frac{r}{c} - \frac{a}{c} \sin\left(\beta_{\mathrm{det}}\right) \cdot n
 \mathcomma
\end{eqnarray}
where the last line represents the far-field approximation.
Here, $\vv{r}_n = na \hat{\vv{e}}_z$
and $\beta_{\mathrm{det}}$ denotes the elevation coordinate of the 
detector position 
$\vv{r}=r ( \cos\beta_{\mathrm{det}} \cos\varphi_{\mathrm{det}} , \cos\beta_{\mathrm{det}} \sin\varphi_{\mathrm{det}} , \sin\beta_{\mathrm{det}} )$ 
($\beta_{\mathrm{det}}=0$ signifies detection perpendicular to the atomic chain and
$\varphi_{\mathrm{det}}$ is the azimuthal angle).

We now transform to the basis of the system's eigenstates by virtue of the expansion
(for details, see \appref{sec:appexpineig})
\begin{eqnarray}
 \label{eq:usefulsigma}
 \sigma^{-}_n &=&
 \sum_k \ket{0} \bra{0} \sigma^{-}_n \ket{k} \bra{k}
 \\ \nonumber
    && ~~~ ~+~
  \sum_k \sum_{K \nu} \ket{k} \bra{k} \sigma^{-}_n \ket{K \nu} \bra{K \nu}
 \\ \nonumber
 &=&
 \frac{1}{\sqrt{M}} \sum_k \ee^{\ii ka n} \hat{S}_{0; k}
 \\ \nonumber
    && ~~~ ~+~ \frac{1}{2 \sqrt{M}} \sum_k \sum_{K \nu} 
      \ee^{\ii (K-k)a n} \bar{\eta}^{(K,\nu)}_{\frac{K}{2}-k} 
      \hat{S}_{k; K \nu} \mathcomma
\end{eqnarray}
where $\hat{S}_{0; k} = \ketsmall{0} \brasmall{k}$ and 
$\hat{S}_{k; K \nu} = \ketsmall{k} \brasmall{K \nu}$.
The quantum number $\nu$ runs over all scattering states ($\nu=p$)
and a possible bound state ($\nu=\mathrm{BS}$).
Assuming the \anfz{harmonic decomposition} \cite{ficek}
for the operators $\hat{S}_{0; k}$ and $\hat{S}_{k; K \nu}$, \ie,
\begin{eqnarray}
 \label{eq:harmdec1}
 \hat{S}_{k;0}\left(t-t_n(\beta_{\mathrm{det}}) \right) &\simeq& 
  \ee^{\ii \frac{\Delta^k_0}{c} a \sin\beta_{\mathrm{det}} \cdot n} \hat{S}_{k;0}\left( t_{\mathrm{ret}} \right) \mathcomma \\
 \label{eq:harmdec2}
 \hat{S}_{K \nu;k}\left(t-t_n(\beta_{\mathrm{det}}) \right) &\simeq& 
  \ee^{\ii \frac{\Delta^{K \nu}_k}{c} a \sin\beta_{\mathrm{det}} \cdot n} \hat{S}_{K \nu;k}\left( t_{\mathrm{ret}} \right) 
 \mathcomma
\end{eqnarray}
(where 
$\Delta^{K \nu}_{k} \equiv \mathrm{Re}(E^{(2)}_{K \nu} - E^{(1)}_{k})$ and
$\Delta^k_0 \equiv \mathrm{Re}(E^{(1)}_k)$ signify the transition energies
and $t_{\mathrm{ret}} \equiv t - r/c$ is the retarded time),
\Eqref{eq:eopfarfield} turns into 
\begin{eqnarray}
 \label{eq:opfarfield4}
 {\hat{\vv{E}}}^{(-)}(\vv{r},t) &=&
   \xi \vv{w}(\vv{r}) \sqrt{M} \times 
   \\ \nonumber
   && \hspace{-1.8cm} \sum_{k} 
		    \Bigg{(} \delta_{ k , \left[ \Delta^k_0 \sin\left(\beta_{\mathrm{det}}\right)/c \right]_{\left( 2\pi / a \right)} } 
			{\hat{S}}_{k;0}\left( t_{\mathrm{ret}} \right) \\ \nonumber
   && \hspace{-1.8cm} ~~~ ~+ \sum_{K \nu} 
			\delta_{ K-k , \left[ \Delta^{K \nu}_k \sin\left(\beta_{\mathrm{det}}\right)/c \right]_{\left( 2\pi / a \right)} } 
			\left( \bar{\eta}^{(K \nu)}_{\frac{K}{2}-k} \right)^*
			{\hat{S}}_{K \nu;k} \left( t_{\mathrm{ret}} \right)
      \Bigg{)}
   \mathperiod
\end{eqnarray}
The Kronecker terms represent a constraint which relates the quantum numbers of 
a one-photon transition to the corresponding observation angle~$\beta_{\mathrm{det}}$ via
\begin{eqnarray}
 \label{eq:qconstraint}
 k &=& \left[ \frac{\Delta^k_0}{c} \sin\beta_{\mathrm{det}} \right]_{\frac{2\pi}{a}} \mathcomma \\
 \label{eq:sconstraint}
 K - k &=& \left[ \frac{\Delta^{K \nu}_k}{c} \sin\beta_{\mathrm{det}} \right]_{\frac{2\pi}{a}} 
 \mathperiod
\end{eqnarray}
These expressions can be understood as specifying the wavenumbers that are transferred to 
the photon field under a given observation angle.
Alternatively, for given $K$, $\nu$, and/or $k$, we can determine the emission direction from these
equations.

Since $\omega_{0} \gg \gamma_0$, the Lamb shifts only 
yield a negligible wavenumber correction in Eqs.~(\ref{eq:qconstraint}) and~(\ref{eq:sconstraint})
and we can use the approximated expressions
\begin{eqnarray}
 \label{eq:qconstraintappr}
 k &\simeq& \left[ k_{\mathrm{at}} \sin\beta_{\mathrm{det}} \right]_{\frac{2\pi}{a}} \mathcomma \\
 \label{eq:sconstraintappr}
 K - k &\simeq& \left[ k_{\mathrm{at}} \sin\beta_{\mathrm{det}} \right]_{\frac{2\pi}{a}} \mathperiod
\end{eqnarray}
To be precise, the decay of single-excitation states 
$\ketsmall{k} \rightarrow \ketsmall{0}$
requires 
$ka = 2 \pi \left[ (a/\lambda_{\mathrm{at}}) \sin \beta_{\mathrm{det}} \right]_1$, 
whereas for 
$\ketsmall{K \nu} \rightarrow \ketsmall{k}$
we have 
$(K-k)a = 2 \pi \left[ (a/\lambda_{\mathrm{at}}) \sin \beta_{\mathrm{det}} \right]_1$ 
(which is valid for both $U=0$ and $\omega_{0} \gg U \gg \gamma_0$).
These expressions are reminiscent of Bragg's law and the emission angles are 
determined by matching the wave numbers transferred to the free-space photon field. 
For the remainder, 
\begin{equation}
 \label{eq:barkeq}
 \bar k = \bar{k}(\vv{r}) = \bar{k}(\beta_{\mathrm{det}}) = \left[ k_{\mathrm{at}} \sin\beta_{\mathrm{det}} \right]_{\frac{2\pi}{a}}
\end{equation}
denotes the wavenumber that can
be detected at $\vv{r}$ (which allows us to set $k= \bar k$ in the first and
$k = K - \bar k$ in the second sum of \Eqref{eq:opfarfield4}, respectively).
In terms of angles, we can rewrite this expression as
\begin{equation}
 \label{eq:barkangle}
 \sin\beta_{\mathrm{det}} = \frac{\bar{k}}{k_{\mathrm{at}}} + \frac{2 \pi}{k_{\mathrm{at}} a} \cdot n
 \mathcomma
\end{equation}
where $n=0,\pm 1,\dots$ (such that $\abssmall{\sin\beta_{\mathrm{det}}} \leq 1$) 
signifies the Bragg order.

From \Eqref{eq:opfarfield4}, we can now construct arbitrary far-field observables.
Specifically, the field--field auto-correlation function can be calculated from
\begin{equation}
 \hat{G}^{(1)}(\vv{r},t,t+\tau) \equiv {\hat{\vv{E}}}^{(-)}(\vv{r},t) ~ {\hat{\vv{E}}}^{(+)}(\vv{r},t+\tau)
 \mathperiod
\end{equation}
Evaluated at $\tau = 0$, we obtain the emitted intensity
\begin{eqnarray}
 \label{eq:fieldcorr}
 && \frac{\hat{G}^{(1)}(\vv{r},t)}{\xi^2 \abs{\vv{w}(\vv{r})}^2 M} =
 \\ \nonumber
 && ~~	      {\hat{S}}_{\bar k; \bar k}\left( t_{\mathrm{ret}} \right)  
	      + \sum_{K \nu \nu^\prime}
	      \bar{\eta}^{(K \nu^\prime)}_{\frac{K}{2}-\bar{k}} \left( \bar{\eta}^{(K \nu)}_{\frac{K}{2}-\bar{k}} \right)^*
	      {\hat{S}}_{K \nu; K \nu^\prime}\left( t_{\mathrm{ret}} \right) 
	  \mathperiod
\end{eqnarray}
For a stationary state, 
the emission spectrum is defined as \cite{orszag}
\begin{eqnarray}
 \label{eq:specdef}
 {S}(\vv{r},\omega) = 
 \lim_{t \rightarrow \infty}
 2 \mathrm{Re}\left[ \int \limits_{0}^{\infty} \mathrm{d} \tau ~ 
 \ee^{\ii \omega \tau}  
 \expvalsmall{ \hat{G}^{(1)}(\vv{r},t,t+\tau) } \right]
 \mathcomma
\end{eqnarray}
where 
$\expvalsmall{ \dots }$ denotes an expectation value.
We employ the quantum regression theorem \cite{mandelwolf}
for the calculation of the auto-correlation function's expectation value
(see \appref{sec:appautocorr} for details).

Furthermore, the intensity correlation 
\begin{equation}
 \hat{G}^{(2)}(\vv{r}_1,\vv{r}_2) \equiv 
    {\hat{\vv{E}}}^{(-)}(\vv{r}_1,t) ~ 
    \hat{G}^{(1)}(\vv{r}_2,t) ~
    {\hat{\vv{E}}}^{(+)}(\vv{r}_1,t)
\end{equation}
for zero time-delay and
two detector positions $\vv{r}_1$ and $\vv{r}_2$ can be obtained from 
\begin{eqnarray}	  
 \label{eq:intcorr}
 && \frac{ \hat{G}^{(2)}(\vv{r}_1,\vv{r}_2) }{ \xi^4 \abs{\vv{w}(\vv{r}_1)}^2 \abs{\vv{w}(\vv{r}_2)}^2 M^2 } =
 \\ \nonumber
 && ~~~~~~
  \sum_{\nu \nu^\prime} 
      \bar{\eta}^{(k_1+k_2 , \nu^\prime)}_{\frac{1}{2}\left(k_1-k_2\right)} \left( \bar{\eta}^{(k_1+k_2, \nu)}_{\frac{1}{2}\left(k_1-k_2\right)} \right)^*
      {\hat{S}}_{k_1+k_2, \nu; k_1+k_2, \nu^\prime}
 \mathcomma
\end{eqnarray}
where $k_1$ and $k_2$ are the wavenumbers detected at $\vv{r}_1$ and $\vv{r}_2$ 
(elevation angles $\beta_1$ and $\beta_2$), respectively.
The normalized intensity correlation
\begin{equation}
 \label{eq:intcorrnorm}
 g^{(2)}(\vv{r}_1,\vv{r}_2) = g^{(2)}(\beta_1,\beta_2) = \frac{ \expvalsmall{\hat{G}^{(2)}(\vv{r}_1,\vv{r}_2)} }{ \expvalsmall{\hat{G}^{(1)}(\vv{r}_1)} \expvalsmall{\hat{G}^{(1)}(\vv{r}_2)} }
\end{equation}
only depends on $\beta_1$ and $\beta_2$ and not on the details of 
the single-dipole patterns $\abs{\vv{w}(\vv{r}_1)}^2$ and $\abs{\vv{w}(\vv{r}_2)}^2$.

\section{Spontaneous Emission Dynamics}
\label{sec:spontemdyn}
In this Section, we investigate the far-field signatures that emerge in the context of 
spontaneous decay, assuming the system has been prepared in a (pure) eigenstate
at time $t=0$.

\subsection{Lindblad Equation}
\label{subsec:lindbladeq}
Based on the knowledge of the dissipative eigenstates, we formulate
a Lindblad equation in order to account for the full dynamics
that includes the decay from the two-excitation submanifold via
the single-excitation subspace to the vacuum.
To this end, we write the density matrix as
\begin{eqnarray}
 \label{eq:densmat}
 \hat{\varrho} &=& \sum_{K \nu} \sum_{K^\prime \nu^\prime} \varrho_{K \nu; K^\prime \nu^\prime} \ketsmall{K \nu} \brasmall{K^\prime \nu^\prime} 
	\\ \nonumber
	&& ~+~ \sum_{k k^\prime} \varrho_{k ; k^\prime} \ketsmall{k} \brasmall{k^\prime} 
	~+~ \varrho_{0; 0} \ketsmall{0} \brasmall{0} \\
 \nonumber
	&& ~+~ \sum_{K \nu} \sum_k \varrho_{K \nu; k}    \ketsmall{K \nu} \brasmall{k} + \mathrm{h.c.} \\
 \nonumber
	&& ~+~ \sum_{K \nu} \varrho_{K \nu; 0}    \ketsmall{K \nu} \brasmall{0} + \mathrm{h.c.} \\
 \nonumber
	&& ~+~ \sum_k \varrho_{k; 0} \ketsmall{k}     \brasmall{0} + \mathrm{h.c.}
	 \mathcomma
\end{eqnarray}
where we utilize the non-dissipative basis (formally realized by setting $\mathrm{Re}(\Gamma_0)=\mathrm{Re}(\Gamma_1)=0$) 
as explained in the introduction of \secref{subsec:eigenstates}.
The dynamics is governed by the Lindblad equation
\begin{equation}
 \label{eq:lindblad}
 \partial_t \hat{\varrho} = \ii \left[ \hat{\varrho} , H^\prime \right] + \mathcal{L}(\hat{\varrho}) \mathcomma
\end{equation}
where 
\begin{eqnarray}
 H^\prime &=& \sum_{K \nu} \mathrm{Re}(E^{(2)}_{K \nu}) ~\hat{S}_{K \nu; K \nu}
	    + \sum_k \mathrm{Re}(E^{(1)}_k)      ~\hat{S}_{k; k} \mathcomma \\
\label{eq:lindbladian}
 \nonumber
 \mathcal{L}(\hat{\varrho}) &=& 
      \sum_{K \nu} \sum_k
      \left[ R_{K \nu; k} \hat{\varrho} R^\dagger_{K \nu; k} 
	    - \frac{1}{2} \left\{ R^\dagger_{K \nu; k} R^{\phantom \dagger}_{K \nu; k} , \hat{\varrho} \right\} \right]
      \\
 && ~+~ 
 \sum_k 
      \left[ R_{k;0} \hat{\varrho} R^\dagger_{k;0} 
	    - \frac{1}{2} \left\{ R^\dagger_{k;0} R^{\phantom \dagger}_{k;0} , \hat{\varrho} \right\} \right]
      \mathperiod
\end{eqnarray}
The dissipators are given as
\begin{eqnarray}
 R_{K \nu; k} &\equiv& \sqrt{\Gamma^{K \nu}_k} \hat{S}_{K \nu; k}
 \mathcomma \\
 R_{k; 0} &\equiv& \sqrt{\Gamma_k} \hat{S}_{k;0}
\mathperiod
\end{eqnarray}
The curly brackets in \Eqref{eq:lindbladian} denote an anti-commutator. 

The resulting equations of motion read
\begin{eqnarray}
 \label{eq:eqstart}
 \partial_t \varrho_{K \nu;K^\prime \nu^\prime} &=& 
 \left[ - \frac{1}{2} \left( \Gamma^{K \nu}_{\mathrm{tot}}  +  \Gamma^{K^\prime \nu^\prime}_{\mathrm{tot}} \right) - \ii \Delta^{K \nu}_{K^\prime \nu^\prime} \right] ~ \varrho_{K \nu; K^\prime \nu^\prime}
 \mathcomma \\
 \partial_t \varrho_{k;k^\prime} &=& 
 \left[ - \frac{1}{2} \left( \Gamma_{k} + \Gamma_{k^\prime} \right) - \ii \Delta^k_{k^\prime} \right] ~ \varrho_{k; k^\prime}
 \\ \nonumber
  && ~ ~+~ \delta_{k, k^\prime} ~ \sum_{K \nu}
    \Gamma^{K \nu}_k \varrho_{K \nu; K \nu}
 \mathcomma \\
 \partial_t \varrho_{0;0} &=&
 \sum_k \Gamma_k \varrho_{k;k} 
 \mathcomma \\
 \partial_t \varrho_{K \nu;k} &=&
 \left[ -\ii \Delta^{K \nu}_k - \frac{1}{2} \left( \Gamma^{K \nu}_{\mathrm{tot}} + \Gamma_k \right) \right] \varrho_{K\nu; k}
 \mathcomma \\
 \partial_t \varrho_{K \nu;0} &=&
 \left[ -\ii \mathrm{Re}(E^{(2)}_{K \nu}) - \frac{1}{2} \Gamma^{K \nu}_{\mathrm{tot}} \right] \varrho_{K\nu; 0}
 \mathcomma \\
 \label{eq:eqend}
 \partial_t \varrho_{k;0} &=&
 \left[ -\ii \Delta^k_0 -\frac{1}{2} \Gamma_k \right] \varrho_{k;0}
 \mathcomma
\end{eqnarray}
where $\Delta^{K \nu}_{K^\prime \nu^\prime} \equiv \mathrm{Re}(E^{(2)}_{K \nu} - E^{(2)}_{K^\prime \nu^\prime})$
and $\Delta^{k}_{k^\prime} \equiv \mathrm{Re}(E^{(1)}_{k} \nolinebreak - \nolinebreak E^{(1)}_{k^\prime})$.
The solution to this set of equations is
\begin{eqnarray}
 \label{eq:sol1}
 \varrho_{K \nu; K^\prime \nu^\prime} (t)
  &=& \ee^{-\ii \Delta^{K \nu}_{K^\prime \nu^\prime} t} 
      \ee^{- \frac{1}{2} \left( \Gamma^{K \nu}_{\mathrm{tot}} + \Gamma^{K^\prime \nu^\prime}_{\mathrm{tot}} \right) t} \varrho_{K \nu; K^\prime \nu^\prime}(0) \mathcomma \\
 \nonumber
 \label{eq:sol2}
 \varrho_{k; k^\prime} (t)
  &=& 
  \\ \nonumber
  && \hspace{-1cm} \delta_{k k^\prime} \sum_{K \nu} \frac{\Gamma^{K \nu}_{k}}{\Gamma^{K \nu}_{\mathrm{tot}} - \Gamma_k} \left( \ee^{-\Gamma_{k} t}  -  \ee^{-\Gamma^{K \nu}_{\mathrm{tot}} t}  \right) \cdot \varrho_{K \nu; K \nu}(0) \\
  && \hspace{-1cm} ~ ~+~ \ee^{-\ii \Delta^{k}_{k^\prime} t}  
	 \ee^{- \frac{1}{2} \left( \Gamma_k + \Gamma_{k^\prime} \right) t} \varrho_{k;k^\prime}(0) \mathcomma \\
 \label{eq:sol3}
 \varrho_{0; 0}(t)
  &=& 1 - \sum_{K \nu} \varrho_{K \nu; K \nu}(t) - \sum_k \varrho_{k; k}(t) \mathcomma \\
 \label{eq:sol4}
 \varrho_{K \nu; k}(t)
  &=& \ee^{-\ii \Delta^{K \nu}_k t} \ee^{-\frac{1}{2} \left( \Gamma^{K \nu}_{\mathrm{tot}} + \Gamma_k \right) t } \varrho_{K \nu; k}(0) \mathcomma \\
 \label{eq:sol5}
 \varrho_{K \nu; 0}(t) 
  &=& \ee^{-\ii \mathrm{Re}(E^{(2)}_{K \nu}) t} \ee^{-\frac{1}{2} \Gamma^{K \nu}_{\mathrm{tot}} t} \varrho_{K \nu; 0}(0) \mathcomma \\
 \label{eq:sol6}
 \varrho_{k; 0}(t) 
  &=& \ee^{-\ii \Delta^k_0 t} \ee^{-\frac{1}{2} \Gamma_k t} \varrho_{k; 0}(0) 
 \mathperiod
\end{eqnarray}

For the initial condition of a pure eigenstate, the coherences are zero at all times.
In addition to that, for $\omega_0 \gg \gamma_0$, the above equations can be reduced to 
\begin{eqnarray}
 \varrho_{K \nu; K \nu} (t)
  &\simeq& \ee^{- 2 \gamma_0 t} \varrho_{K \nu; K \nu}(0) \mathcomma \\
  \nonumber
 \varrho_{k; k} (t)
  &\simeq& \sum_{K \nu} 2 b^{(K \nu)}_k \left( \ee^{-\gamma_0 t}  -  \ee^{-2 \gamma_0 t}  \right) ~\varrho_{K \nu; K \nu}(0) 
  \\ 
  && ~ ~+~ \ee^{- \gamma_0 t} ~\varrho_{k;k}(0)
 \mathperiod
\end{eqnarray}
To be precise, for an initial single-excitation state, \ie,
$\varrho_{k;k}(0)=1$ (for a single $k$), we simply find
\begin{equation}
 \label{eq:sol1ex}
 \varrho_{k;k}(t) \simeq \ee^{-\gamma_0 t}
 \mathperiod
\end{equation}
Similarly, for an initial two-excitation state ($\varrho_{K \nu; K \nu}(0) \nolinebreak = \nolinebreak 1$), 
the dynamics is
\begin{eqnarray}
 \label{eq:sol2exa}
 \varrho_{K \nu; K \nu} (t)
  &\simeq& \ee^{- 2 \gamma_0 t} \mathcomma \\
  \label{eq:sol2exb}
 \varrho_{k; k} (t)
  &\simeq& 2 b^{(K \nu)}_k \left( \ee^{-\gamma_0 t}  -  \ee^{-2 \gamma_0 t}  \right)
  \\ \nonumber
  &=&  \abs{\bar{\eta}^{(K \nu)}_{\frac{K}{2}-k}}^2 \left( \ee^{-\gamma_0 t}  -  \ee^{-2 \gamma_0 t}  \right)
 \mathperiod
\end{eqnarray}
In the next Section, we answer the question of
how this \anfz{internal} dynamics of the atomic system translates to signatures that can be detected in the 
optical far field.

\subsection{Emitted Intensity}
The intensity which is emitted by an initial eigenstate and detected in the far field 
can be obtained from the expectation value of \Eqref{eq:fieldcorr} via 
${G}^{(1)}(\vv{r},t) \equiv \expvalsmall{\hat{G}^{(1)}(\vv{r},t)} = \mathrm{tr}[\hat{G}^{(1)} \hat{\varrho}]$, \ie,
\begin{equation}
\frac{{G}^{(1)}(\vv{r},t)}{\xi^2 \abs{\vv{w}(\vv{r})}^2 M} =
	      {\varrho}_{\bar k; \bar k}\left( t_{\mathrm{ret}} \right)  
	      + 
	      \abs{\bar{\eta}^{(K \nu)}_{\frac{K}{2}-\bar{k}}}^2
	      {\varrho}_{K \nu; K \nu}\left( t_{\mathrm{ret}} \right) 
	  \mathperiod
\end{equation}

According to \Eqref{eq:sol1ex}, an initial single-excitation state $\ketsmall{k}$ 
yields the far-field intensity
\begin{equation}
\label{eq:G11exc}
\frac{{G}^{(1)}(\vv{r},t)}{\xi^2 \abs{\vv{w}(\vv{r})}^2 M}  =
  	      \ee^{-\gamma_0 t_{\mathrm{ret}}} \delta_{k , \left[ k_{\mathrm{at}} \sin_{\beta_{\mathrm{det}}} \right]_{(2\pi/a)}}
	  \mathcomma
\end{equation}
which can be observed for elevation angles (see also \Eqref{eq:barkangle})
\begin{equation}
 \label{eq:singleexcangle}
 \sin\beta_{\mathrm{det}} = \frac{k}{k_{\mathrm{at}}} + \frac{2 \pi}{k_{\mathrm{at}} a} \cdot n
 \mathperiod
\end{equation}
The spontaneous decay of a two-excitation state $\ketsmall{K \nu}$ according to
Eqs.~(\ref{eq:sol2exa}) and~(\ref{eq:sol2exb}) results in 
\begin{eqnarray}
\label{eq:emittintspontem}
 \nonumber
\frac{{G}^{(1)}(\vv{r},t)}{\xi^2 \abs{\vv{w}(\vv{r})}^2 M}  &=&
  \abs{\bar{\eta}^{(K \nu)}_{\frac{K}{2}-\bar{k}}}^2 \left( \ee^{-\gamma_0 t}  -  \ee^{-2 \gamma_0 t}  \right)
  + 
 \abs{\bar{\eta}^{(K \nu)}_{\frac{K}{2}-\bar{k}}}^2
   \ee^{- 2 \gamma_0 t}
 \\ 
 &=&
     \abs{\bar{\eta}^{(K \nu)}_{\frac{K}{2}-\bar{k}}}^2
     \ee^{-\gamma_0 t_{\mathrm{ret}}} 
\mathcomma
\end{eqnarray}
which is mono-exponential even though the overall decay is a two-step process 
($\ketsmall{K \nu} \rightarrow \ketsmall{\bar{k}} \rightarrow \ketsmall{0}$).
Hence, we observe a mono-exponential decay at a rate $\gamma_0$ for 
scattering states ($\nu=p$), bound states ($\nu=\mathrm{BS}$),
as well as for single-excitation states (\Eqref{eq:G11exc}).
In other words, the temporal decay as such cannot serve as a characteristic fingerprint 
for identifying a specific state.

We therefore now turn to the discussion of the angle dependent
emission pattern.
According to \Eqref{eq:emittintspontem}, the far-field intensity as a function of the 
detection angle $\beta_{\mathrm{det}}$ normalized to the single-dipole pattern and the atom
number (and, for simplicity, evaluated at a fixed time $t_{\mathrm{ret}}=0$) is completely 
determined by the
momentum distribution of the eigenstate's relative wavefunction.
In other words, the quantity ${G}^{(1)}(\vv{r}(\beta_{\mathrm{det}}))/{\xi^2 \abs{\vv{w}(\vv{r(\beta_{\mathrm{det}})})}^2 M}$ is in 
principle just another way of plotting the momentum distribution 
$\abssmall{\bar{\eta}^{(K \nu)}_{K/2-\bar{k}}}^2$.
Hence, Figs.~\ref{fig:momdists} already contain the information about the emission pattern
but we have to analyze this information as a function of the detection angle 
$\beta_{\mathrm{det}}$ 
(rather than as a function of $q=K/2-\bar{k}$).

\subsubsection{Emission Pattern of Scattering States}
\begin{figure*}[thp]
 \centering
  \includegraphics[width=\textwidth]{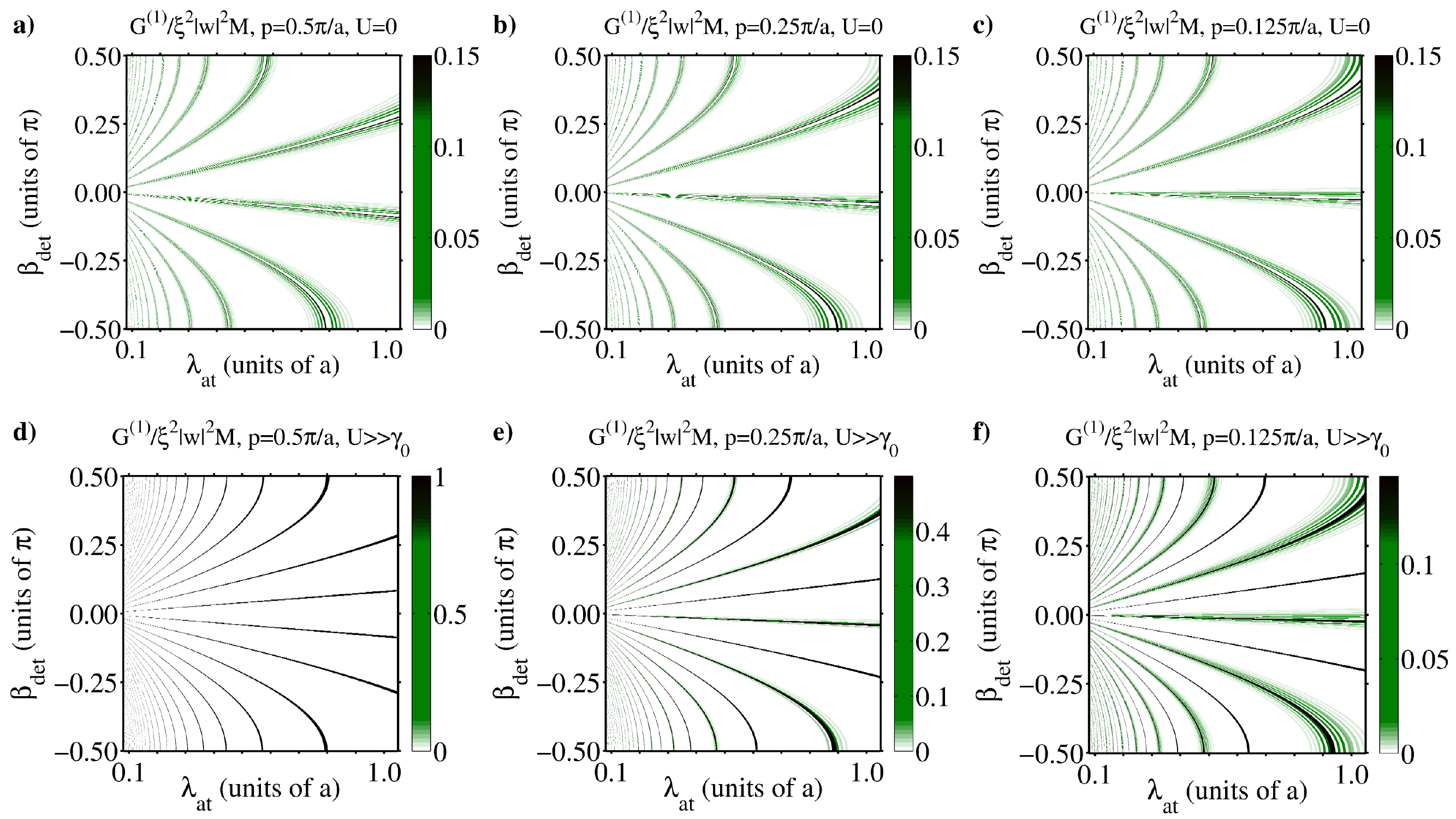}
  \caption{\label{fig:emissionpattern1}
	    (color online).
	    Spontaneous emission patterns according to \Eqref{eq:emittintspontem}
	    for an initial state $\ketsmall{K=0, p}$  
	    as a function of the detection angle $\beta_{\mathrm{det}}$
	    and the emission wavelength $\lambda_{\mathrm{at}}$ in units of $a$
	    ($M=101$).
	    Top row: $U=0$, bottom row: $U \gg \gamma_0$.
	    From left to right: $pa/\pi=1/2, 1/4, 1/8$.
	    Note the different scales of the colorbars.
	    }
\end{figure*}
In \figref{fig:emissionpattern1}, we depict the normalized emission pattern 
emerging from the decay of an initial scattering state $\ketsmall{K=0, p}$
for the two cases $U=0$ and $U \gg \gamma_0$ and for different relative 
wavenumbers $p$.
We visualize this pattern as a function of the detection angle $\beta_{\mathrm{det}}$ 
and use the emission wavelength $\lambda_{\mathrm{at}}$ (in units of $a$) as a parameter.

Similarly to \secref{subsubsec:dirbgetc}, we begin the discussion with 
the case of non-interacting atoms ($U=0$, a)--c) in \figref{fig:emissionpattern1}).
We notice that the \anfz{middle regions} of the emission peaks are zero.
This is a consequence of the background fluorescence which is the relevant 
decay mechanism for $U=0$.
These \anfz{dark} middle regions (which are surrounded by non-zero contributions) 
correspond to the direct channel (\ie, the case $\pm q=p$ 
in \secref{subsubsec:dirbgetc}) which is completely suppressed for $U=0$.
This observation is unique to all relative wavenumbers $p$ for $U=0$
(\figref{fig:emissionpattern1}a)--c))
since the momentum distributions in the case of non-interacting atoms are
featureless with respect the relative wavenumber (see the discussion in \secref{subsubsec:dirbgetc}).
In contrast to this, for $U \gg \gamma_0$, we found that the momentum distributions
can be qualitatively different for different relative wavenumbers since a 
combination of background and direct fluorescence can occur.
These properties manifest themselves in the emission patterns as depicted
in \figref{fig:emissionpattern1}d)--f).
For $p=\pi/2a$ (\figref{fig:emissionpattern1}d)), 
we observe single peaks as a function of the emission angle.
This is a direct consequence of the fact that the momentum distribution for $p=\pi/2a$
in the regime $U \gg \gamma_0$ exclusively exhibits the direct channel as an allowed
decay mechanism.
Hence, we observe emission peaks at exactly those angles which would correspond to the 
aforementioned \anfz{dark} middle regions for $U=0$.
However, as we choose relative wavenumbers away from $p=\pi/2a$
(\eg, $p=\pi/4a$ in \figref{fig:emissionpattern1}e) and 
$p=\pi/8a$ in \figref{fig:emissionpattern1}f)), light emitted via background fluorescence
becomes important.
Especially in \figref{fig:emissionpattern1}f), we can see that for $p=\pi/8a$
the emission pattern almost looks as in the case of non-interacting atoms (\figref{fig:emissionpattern1}c)),
except for the \anfz{middle region} of the emission peaks that stems from some remaining contribution 
of the direct fluorescence.
The relative intensities between direct and background fluorescence can be inferred
from the corresponding momentum distributions in Figs.~\ref{fig:momdists}b) and~c).

Based on \Eqref{eq:barkangle} 
we can also discuss the 
case where the bare atomic emission wavelength is much larger than the distance between 
the atoms, \ie, ${\lambda_{\mathrm{at}}}/{a} \rightarrow \infty$,
as well as the case of an isolated atom (${\lambda_{\mathrm{at}}}/{a} \rightarrow 0$).
However, one should keep in mind that our tight-binding model becomes invalid when
${\lambda_{\mathrm{at}}}/{a} \rightarrow \infty$ and one would have to sum up all 
dipole--dipole coupling terms for the eigenproblem in \secref{subsec:eigenstates} 
(not just the coupling to nearest neighbors).
From \Eqref{eq:barkangle}
we can deduce that,
in the limit ${\lambda_{\mathrm{at}}}/{a} \rightarrow \infty$,
the emission angles can only remain real-valued for $n=0$,
which means that only a single Bragg order would be observable.
In contrast to this, 
the opposite limit of an isolated atom (${\lambda_{\mathrm{at}}}/{a} \rightarrow 0$)
results in a continuum of possible emission angles as infinitely many Bragg orders are allowed.

\begin{figure}[thp]
 \centering
  \includegraphics[width=0.4\textwidth]{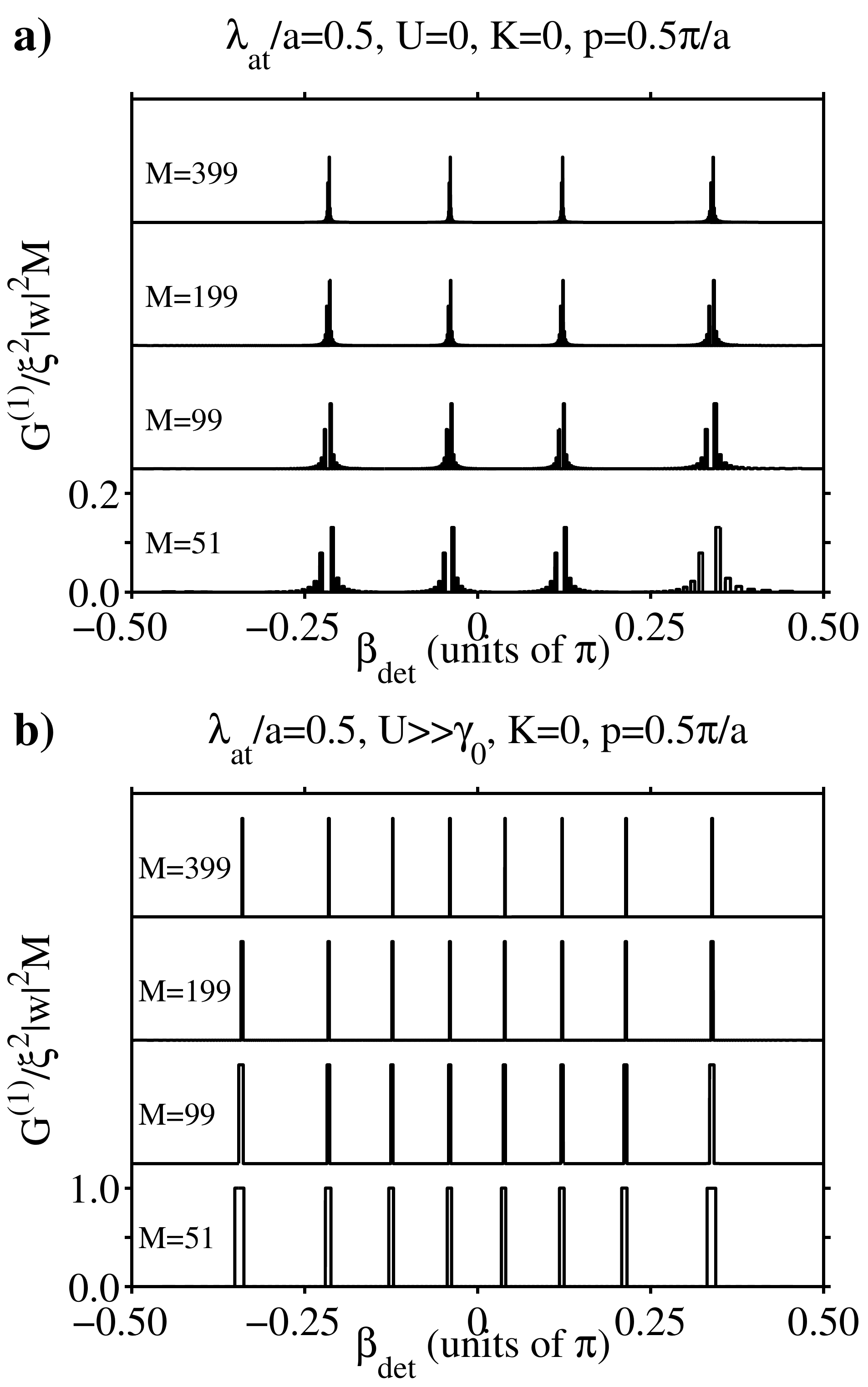}
  \caption{\label{fig:emissionpattern2}
	    Spontaneous emission patterns as in \figref{fig:emissionpattern1} 
	    for an initial scattering state $\ketsmall{K=0,p=\pi/2a}$
	    but for a fixed value of the emission wavelength $\lambda_{\mathrm{at}}/a=0.5$.
	    In essence, this figure represents a cut of 
	    Figs.~\ref{fig:emissionpattern1}a) and~d)
	    along $\lambda_{\mathrm{at}}/a=0.5$.
	    We have plotted the angle-dependent emission pattern 
	    for various atom numbers $M=51,99,199,399$.
	    Note that the $x$-axis is discrete and that we have plotted
	    each data point with a width $\delta\beta_{\mathrm{det}}(M)$
	    with respect to the $x$-axis.
	    }
\end{figure}
Finally, we utilize \figref{fig:emissionpattern2} to conclude the discussion with some
remarks on the width of the emission peaks.
Remember that the wavenumber spacing is $\delta k \equiv 2\pi/aM$.
Since the allowed wavenumbers are discrete, the resulting emission angles
are also discrete.
As an estimation for the resulting spacing of the detection angles, 
we can employ \Eqref{eq:sconstraintappr} and assume two transferred 
wavenumbers differing by $\delta k$, resulting in an angle difference of
$\delta \beta_{\mathrm{det}} \sim \delta k / k_{\mathrm{at}} = \lambda_{\mathrm{at}}/aM$.
However, in the limit of many atoms, 
the angles become infinitely sharp.
Hence, also the aforementioned \anfz{middle region} associated to the background fluorescence
becomes infinitesimally small.
Figure~\ref{fig:emissionpattern2} illustrates how the spacing of the discrete emission angles
affects the emission pattern of an initial scattering state $\ketsmall{K=0,p=\pi/2a}$ 
as the number of atoms $M$ increases.
As before, we have chosen the value of $p=\pi/2a$ for the relative wavenumber in \figref{fig:emissionpattern2} 
to highlight the differences between the background and direct fluorescence channels
for $U=0$ and $U \gg \gamma_0$, respectively.
For instance, note how the \anfz{middle region} of the emission peaks for $U=0$ 
becomes smaller as $M$ increases.

\subsubsection{Emission Pattern of Bound States}
\label{subsubsec:empattbs}
For a bound state $\ketsmall{K, \mathrm{BS}}$ in the presence of
strong atom--atom interactions $U \gg \gamma_0$, the far-field intensity can 
be written as (\cf Eqs.~(\ref{eq:etaUlargeBS}) and~(\ref{eq:emittintspontem}))
\begin{equation}
\label{eq:emittintBS}
\frac{{G}^{(1)}(\vv{r},t)}{\xi^2 \abs{\vv{w}(\vv{r})}^2}  =
  4 \cos^2\left( \frac{Ka}{2} - \frac{2\pi a}{\lambda_{\mathrm{at}}} \left[ \sin{\beta_{\mathrm{det}}} \right]_{\frac{\lambda_{\mathrm{at}}}{a}} \right)
  \ee^{-\gamma_0 t_{\mathrm{ret}}}
  \mathperiod
\end{equation}
Note that this expression does not depend on the number of atoms $M$,
which can be interpreted as a consequence of the bound state being spatially localized
(rather than delocalized which would cover all atoms).
\begin{figure*}[thp]
 \centering
 \includegraphics[width=\textwidth]{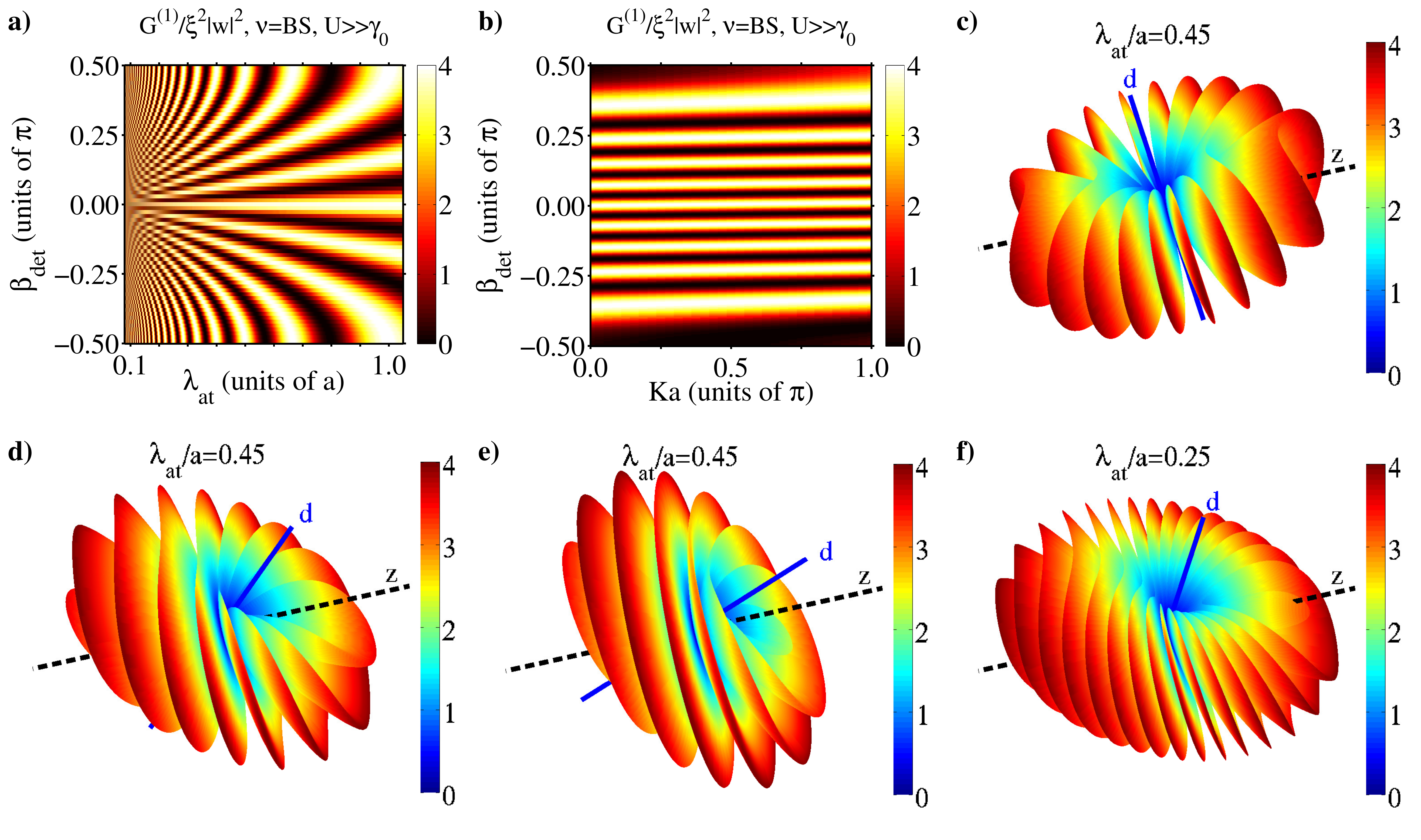}
  \caption{\label{fig:emissionpattern3}
	    (color online).
	    Spontaneous emission patterns according to \Eqref{eq:emittintBS}
	    (for $t_{\mathrm{ret}}=0$)
	    emerging from an initial bound state $\ketsmall{K, \mathrm{BS}}$.
	    a) Emission pattern (for $K=0$) normalized to the single-dipole pattern 
		as a function of the detection angle $\beta_{\mathrm{det}}$
		and the emission wavelength $\lambda_{\mathrm{at}}$ (in units of $a$).
		As $\lambda_{\mathrm{at}}/a$ increases, less Bragg orders become visible 
		and the width of the emission peaks increases.
	    b) Emission pattern as in a) but for $\lambda_{\mathrm{at}}/a = 0.45$
		as a function of the detection angle 
		and the center-of-mass wavenumber $K$.
	    The full emission pattern taking into account the single-dipole pattern 
	    is shown in c)--f) (we have plotted $G^{(1)} r^2 / \xi^2 \abssmall{d}^2$ 
	    for $K=0$ as a 3$D$ spherical plot, the colorbar indicates the emission strength).
	    The patterns exhibit a toroidal-like structure
	    (which is a remnant of the single-dipole pattern) 
	    with lobes resulting from the properties of the bound state's momentum distribution.
	    The angle between the single-atom dipole moment $\vv{d}$ (denoted by the blue line) 
	    and the atomic chain (dashed black line) is 
	    c) $\theta=\pi/2$,
	    d) $\theta=\pi/4$,
	    e) $\theta=\pi/8$, and
	    f) $\theta=\pi/3$.
	    }
\end{figure*}
\begin{figure*}[thp]
 \centering
  \includegraphics[width=\textwidth]{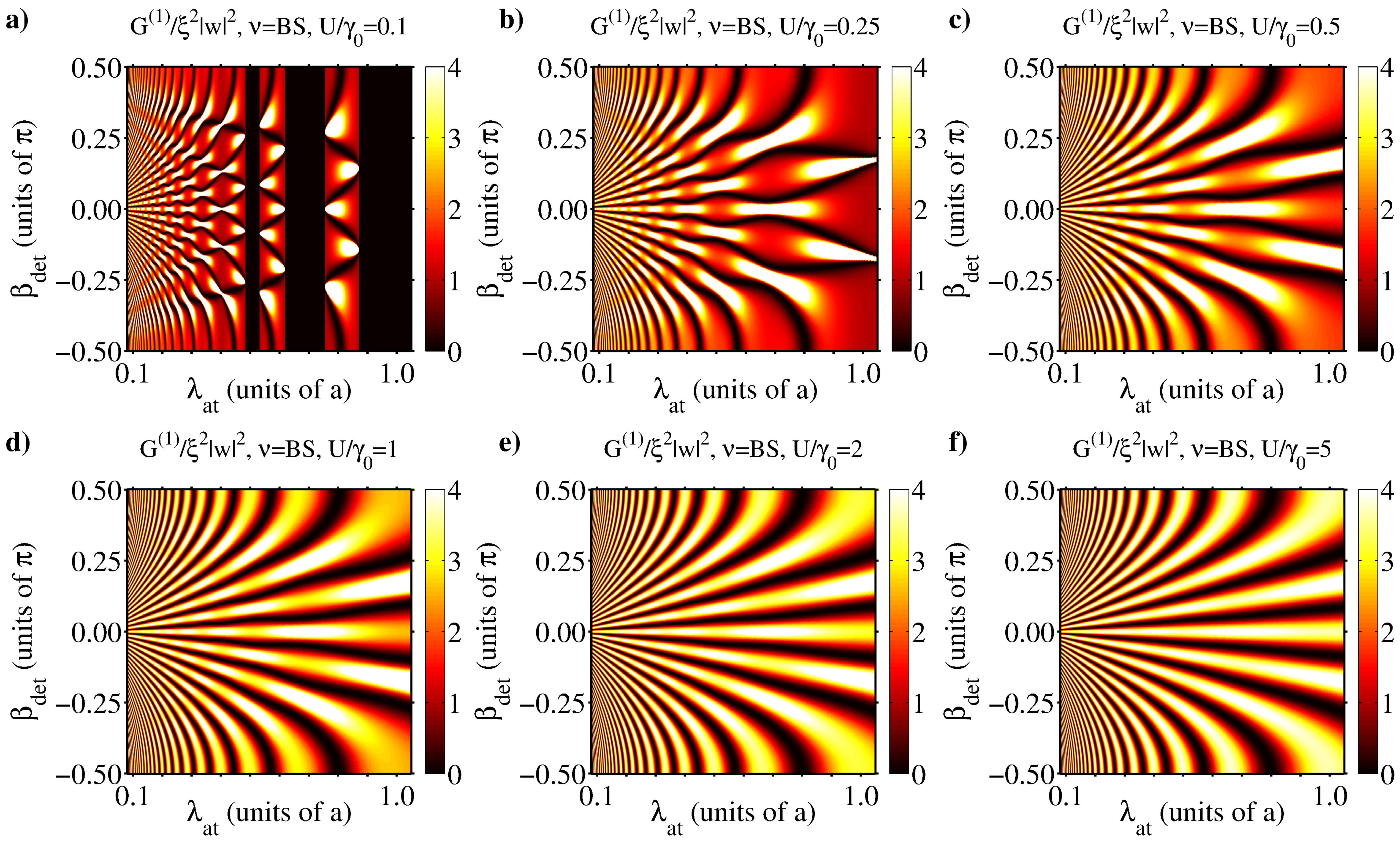}
  \caption{\label{fig:emissionpattern4}
	    (color online).
	    Normalized emission pattern emerging from a bound state $\ketsmall{K=0, \mathrm{BS}}$
	    as in \figref{fig:emissionpattern3}, but for different strengths
	    of the atom--atom interaction $U$
	    (atomic dipole moments perpendicular to the atomic chain).
	    Regions that are \anfz{dark} for any angle correspond to parameters
	    for which the criterion for the existence of a bound state is not satisfied.
	    }
\end{figure*}

Figure~\ref{fig:emissionpattern3} displays the bound state's emission pattern,
which can be regarded as a distinct feature for proofing the existence 
of a bound state on the lattice.
The bound state sets a minimal spatial scale given by the relative wavefunction's
spatial extent. 
For $U \gg \gamma_0$, this is simply the lattice constant $a$.
Therefore, the momentum distribution covers a finite window in momentum space
(on the order of $\delta k \sim 2\pi/a$), 
which translates into emission peaks having a finite width
$\delta \beta_{\mathrm{det}} \sim \lambda_{\mathrm{at}}/a$.
This width is independent of the number of atoms $M$
and it is larger than the emission angle spacing due to a finite lattice
(see discussion in the context of scattering states).
This property is in stark contrast
to the sharp peaks that would be observed for delocalized
single-excitation states or for two-excitation scattering states.
In \figref{fig:emissionpattern4}, we additionally display the influence of the strength 
of the atom--atom interaction~$U$, where
we utilize \Eqref{eq:emittintspontem} together with \Eqref{eq:etaBSred}.
Already moderate values of $U/\gamma_0$ realize the result we obtained for $U \gg \gamma_0$
(\eg, see $U/\gamma_0=5$ in \figref{fig:emissionpattern4}), justifying our previous assumptions.
Note that \figref{fig:emissionpattern4}a) depicts the special situation where the 
criterion~(\ref{eq:critforexist}) for the existence of a bound state is not fulfilled 
for certain of the shown values of $\lambda_{\mathrm{at}}/a$.
Consequently, no light emission from a bound state can be observed in these regions 
which are \anfz{dark} across all detection angles $\beta_{\mathrm{det}}$.
For $K=0$ and $U>0$ criterion~(\ref{eq:critforexist}) simply reads $\abs{\mathrm{Im}(\Gamma_1)} < U$.
Since the quantity $\mathrm{Im}(\Gamma_1)$ displays an oscillatory behavior as a function 
of $k_{\mathrm{at}} a = 2 \pi a / \lambda_{\mathrm{at}}$ 
(\cf Eqs.~({\ref{eq:fullrates}})--({\ref{eq:fullrateslast}})), 
a \anfz{bright} region in between two \anfz{dark} regions 
can be observed (as a function of $\lambda_{\mathrm{at}}/a$) in \figref{fig:emissionpattern4}a).

In conclusion, the angle-dependent far-field spontaneous emission pattern provides a
distinct signature for identifying and studying the individual single- and two-excitation
eigenstates.
Moreover, without the need to address and/or manipulate single atoms, we gain access to the
relative wavefunction's complete momentum distribution (see also \Eqref{eq:emittintspontem}).
This is achieved by tuning the argument $q = K/2 - \bar{k}(\beta_{\mathrm{det}})$
across the first Brillouin zone.
Given a value for $K$, this means we would need to sweep over the 
detection angles $\beta_{\mathrm{det}}$ for realizing different values of
$\bar k = \left[ k_{\mathrm{at}} \sin\beta_{\mathrm{det}} \right]_{2\pi/a}$.

\section{Steady-State Signatures under the Influence of a Weak Driving Field}
\label{sec:steadystatesigs}
All considerations in \secref{sec:spontemdyn} assumed the system to be 
prepared in a pure eigenstate at time $t=0$ and the far-field intensity
discussed is a consequence of the system's evolution in the absence of external 
driving fields.
The pulsed excitation of x-ray quantum optical systems~\cite{xraybook} already approximately realizes these conditions. 
For example, in the archetype setup of ${}^{57}$Fe M\"ossbauer nuclei driven by synchrotron radiation, the exciting synchrotron pulse has a duration of order 10-100ps, whereas the natural lifetime of the nuclei is about 140ns. Our study is of high significance for this field, since state-of-the-art experiments operate in the single-excitation regime, and have the capability of probing individual eigenstates via a delicate choice of the sample geometry~\cite{exp4,exp6,exp7,preprint}.  It would be highly desirable to make the double-excitation states studied in our paper accessible to those experiments.
However, since such constraints may be difficult with regard to experimental realizations in general, 
we now turn to the discussion of the system's response to an external driving field.
In particular, we focus on the steady state that emerges when the system is probed 
optically in a very simple way. 

\subsection{Rate Equations}
Specifically, we consider a weak (\ie, strongly attenuated) 
incoherent driving field (\eg, pseudothermal light \cite{meschedebuch}) with a 
spatial plane-wave pattern.
Choosing the external field to be incoherent greatly simplifies the theoretical
description as we can resort to a set of rate equations rather than solving
the master equation with all coherences (see Eqs.~(\ref{eq:ssI}) and~(\ref{eq:ssII}) below).
Let $\abssmall{\mathcal{P}_n}^2$ be the pump rate of the driving field's 
$n$-th plane wave component whose spatial projection on the direction of the 
atomic chain is of the form $\exp({\ii k^{(z)}_n z})$, 
where $k^{(z)}_n$ signifies
the $z$-component of the driving field's wavevector $\vv{k}_n$ 
(magnitude $\abssmall{\vv{k}_n}=k_L =k_{\mathrm{at}}= 2\pi / \lambda_{\mathrm{at}}$).
This external wavenumber \anfz{imprints} the \anfz{internal} wavenumber
\begin{equation}
 k_n = \left[ k^{(z)}_n \right]_{\frac{2\pi}{a}} = \left[ k_L \sin{\beta^{(n)}_{\mathrm{exc}}} \right]_{\frac{2\pi}{a}}
\end{equation}
on the atomic system, where $\beta^{(n)}_{\mathrm{exc}}$ signifies the excitation angle 
($\vv{k}^{(n)}=k_L ( \cos{\beta^{(n)}_{\mathrm{exc}}} , 0 , \sin{\beta^{(n)}_{\mathrm{exc}}})$,
see also \figref{fig:overview}b)).
A weak drive with $\abssmall{\mathcal{P}_n}^2 \ll \gamma_0$ allows us to work in the 
truncated Hilbert space of at most two excitations.
Note that in practice the pump rates actually depend on the external field's angle of incidence 
since ultimately the projection of the pump's linearly polarized electric field 
on the atomic dipole moments is of relevance.
There is especially no coupling for angles where the dipole moment and the electric field are
perpendicular to each other.
In the following, this specific dependence of the pump rate on the angle of incidence
will not be taken into account explicitly. 
Rather, we assume the external field's parameters to be tunable such that the pump rate 
is adjusted for each angle and we effectively get angle-independent driving terms.

We can then extend Eqs.~(\ref{eq:eqstart})--(\ref{eq:eqend}) to account for the 
external driving field, ultimately yielding a set of coupled rate equations for the 
diagonal elements $N_m \equiv \varrho_{m;m}$ of the density matrix. 
The steady state can be determined through (see \appref{sec:eqominc} for details)
\begin{eqnarray}
 \label{eq:ssI}
 N_{K \nu} 
 &=&
 \sum_n \frac{\abs{\mathcal{P}_n}^2}{\Gamma^{K \nu}_{\mathrm{tot}}} \abs{\bar{\eta}^{(K,\nu)}_{\frac{K}{2} - k_n}}^2 N_{K-k_n}
 \mathcomma
 \\ 
 \label{eq:ssII}
 N_{k} 
 &=&
 \sum_n \frac{\abs{\mathcal{P}_n}^2}{\tilde{\Gamma}_k} \delta_{k k_n} \left( 1- \sum_k N_k - \sum_{K \nu} N_{K \nu} \right)
 \\ \nonumber
 && ~~~ ~+~ \sum_{K \nu} \frac{\Gamma^{K \nu}_k}{\tilde{\Gamma}_k} N_{K \nu}
 \mathperiod
\end{eqnarray}

We begin with the investigation of the atomic system's steady state that emerges 
for the simple case where the external field exhibits only a single spatial Fourier component.
In other words, we investigate the system's response to a single external pump.

\subsection{Single-Pump Setup}
\label{subsec:spump}
\begin{figure}[t]
 \centering
 \includegraphics[width=0.35\textwidth]{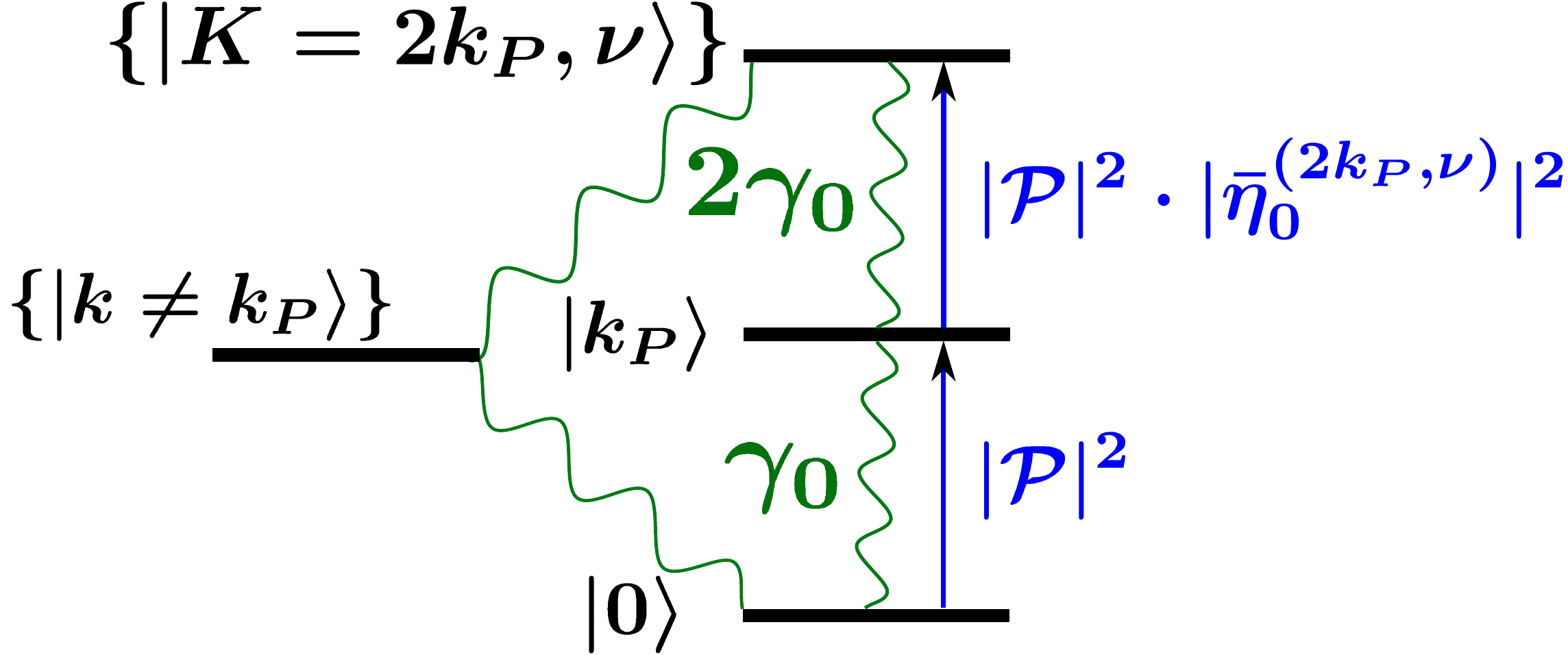}
 \caption{\label{fig:singlepumpsketch}
	    Driving the
	    1D lattice of two-level atoms 
	    with an incoherent drive as sketched in \figref{fig:overview}b),
	    reduces the Hilbert space to the levels shown here.
	    Since the external pump \anfz{imprints} the wave number $k_P$ on the atomic system 
	    the relevant Hilbert space comprises only two-excitation states with $K = 2 k_P$. 
	    }
\end{figure}
We envision an external driving field with a single spatial Fourier component
which \anfz{imprints} the wavenumber 
$k_P=[ k_L \sin{\beta^{(n)}_{\mathrm{exc}}} ]_{{2\pi}/{a}}$ on the 
atomic system (see Figs.~\ref{fig:overview}b) and~\ref{fig:singlepumpsketch}).
The pump rate is $\abssmall{\mathcal{P}}^2$ 
and for a weak pump we have $\Xi \equiv \abssmall{\mathcal{P}}^2 / \gamma_0 \ll 1$.
The steady-state occupation numbers according to 
Eqs.~(\ref{eq:ssI}) and~(\ref{eq:ssII}) are (see \appref{subsec:appsingpump} for details)
\begin{eqnarray}
 \label{eq:ssIappr}
 N_{k} &\simeq& 
  \Xi \delta_{k k_P} 
   + \Xi^2 \sum_{\nu} b^{(2 k_P, \nu)}_k \abs{\bar{\eta}^{(2 k_P, \nu)}_{0}}^2
   \\ \nonumber
   &=& 
  \Xi \delta_{k k_P} 
   + \frac{\Xi^2}{2} \sum_{\nu} \abs{\bar{\eta}^{(2 k_P, \nu)}_{k_P-k}}^2 \abs{\bar{\eta}^{(2 k_P, \nu)}_{0}}^2
  \mathcomma \\  
 \label{eq:ssIIappr}
 N_{K \nu} &\simeq& 
  \frac{\Xi^2}{2} 
  \abs{\bar{\eta}^{(2 k_P, \nu)}_{0}}^2
  \delta_{K, 2 k_P}
 \mathcomma
\end{eqnarray}
where we have assumed $\omega_0 \gg \gamma_0 \gg \abssmall{\mathcal{P}}^2$.
The external pump only directly drives single-excitation
states $\ketsmall{k=k_P}$ such that states $\ketsmall{k \neq k_P}$ are populated via
spontaneous emission of a two-excitation state $\ketsmall{K=2k_P, \nu}$,
which is of order $\Xi^2$. 
Furthermore, only two-excitation states with $K=2k_P$ are excited 
(see \figref{fig:singlepumpsketch}).

Even though the pump only excites certain wavenumbers, the 
steady state that results from this simple, incoherent excitation scheme is 
still a mixed state.
In the following, we analyze the signatures that emerge from this mixed state 
and investigate how they can be used for inferring information about the atomic system 
from the far field.

\subsubsection{Far-Field Intensity}
According to \Eqref{eq:fieldcorr}, 
the far-field intensity that emerges in the steady state is given by 
\begin{eqnarray}
 \frac{{G}^{(1)}(\vv{r})}{\xi^2 \abs{\vv{w}(\vv{r})}^2 M} 
 =  N_{\bar{k}} + \sum_{\nu} \abs{\bar{\eta}^{(2 k_P, \nu)}_{k_P-\bar{k}}}^2 N_{2 k_P, \nu} 
 \mathcomma
\end{eqnarray}
where, as before, $\bar k = [ k_{\mathrm{at}} \sin\beta_{\mathrm{det}} ]_{{2\pi}/{a}}$
(and $\beta_{\mathrm{det}}$ is the elevation angle of the detector position $\vv{r}$).
The intensity detected in the far field is thus always a sum of a 
contribution from a single-excitation state (first term),
two-excitation scattering states, 
\emph{and} a two-excitation bound state (if it exists).
Upon insertion of the steady-state solutions~(\ref{eq:ssIappr})--(\ref{eq:ssIIappr}),
we arrive at
\begin{equation}
 \label{eq:g1full}
  \frac{{G}^{(1)}(\vv{r})}{\xi^2 \abs{\vv{w}(\vv{r})}^2 M}  
   = \Xi \cdot \left(
  \delta_{\bar{k} k_P} 
   + \Xi \sum_{\nu} \abs{\bar{\eta}^{(2 k_P, \nu)}_{0}}^2 \abs{\bar{\eta}^{(2 k_P, \nu)}_{k_P-\bar{k}}}^2
   \right)
 \mathperiod
\end{equation}
A contribution that scales linearly with the pump power can only be observed for 
wavenumbers that match the wavenumber excited by the pump field ($\bar k=k_P$).
Varying the pump-power would in principle allow to identify and separate
the linear and quadratic terms in \Eqref{eq:g1full}.

Performing the sum over $\nu$ in \Eqref{eq:g1full} (see \appref{sec:appetasums} for details)
yields the explicit expression
\begin{eqnarray}
 \label{eq:g1full2}
  && \frac{{G}^{(1)}(\vv{r})}{\xi^2 \abs{\vv{w}(\vv{r})}^2 M}  
   = \Xi \delta_{\bar{k} k_P} 
   \\ \nonumber
   && ~+~ \Xi^2
   \begin{cases}
    \frac{1}{3} \delta_{\bar{k} k_P} 
    &
    \mathrm{for~}U=0
    \\
    \frac{1}{3} \delta_{\bar{k} k_P} + \frac{16}{M^2} \cos^2\left[\left(\bar{k}-k_P\right)a\right]
    &
    \mathrm{for~}U \gg \gamma_0
   \end{cases}       
 \mathperiod
\end{eqnarray}
Hence, for detected wavenumbers that are different from the wavenumber enforced by the pump
(\ie, $\bar{k} \neq k_P$), only light originating from the bound state contributes
(see also \appref{sec:appetasums} for details).
However, this contribution is suppressed in the limit of many atoms $M \gg 1$ 
(scaling as ${G}^{(1)} \propto 1/M$).
The only difference between the two cases $U=0$ and $U \gg \gamma_0$ with respect 
to the far-field intensity for this single-pump setup is thus a small correction 
to the $M\gg1$-limit that comes from the bound state.
Furthermore, the quantity~(\ref{eq:g1full2}) is a result of a sum over different states
since the steady state is a mixed state.
We therefore now turn to a frequency-specific far-field quantity---the emission spectrum.

\subsubsection{Emission Spectrum}
\label{subsubsec:emspecspump}
Using the definition~(\ref{eq:specdef}),  the emission spectrum reads
(see \appref{sec:appautocorr} for details on the calculation of the field--field autocorrelation's
expectation value via the quantum regression theorem)
\begin{eqnarray}
 \label{eq:specfinal}
 && \frac{S(\vv{r},\omega) \gamma_0}{2 \xi^2 \abs{\vv{w}(\vv{r})}^2 M} =
 \\ \nonumber
  && ~~~ \Xi 
  \Bigg{[} \frac{2}{\left( 2 \frac{\omega-\omega_0}{\gamma_0} \right)^2 + 1 } 
      \left( \delta_{\bar{k} k_P}  + \frac{\Xi}{2} \sum_{\nu} \abs{\bar{\eta}^{(2 k_P, \nu)}_{k_P-\bar{k}}}^2 \abs{\bar{\eta}^{(2 k_P, \nu)}_{0}}^2 \right) 
    \\ \nonumber
    && ~~~~~+~ \frac{\Xi}{3} \frac{1}{\left( \frac{2}{3} \frac{\omega-\omega_0}{\gamma_0} \right)^2 + 1 } 
    \sum_{p}
    \abs{ \bar{\eta}^{(2 k_P,p)}_{k_P-\bar{k}} }^2
    \abs{\bar{\eta}^{(2 k_P, p)}_{0}}^2 
    \\ \nonumber
    && ~~~~~+~ \frac{\Xi}{3} \frac{1}{\left( \frac{2}{3} \frac{\omega-\omega_0-U}{\gamma_0} \right)^2 + 1 } 
    \abs{ \bar{\eta}^{(2 k_P,\mathrm{BS})}_{k_P-\bar{k}} }^2
    \abs{\bar{\eta}^{(2 k_P, \mathrm{BS})}_{0}}^2 
    ~~~ \Bigg{]}
 \mathperiod
\end{eqnarray}
Again, the light emitted by the driven single-excitation state $\ketsmall{k_P}$ scales
linearly with the pump power ($\propto \Xi$), whereas all other contributions scale quadratically ($\propto \Xi^2$).
Specifically, the contributions around $\omega=\omega_0$ (first and second Lorentzian)
contain light from single-excitation states (first line) \emph{and}
two-excitation scattering states (see the sum over the relative 
wavenumber $p$ in the second line).
The contribution around $\omega=\omega_0 + U$ (third Lorentzian) 
is exclusively due to the bound state.

For $U=0$, the last line in \Eqref{eq:specfinal} would be absent.
For $U \gg \gamma_0$, we can exploit the bound states' separation in 
energy from the band of scattering states (\cf \figref{fig:disprels}).
In other words, since $U \gg \gamma_0$, the corresponding emission spectrum 
at the frequency $\omega=\omega_0+U$ has practically no overlap to transitions 
around $\omega=\omega_0$.
We may then extract the following information
around the resonance $\omega = \omega_0 + U$:
\begin{eqnarray}
 \frac{S(\vv{r},\omega=\omega_0+U) \gamma_0}{2 \xi^2 \abs{\vv{w}(\vv{r})}^2 M} &=&
   \frac{\Xi^2}{3} 
    \abs{ \bar{\eta}^{(2 k_P,\mathrm{BS})}_{k_P-\bar{k}} }^2
    \abs{\bar{\eta}^{(2 k_P, \mathrm{BS})}_{0}}^2 
    \\ \nonumber
 &=&
 \frac{\Xi^2}{3} \cdot \frac{16}{M^2} \cos^2\left[ \left( \bar{k}-k_P \right) a \right]
 \mathperiod
\end{eqnarray}
When compared with \Eqref{eq:g1full2}, we see that this is the \anfz{pure} far-field signature for the
existence of a two-body bound state on the lattice.
Moreover, 
if we imagine the emission spectrum being measured at 
$\vv{r}$ (elevation $\beta_{\mathrm{det}}$) and, for convenience, 
normalize this signal to the value recorded at a fixed direction 
$\vv{r}^\prime$ ($\abssmall{\vv{r}}=\abssmall{\vv{r}^\prime}=r$, 
elevation $\beta_{\mathrm{exc}}$) in the $y$-$z$ plane, 
we have
\begin{eqnarray}
 \label{eq:Sbssig}
 \frac{S(\vv{r},\omega=\omega_0+U)}{S(\vv{r}^\prime,\omega=\omega_0+U)}
 \cdot 
 \frac{\abs{\vv{d}}^2/r^2}{\abs{\vv{w}(\vv{r})}^2}
 &=&
 \cos^2\left[ \left(k_P - \bar{k}\right) a \right]
 \\ \nonumber
 && \hspace{-3cm} = 
 \cos^2\left( k_P a - k_{\mathrm{at}} a \left[\sin\beta_{\mathrm{det}}\right]_{\frac{2 \pi}{a}} \right)
 \mathperiod
\end{eqnarray}

This is the same signature as obtained in the context of
spontaneous emission from a pure eigenstate (see \Eqref{eq:emittintBS}),
even though here the external probing field is incoherent and weak. 
The discussion from \secref{subsubsec:empattbs} therefore applies to this 
result as well.
Hence, \Eqref{eq:Sbssig} does not only represent an explicit far-field feature 
for the existence of a bound state on a lattice. 
This expression can, analogous to the discussion of \Eqref{eq:emittintBS}, be utilized
to extract the relative wave function's complete momentum distribution
by tuning the argument $k_P-\bar{k}$ across the first Brillouin
zone. 
To this end, one could, for instance, vary the detection angle 
$\beta_{\mathrm{det}}$ while keeping the excitation angle $\beta_{\mathrm{exc}}$ fixed. 
The spectrum only needs to be recorded at a single frequency $\omega=\omega_0+U$.

Conversely, around the frequency $\omega = \omega_0$ we can extract
\begin{eqnarray}
 \nonumber
 && \frac{S(\vv{r},\omega=\omega_0) \gamma_0}{2 \xi^2 \abs{\vv{w}(\vv{r})}^2 M} =
  2 \Xi \delta_{\bar{k} k_P}
  + \frac{4}{3} \Xi^2 \sum_{p} \abs{\bar{\eta}^{(2 k_P, p)}_{k_P-\bar{k}}}^2 \abs{\bar{\eta}^{(2 k_P, p)}_{0}}^2
  \\ 
  && ~~~ 
  ~+~ \Xi^2 \abs{ \bar{\eta}^{(2 k_P, \mathrm{BS})}_{k_P-\bar{k}} }^2 \abs{\bar{\eta}^{(2 k_P, \mathrm{BS})}_{0}}^2
 \mathcomma
\end{eqnarray}
which can be rewritten performing the sums over $\nu$ and $p$ as
(see \appref{sec:appetasums} for details)
\begin{eqnarray}
 \nonumber
 && \frac{S(\vv{r},\omega=\omega_0) \gamma_0}{2 \xi^2 \abs{\vv{w}(\vv{r})}^2 M} =
  2 \Xi \delta_{\bar{k} k_P}
  + \frac{4}{9} \Xi^2 \delta_{\bar{k} k_P}
  \\ 
  && ~~~ 
  ~+~ \Xi^2
  \begin{cases}
   0
   &
   \mathrm{for~}U=0   
   \\
   \frac{16}{M^2} \cos^2\left[\left( \bar{k} - k_P \right)a\right]
   &
   \mathrm{for~}U \gg \gamma_0   
  \end{cases}
 \mathperiod
\end{eqnarray}
Similar to the discussion of \Eqref{eq:g1full2},
the contribution from the bound state only appears for detected
wavenumbers $\bar{k} \neq k_P$ as a correction to the $M\gg1$-limit.
However, also here, we cannot infer more specific properties on the nature of the
scattering states from the far field.
This prompts for an extension of this simple single-pump excitation scheme 
to the case of an external field with two Fourier components.

\subsection{Two-Pump Setup}
\label{subsec:twopump}
\begin{figure}[t]
 \centering
  \includegraphics[width=0.48\textwidth]{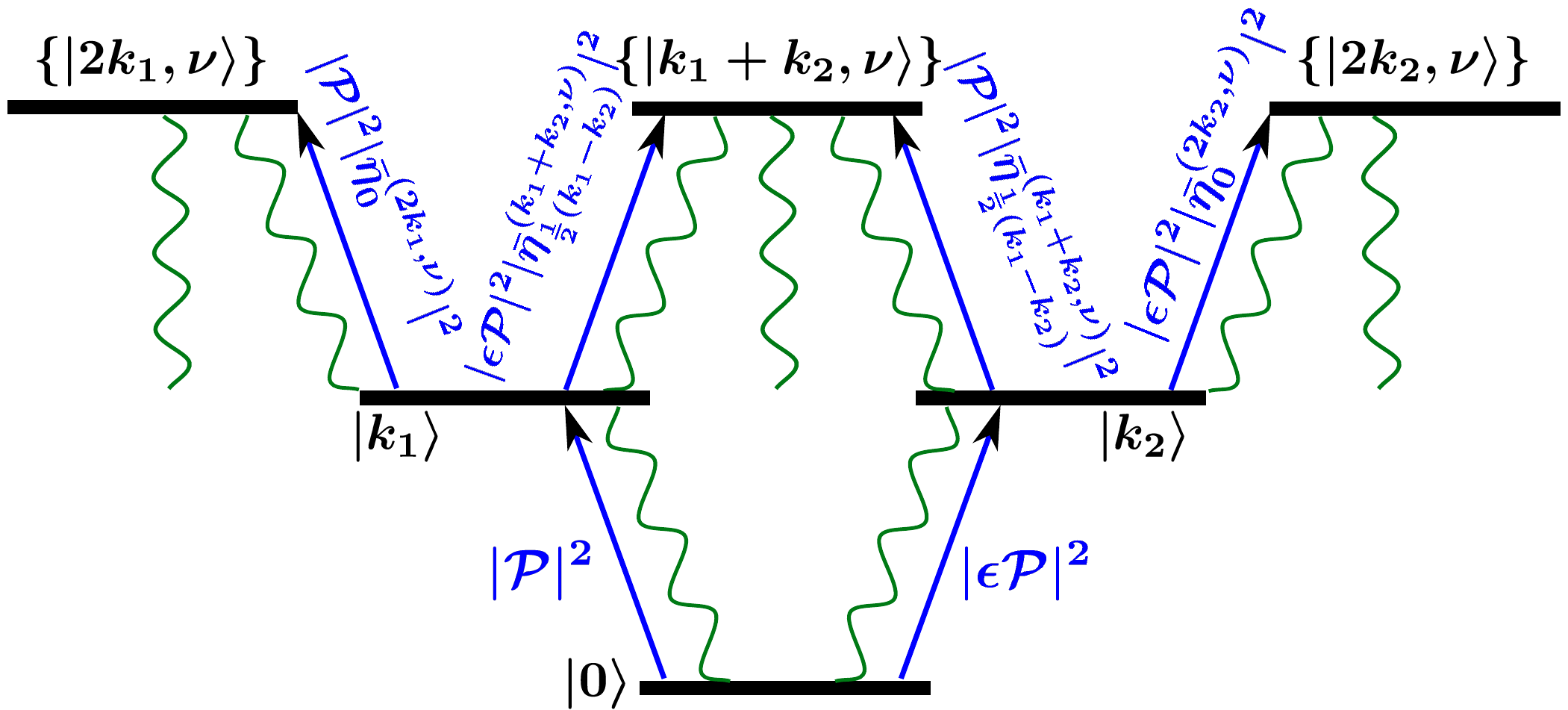}
  \caption{\label{fig:2psetup}
	    An external, incoherent field with two spatial Fourier components
	    (as depicted in \figref{fig:overview}c))
	    drives the atomic system and \anfz{imprints}
	    the wavenumbers $k_1$ and $k_2$
	    (the ratio of the driving fields' pump rates is $\epsilon^2$).
	    The reduced level scheme for this two-pump excitation setup comprises
	    single-excitation states $\ketsmall{k_1}$ and $\ketsmall{k_2}$,
	    and two-excitation states $\ketsmall{2k_1, \nu}$, $\ketsmall{2k_2, \nu}$,
	    and $\ketsmall{k_1+k_2, \nu}$.
	    The figure shows the situation for the case $k_1 \neq k_2$.
	    Single-excitation levels that are only populated via spontaneous emission 
	    from two-excitation states are not shown here since they would only be relevant 
	    for detection angles $\beta^{(1)}_{\mathrm{det}} \neq \beta_1, \beta_2$ 
	    (which we do not consider).}
\end{figure}
In this Section, 
we investigate the atomic system's far-field response along the same lines as in \secref{subsec:spump}
but for a two-pump setup (see \figref{fig:overview}c)).
We thus imagine two plane wave components with wavevectors $\vv{k}^{(1)}$ and $\vv{k}^{(2)}$, 
respectively ($\abssmall{\vv{k}^{(1)}}=\abssmall{\vv{k}^{(2)}}=k_L=k_{\mathrm{atom}}$).
The corresponding wavenumbers that are \anfz{imprinted} on the atomic system are 
$k_i = \left[ k_{\mathrm{atom}} \sin\beta_{i} \right]_{2 \pi/a}$ 
($i=1,2$),
where $\beta_i$ signify the excitation angles (angles of incidence).
We choose the corresponding pump rates according to 
$\abssmall{\mathcal{P}_1}^2 \equiv \abssmall{\mathcal{P}}^2$ and 
$\abssmall{\mathcal{P}_2}^2 \equiv \epsilon^2 \abssmall{\mathcal{P}}^2$.

In \secref{subsec:spump}, we have revealed that the signatures 
at detection angles corresponding to wavenumbers that are not directly driven by the external pump 
(\ie, $\bar{k} \neq k_P$ in \secref{subsec:spump}, see also \figref{fig:singlepumpsketch}) only provide 
information about the bound state.
Conversely, at detection angles corresponding to the wavenumber that is directly driven by the pump
(\ie, $\bar{k} = k_P$ in \secref{subsec:spump}), the single-pump setup failed to provide 
specific details about the scattering states' properties.

In the following, we will show how a two-pump setup can be exploited to gain further insight.
To this end, we will, unlike in \secref{subsec:spump}, exclusively focus on an out-of-plane detection scheme 
where the detection angles coincide with the excitation angles, \ie, 
$\beta^{(i)}_{\mathrm{det}} = \beta_{i}$.
This choice considerably simplifies the relevant expressions 
for the steady-state occupation numbers (which can be found in \appref{subsec:apptwopump}).
Furthermore, it allows us to only consider those wavenumbers in the description that are imposed
by the external field (\ie, $\bar{k}_i = k_i$).

\subsubsection{Far-Field Intensity}
\label{subsubsec:ffinttwopumps}
The far-field intensity measured at a detector position $\vv{r}_1$
out-of-plane (elevation $\beta_1$, detected wavenumber $k_1$) can be
determined from \Eqref{eq:fieldcorr}, yielding 
\begin{eqnarray}
 && \frac{{G}^{(1)}(\vv{r}_1)}{\xi^2 \abs{\vv{w}(\vv{r}_1)}^2 M}  =  
 N_{k_1}  +  \sum_{\nu} \abs{\bar{\eta}^{(2 k_1, \nu)}_{0}}^2 N_{2 k_1, \nu}
 \\ \nonumber
 && ~~~~~ ~~~~ + \left( 1-\delta_{k_1 k_2} \right)~
  \sum_{\nu} \abs{\bar{\eta}^{(k_1+k_2, \nu)}_{\frac{1}{2} \left( k_1-k_2 \right)}}^2 N_{k_1+k_2, \nu}  
 \mathperiod
\end{eqnarray}
Inserting the steady-state occupation numbers~(\ref{eq:2pssbegin})--(\ref{eq:2pssend})
leads to 
\begin{eqnarray}
 \label{eq:G1twopumpcompl}
 && \frac{{G}^{(1)}(\vv{r}_1)}{\xi^2 \abs{\vv{w}(\vv{r}_1)}^2 M} =
 \Xi 
 \left( 1 + \epsilon^2 \delta_{k_1 k_2} \right)
 \\ \nonumber
 && ~~~~
 ~+~ \Xi^2 \sum_{\nu} \abs{\bar{\eta}^{(2 k_1, \nu)}_{0}}^4
 + 2 \epsilon^2 \Xi^2 \sum_{\nu} \abs{\bar{\eta}^{(k_1+k_2, \nu)}_{\frac{1}{2}\left(k_1-k_2\right)}}^4
 \\ \nonumber
 && ~~~~
 ~+~ \epsilon^4 \frac{\Xi^2}{2} \left( 1 + \delta_{k_1 k_2} \right)  
      \sum_{\nu} \abs{\bar{\eta}^{(2 k_2, \nu)}_{k_2-k_1}}^2 \abs{\bar{\eta}^{(2 k_2, \nu)}_{0}}^2
 \mathperiod
\end{eqnarray}
For $\epsilon=0$ (\ie, zero second field) we simply recover 
\Eqref{eq:g1full} (for the case $\bar k = k_P$).

Note that \Eqref{eq:G1twopumpcompl} contains linear and non-linear terms in the 
dimensionless pump power $\Xi$.
In order to investigate the dependence on the parameters that can be tuned
externally, \ie, $\beta_1$, $\beta_2$, and $\epsilon$,
we start from the quantity
\begin{equation}
 \bar{G}^{(1)}_{\mathrm{NL}}(\beta_1, \beta_2, \epsilon)
 =
 \left[ \frac{{G}^{(1)}(\vv{r}_1)}{\xi^2 \abs{\vv{w}(\vv{r}_1)}^2 M} 
 - \Xi 
 \left( 1 + \epsilon^2 \delta_{k_1 k_2} \right)
 \right]
 \cdot
 \frac{1}{\Xi} 
 \mathcomma
\end{equation}
which is the rescaled, nonlinear part of the emitted intensity (low-power limit subtracted).
In particular,
we focus on the relative difference
of this quantity for zero and non-zero second driving field.
In other words, we define
\begin{eqnarray}
 \label{eq:G1nlpart}
 \nonumber
 && \delta \bar{G}^{(1)}_{\mathrm{NL}}(\beta_1, \beta_2, \epsilon)
 \equiv
 \frac{\bar{G}^{(1)}_{\mathrm{NL}}(\beta_1, \beta_2, \epsilon)-\bar{G}^{(1)}_{\mathrm{NL}}(\beta_1, \beta_2, \epsilon=0)}{\bar{G}^{(1)}_{\mathrm{NL}}(\beta_1, \beta_2, \epsilon=0)}
 \\
 && ~~~ = 
 \delta_{q 0} ~\epsilon^2 \left( \epsilon^2 + 2 \right)
 \\ \nonumber
 && ~~~~~ ~+~
 \left( 1-\delta_{q 0} \right) ~\epsilon^2 
    \Bigg{(}	  2 \sum_{\nu} \abs{\bar{\eta}^{(k_1+k_2, \nu)}_{q}}^4 
    \\ \nonumber
 && ~~~~~~~~~~~ ~+~
    \frac{\epsilon^2}{2} \sum_{\nu} \abs{\bar{\eta}^{(2 k_2, \nu)}_{2q}}^2 \abs{\bar{\eta}^{(2 k_2, \nu)}_{0}}^2  
    \Bigg{)}
 \\ \nonumber
 && ~~~~~~~~~~~~~~ ~\times~ \frac{1}{\sum_{\nu} \abs{\bar{\eta}^{(2 k_1, \nu)}_{0}}^4}
 \mathcomma
\end{eqnarray}
where $q \equiv (k_1-k_2)/2$.
Explicitly performing the sums (see \appref{sec:appetasums}), we arrive at
\begin{eqnarray}
 \label{eq:deltag1}
 && \delta \bar{G}^{(1)}_{\mathrm{NL}}(\beta_1, \beta_2, \epsilon) / \epsilon^2
 = 
 \delta_{q 0}~\left( \epsilon^2+2 \right)
 \\ \nonumber
 && ~~~ ~+~ \left( 1 - \delta_{q 0} \right)
 \begin{cases}
  4 & \mathrm{for}~U=0 \\
  \frac{\frac{4}{3}\cos^4(qa) + 4 \sin^4(qa) + \frac{32}{M^2} \cos^4(qa)}{\frac{1}{3} + \frac{16}{M^2}}
  & \mathrm{for}~U \gg \gamma_0
 \end{cases}
 \mathcomma \\
 \nonumber
 && \overset{M \gg 1}{\simeq}
 \delta_{q 0}~\left( \epsilon^2+2 \right)
 \\ \nonumber
 && ~~~ ~+~ 4 \left( 1 - \delta_{q 0} \right)
 \begin{cases}
  1 & \mathrm{for}~U=0 \\
  \cos^4(qa) + 3 \sin^4(qa) 
  & \mathrm{for}~U \gg \gamma_0
 \end{cases}
 \\ \nonumber
 && ~~~ ~+~ \left( 1 - \delta_{q 0} \right)
 \begin{cases}
  0 & \mathrm{for}~U=0 \\
  \frac{96}{M^2} \cos^4(qa)
  & \mathrm{for}~U \gg \gamma_0
 \end{cases}
 \mathperiod
\end{eqnarray}

\begin{figure}[t]
 \centering
  \includegraphics[width=0.48\textwidth]{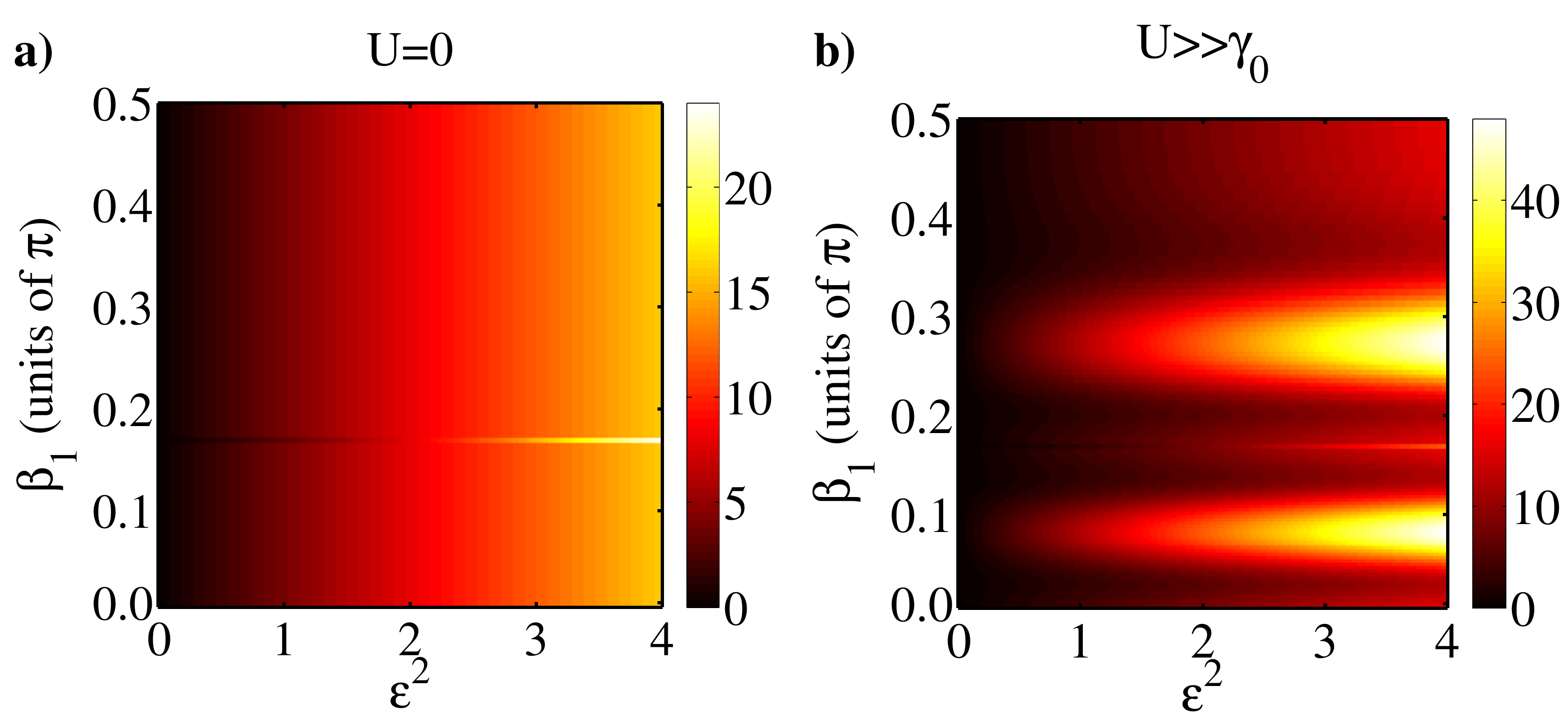}
  \caption{\label{fig:nonlinresponse}
	    (color online).
	    Relative difference of the normalized far-field intensity for zero and non-zero second driving field,
	    \protect\ie, $\delta \bar{G}^{(1)}_{\mathrm{NL}}(\beta_1, \beta_2, \epsilon)$ according 
	    to \Eqref{eq:deltag1} (in the limit of many atoms).
	    We have chosen $\lambda_{\mathrm{atom}} / a = 0.5$ and $\beta_2=\arcsin(\lambda_{\mathrm{at}}/a)$.
	    The signatures for the two cases of non-interacting ($U=0$, a)) 
	    and strongly interacting atoms ($U \gg \gamma_0$, b)) are clearly different.
	    See the main text for the identification and discussion of the individual features.}
\end{figure}
Also here, signatures emerging from a bound state 
(last line in \Eqref{eq:deltag1}, see \appref{sec:appetasums} for details) appear
as a small correction to the $M \gg 1$-limit.
In \figref{fig:nonlinresponse}, 
we plot the quantity~(\ref{eq:deltag1}) (for $\lambda_{\mathrm{atom}} / a = 0.5$)
in the limit of many atoms (such that the last line in \Eqref{eq:deltag1} vanishes).
For simplicity, we choose the second driving field to impinge under an angle such that it excites a zero wavenumber 
mode, \eg, $\beta_2=\arcsin(\lambda_{\mathrm{at}}/a)$ (which translates into $k_2=0$ and, therefore, $q=k_1/2$).
The individual peaks in \figref{fig:nonlinresponse} as a function 
of the angle $\beta_1$ can be identified as follows.
According to \Eqref{eq:sconstraintappr}, we have
\begin{equation}
 \label{eq:simpleeq}
 \beta_1 = \arcsin \left[ \frac{\lambda_{\mathrm{at}}}{a} \left( \frac{qa}{\pi} + n \right) \right]
 ~~~~~(n=0,\pm 1,\dots)
 \mathperiod
\end{equation}

The level scheme for the reduced Hilbert space of the driven states (as shown in \figref{fig:2psetup}) 
effectively reduces to the situation of a single pump (\cf \figref{fig:singlepumpsketch}) 
when $k_1=k_2$, which is realized here for $q=0$.
From the properties of the momentum distributions~(\ref{eq:etaU0})--(\ref{eq:etaUlargeBS}),
we know that for both $U=0$ and $U \gg 0$ only the background fluorescence
contributes at the argument $q=0$.
Employing \Eqref{eq:simpleeq} for $\lambda_{\mathrm{at}}/a=0.5$,
the first Bragg order ($n=1$) of this background fluorescence is the peak visible 
at $\beta_1 / \pi = 1/6 \simeq 0.17$ in both Figs.~\figref{fig:nonlinresponse}a) and~b).
In contrast to this, 
the decay channel for the direct fluorescence, 
which only occurs for $U \gg \gamma_0$ (\figref{fig:nonlinresponse}b)),
has its maximum contribution for $q=\pi/2a$.
The corresponding $n=0$-order for $q=\pi/2a$ yields the peak at $\beta_1 / \pi \simeq 0.08$
\figref{fig:nonlinresponse}b), 
whereas the first Bragg order ($n=1$) results in $\beta_1 / \pi \simeq 0.27$.

On the whole, the additional degrees of freedom in the context of a two-pump setup (\ie, two angles
and a relative strength between the two pumps) open up a way to reveal the scattering states'
properties of direct and background fluorescence from a suitably normalized far-field intensity.
This possibility is absent if one had only a single pump (as discussed in \secref{subsec:spump}).
However, in contrast to the two-body bound states, 
the band of scattering states is quasi-degenerate (see \figref{fig:disprels})
such that scattering states with different relative wavenumbers overlap spectrally.
As a result, it is not possible to extract the full momentum distribution of an individual
scattering state. 
Instead, the far-field signature is a superposition from scattering states with different
relative wavenumbers (see the sum over $\nu$ in \Eqref{eq:G1nlpart}).
Signatures from a bound state are suppressed as $1/M^2$
in the limit of many atoms.

\subsubsection{Emission Spectrum around $\omega=\omega_0+U$}
Analogous to \secref{subsubsec:emspecspump}, 
the steady-state emission spectrum for the two-pump scheme 
recorded at the position $\vv{r}_1$ around the frequency $\omega=\omega_0+U$ reads
(see \appref{sec:appautocorr} for details)
\begin{equation}
 \frac{S(\vv{r}_1,\omega=\omega_0+U) \gamma_0}{2 \xi^2 \abs{\vv{w}(\vv{r}_1)}^2 M} = 
 \frac{16 \Xi^2}{3 M^2} 
 \left[
 \left( 1+\epsilon^4 \delta_{q 0} \right) 
      + 2 \epsilon^2 \cos^4(qa)
 \right]
 \mathperiod
\end{equation}
Along the same lines as in the previous section, we define 
the relative difference for zero and non-zero second driving field as
\begin{eqnarray}
 \label{eq:bssigtwopumps}
 && \delta {S}(\beta_1, \beta_2, \epsilon, \omega=\omega_0+U)
 \equiv
 \\ \nonumber
 && 
 ~~~
 \frac{ S(\vv{r_1}, \omega=\omega_0+U) \vert_{\epsilon \neq 0} 
	- S(\vv{r_1}, \omega=\omega_0+U) \vert_{\epsilon = 0} }
	{S(\vv{r_1}, \omega=\omega_0+U) \vert_{\epsilon = 0}}
 \\ \nonumber
 && 
 ~~~
 =
 \epsilon^2 \left( \epsilon^2 \delta_{q 0} + 2 \cos^4(qa) \right)
 \mathperiod
\end{eqnarray}

\begin{figure}[t]
 \centering
 \includegraphics[width=0.48\textwidth]{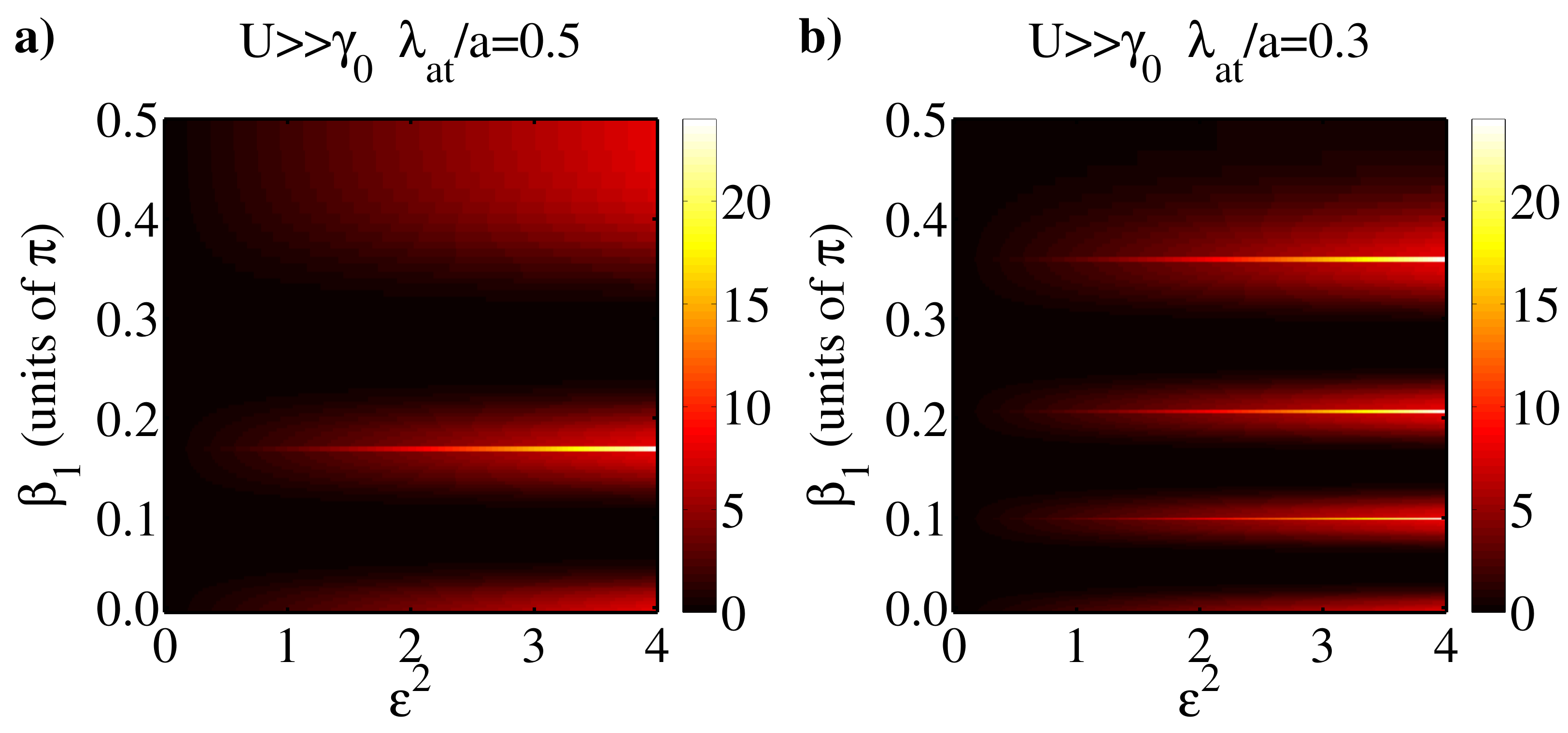}
  \caption{\label{fig:bssigtwopumps}
	    (color online).
	    Spectral signature 
	    $\delta {S}(\beta_1, \beta_2, \epsilon, \omega=\omega_0+U)$
	    of a bound state as a function of the detection angle
	    according to \Eqref{eq:bssigtwopumps}.
	    As the bound states' momentum distribution has a maximum at 
	    the argument $q=0$ (see \secref{subsubsec:dirbgetc}),
	    we observe peaks around (see also the discussion in \secref{subsubsec:ffinttwopumps})
	    $\beta_1 / \pi \simeq 0, 0.17, 0.5$
	    (for $\lambda_{\mathrm{at}}/a=0.5$, a)).
	    For $\lambda_{\mathrm{at}}/a=0.3$ (b)),
	    we have $\beta_1 / \pi \simeq 0, 0.10, 0.20, 0.36$.}
\end{figure}
This expression represents an approach which is complimentary to 
the signature~(\ref{eq:emittintBS}) we obtained in the 
context of spontaneous emission or the emission spectrum~(\ref{eq:Sbssig})
for a single-pump setup.
Unlike in Eqs.~(\ref{eq:emittintBS}) and~(\ref{eq:Sbssig})
where we relied on the collection of spontaneously emitted light,
the bound state signature here is a consequence of collected photons 
which correspond to transitions directly driven by the external pumps 
(see the introduction of \secref{subsec:twopump} for the choice of the detector positions).

In \figref{fig:bssigtwopumps}, we 
visualize the bound state signature according to \Eqref{eq:bssigtwopumps}.
Similarly to \figref{fig:nonlinresponse},
the positions of the peaks can be determined according to \Eqref{eq:simpleeq}.
The bound state's momentum distribution~(\ref{eq:etaUlargeBS}) has a maximum
at the argument $q=0$.
Employing \Eqref{eq:simpleeq} for $\lambda_{\mathrm{at}}/a=0.5$, this leads to 
the peaks around 
$\beta_1/\pi = 0$ ($n=0$),
$\beta_1/\pi \simeq 0.17$ ($n=1$), and
$\beta_1/\pi \simeq 0.5$ ($n=2$),
which can be seen in \figref{fig:bssigtwopumps}a).
Likewise, for a smaller wavelength $\lambda_{\mathrm{at}}/a=0.3$
more Bragg orders become visible at 
$\beta_1/\pi = 0$ ($n=0$),
$\beta_1/\pi \simeq 0.10$ ($n=1$),
$\beta_1/\pi \simeq 0.20$ ($n=2$), and
$\beta_1/\pi \simeq 0.36$ ($n=3$)
in \figref{fig:bssigtwopumps}b).

In the next section, we conclude our investigations on the two-pump setup 
with some thoughts on the measurement of consecutive photon counts.
Within this framework, we analyze, as a complement to the previous studies, 
what can be learned from the intensity correlations with respect to the detection
angle.

\subsubsection{Intensity Correlations}
We envision a coincidence detection scheme, where two detectors
are positioned out-of-plane ($y$-$z$ plane) at $\vv{r}_1$ and 
$\vv{r}_2$, respectively.
The two detectors have the same distance to the origin 
(\ie, $\abssmall{\vv{r}_1}=\abssmall{\vv{r}_2}$) and we aim at events with 
zero time delay ($\tau=0$).
Note that the detection angles are chosen such that they are equal to the
excitation angles (see introduction of \secref{subsec:twopump}).

According to \Eqref{eq:intcorr}, the intensity correlation function is
\begin{eqnarray}	  
\nonumber
&& \frac{ {G}^{(2)}(\vv{r}_1,\vv{r}_2) }{ \xi^4 \abs{\vv{w}(\vv{r}_1)}^2 \abs{\vv{w}(\vv{r}_2)}^2 M^2 } =
  \sum_{\nu} \abs{ \bar{\eta}^{(k_1+k_2, \nu)}_{\frac{1}{2}\left(k_1-k_2\right)} }^2
      N_{k_1+k_2, \nu; k_1+k_2, \nu}
  \\ \nonumber
 && ~~~~~ =
 \Xi^2 \left( \delta_{k_1 k_2} \frac{\left( 1+\epsilon^2 \right)^2}{2} + \left( 1-\delta_{k_1 k_2 }\right) \epsilon^2 \right) 
 \sum_{\nu} \abs{ \bar{\eta}^{(k_1+k_2, \nu)}_{\frac{1}{2}\left(k_1-k_2\right)} }^4
 \mathperiod
\end{eqnarray}
To arrive at the normalized correlation function, we use 
\begin{eqnarray}
 \frac{{G}^{(1)}(\vv{r}_1)}{\xi^2 \abs{\vv{w}(\vv{r}_1)}^2 M} &=& 
 \Xi \left( 1 + \delta_{k_1 k_2} \epsilon^2 \right) + \mathcal{O}(\Xi^2)
 \mathcomma
 \\
 \frac{{G}^{(1)}(\vv{r}_2)}{\xi^2 \abs{\vv{w}(\vv{r}_2)}^2 M} &=& 
 \Xi \left( \epsilon^2 + \delta_{k_1 k_2}  \right) + \mathcal{O}(\Xi^2)
 \mathcomma
\end{eqnarray}
leading to
\begin{eqnarray}
 && 
 \frac{{G}^{(1)}(\vv{r}_1)}{\xi^2 \abs{\vv{w}(\vv{r}_1)}^2 M}
 \cdot 
 \frac{{G}^{(1)}(\vv{r}_2)}{\xi^2 \abs{\vv{w}(\vv{r}_2)}^2 M}
 =
 \\ \nonumber
 && 
 ~~~~~~
 \Xi^2 \left[ \delta_{k_1 k_2} \left(1+\epsilon^2\right)^2 + \left( 1-\delta_{k_1 k_2}\right) \epsilon^2 \right]
 + \mathcal{O}(\Xi^3)
 \mathperiod
\end{eqnarray}
According to \Eqref{eq:intcorrnorm}, the normalized correlation function 
reads (taking into account orders up to $\mathcal{O}(\Xi^2)$ for both the nominator and the denominator)
\begin{eqnarray}
 g^{(2)}(\beta_1, \beta_2)
 &=&
  \left( 1 - \frac{1}{2} \delta_{k_1 k_2}  \right)
  \sum_{\nu} \abs{ \bar{\eta}^{(k_1+k_2,\nu)}_{\frac{1}{2}\left( k_1-k_2 \right)} }^4
 \mathperiod
\end{eqnarray}
Performing the sum over $\nu$ (see \appref{sec:appetasums} for details),
we finally arrive at
\begin{eqnarray}
\label{eq:g2final}
 && g^{(2)}(\beta_1,\beta_2) =
 \frac{1}{6} \delta_{q 0} 
 \\ \nonumber
 && ~~~ ~+~
 \frac{2}{3}
 \left( 1- \delta_{q 0} \right)
 \begin{cases}
  1
  & \mathrm{for~}U=0
  \\
  \cos^4(qa) + 3 \sin^4(qa) 
  & \mathrm{for~}U \gg \gamma_0
 \end{cases}
 \\ \nonumber
 && ~~~ ~+~
 \begin{cases}
  0
  & \mathrm{for~}U=0
  \\
  \frac{16}{M^2} \left( 1 -\frac{1}{2} \delta_{q0} \right) \cos^4(qa)
  & \mathrm{for~}U \gg \gamma_0
 \end{cases}
 \mathcomma
\end{eqnarray}
where again $q \equiv (k_1-k_2)/2$ (which depends on $\beta_1$ and $\beta_2$).
This correlation function measures two photon counts with
zero time delay at the detection angles $\beta_1$ and $\beta_2$.
Note that our two-pump setup is such that excitation and detection angles 
coincide. 
Hence, varying $\beta_1$ and $\beta_2$ represents a simultaneous change of 
both the detection and excitation angles.

\begin{figure}[t]
 \centering
  \includegraphics[width=0.48\textwidth]{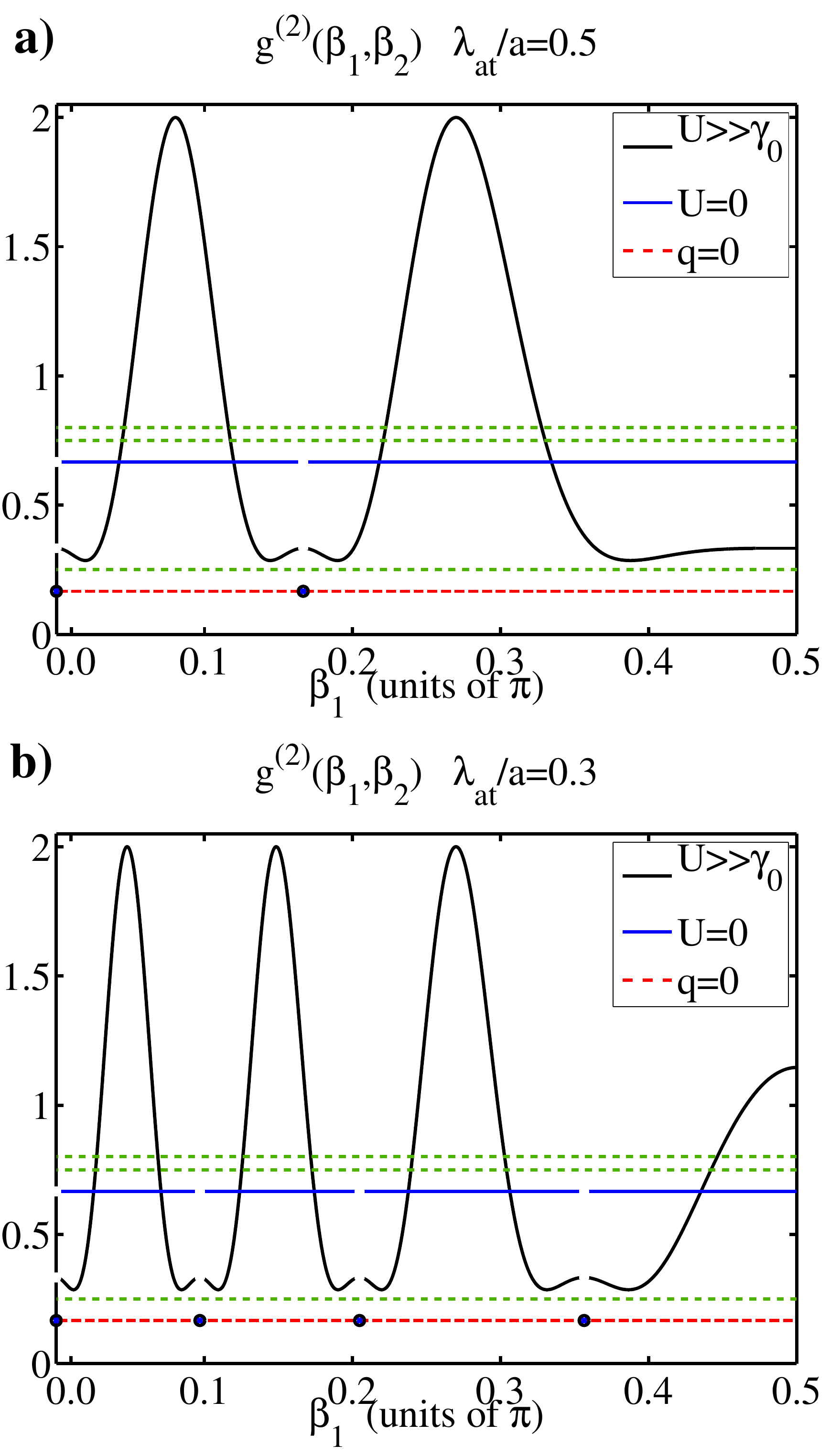}
  \caption{\label{fig:g2plot}
	    (color online).
	    Normalized photon--photon correlation function 
	    $g^{(2)}(\beta_1,\beta_2=\arcsin(\lambda_{\mathrm{at}}/a))$ according to \Eqref{eq:g2final}
	    in the limit of many atoms (such that the last line in \Eqref{eq:g2final} vanishes).
	    For $U=0$ (blue line), the correlation function exhibits a constant value
	    of $2/3$, except for angles where $q=0$ is realized (red crosses). 
	    For $U \gg \gamma_0$ (black line), the $g^{(2)}$-function varies smoothly
	    with respect to the angle $\beta_1$,
	    but is also interrupted when $q=0$ (red crosses).
	    At $q=0$, the correlation function always has a value of $1/6$ (denoted by the red dashed line)
	    for both $U=0$ and $U \gg \gamma_0$.
	    The green dashed lines indicate reference values from Refs.~\cite{wiegand,ficekPRA,commentficek,cordes,machnikowski},
	    referring to a two-atom system (see text for discussion).}
\end{figure}
As before, the contribution from a bound state (last line in \eqref{eq:g2final}) 
appears as a correction to the $M\gg1$-limit and is suppressed as $1/M^2$.
Still, this correlation function provides some very characteristic features for the
two cases of non-interacting atoms ($U=0$) and strong atom--atom interactions ($U \gg \gamma_0$).
For $U=0$, the intensity correlation is essentially flat. 
This can again be seen as a consequence of the momentum distribution being featureless
for $U=0$ (\cf \figref{fig:momdists}b)).
The correlation function only jumps between the values
$1/6$ and $2/3$ if the atomic system's level scheme undergoes the sudden change 
from an effective single-pump setup ($q=0$, degenerate pump fields) 
to a two-pump setup ($q \neq 0$).
In contrast to this, for $U \gg \gamma_0$, we observe a continuous angle-dependence
because the underlying relevant momentum distributions also exhibit contributions 
from the decay channel of direct fluorescence (\cf \figref{fig:momdists}c)).
Also here, we can see jumps to the value of $1/6$ that occurs at angles where the pump fields
are degenerate ($q=0$).
We depict these findings in \figref{fig:g2plot} for $\lambda_{\mathrm{at}}/a=0.5$ 
and $\lambda_{\mathrm{at}}/a=0.3$.
The features of \figref{fig:g2plot}a) are visible at the same angles as in \figref{fig:nonlinresponse},
which can be calculated from \Eqref{eq:simpleeq}.
When compared to \figref{fig:g2plot}a), \figref{fig:g2plot}b) features more peaks because the smaller
wavelength of $\lambda_{\mathrm{at}}/a=0.3$ allows for the observation of more Bragg orders.

The sharp jumps in \figref{fig:g2plot} due to the \anfz{collapse} from a two-pump scheme
to the Hilbert space accessible by a single pump refer to the case of infinitely sharp
excitation and detection angles.
In practice, finite apertures will eventually lead to a smoothening of the curves 
(imagine averaging over a small angular window).

We conclude the discussion of the zero time delay intensity correlation function
by referring to results which have been obtained for the case of two atoms.
Generally, as a consequence of spatial interference effects as pointed out in Refs.~\cite{wiegand,ficekPRA}, 
the values of the $g^{(2)}$-function depend on the direction of observation and on the distance between 
the two atoms.
In the case of a weak driving field,
the zero-time delay intensity correlation $g^{(2)}(0)$ for the same detection geometry as described 
in this section (the two atoms are aligned along the $z$-axis) has the value $g^{(2)}(0)=1/4$ \cite{wiegand,ficekPRA,commentficek}
(see the corresponding green dashed line in \figref{fig:g2plot}).
Interestingly, $g^{(2)}(0)=3/4$ would be observed in the limit of a strong driving field
in the case of zero detuning \cite{cordes}.
This value would drop down to $g^{(2)}(0)=1/4$ if the driving field is tuned far off-resonant~\cite{cordes}.
In the context of spontaneous emission from two initially excited and independent (\ie, uncoupled) atoms,
the zero-time correlation function yields a value of $g^{(2)}(0)=4/5$ \cite{machnikowski}.
Likewise, going beyond two atoms, the spontaneous emission from $M$ independent atoms (that are initially all in the excited state)
results in $g^{(2)}(0)=1-1/M$ \cite{machnikowski}, which is exactly the same expression one would also
obtain for an $M$-photon Fock state of a single-mode radiation field \cite{orszag}.

\section{Conclusion}
\label{sec:conclusion}
In conclusion,
we have analyzed the characteristic, angle-dependent far-field signatures emerging 
from the excitation of collective few-excitation atomic states in the context of a 
one-dimensional lattice of (interacting) two-level atoms.
Compared to state-of-the-art experimental techniques that rely on single-atom addressability and/or 
manipulation, the schemes presented in this paper represents an alternative approach for 
studying collective phenomena in a quantum-optical context.

In particular, we started in \secref{sec:fundamentals} with the static properties
of the underlying model system.
This included an in-depth discussion of the dissipative eigenstates in the submanifold
of one and two atomic excitations.
We put special emphasis on the distinction between the two classes of possible two-excitation 
states, \ie, two-excitation scattering states (which can be thought of as two colliding spin waves) 
and two-body bound states whose relative wavefunction is spatially localized 
(with respect to the relative coordinate of the two excitations).
We then showed that characteristic key quantities such as the collective dipole moments, 
branching ratios, and, most importantly, the momentum distribution of the collective states' 
relative wavefunctions are intimately linked to each other.
Moreover, an expansion of the operator for the electric far field in terms of
the collective atomic eigenbasis allowed us to construct operators for the 
emitted intensity, the emission spectrum, and an intensity correlation function.

In \secref{sec:spontemdyn}, we continued with the angle-dependent far-field pattern
emerging from the spontaneous emission of the system's eigenstates.
We especially discussed the characteristic differences of the light emitted by 
two-body scattering states and bound states, arguing that the emission patterns 
can serve as a distinct fingerprint for identifying and proofing the existence
of certain eigenstates on the lattice.
In the course of the discussion, 
we also contrasted the differences between the two cases of non-interacting atoms 
and strong atom--atom interactions. Further, we pointed out how one could in principle extract the relative wave function's complete momentum distribution from the far-field pattern. This is particularly appealing for the case of strong interactions, where a tightly confined two-body bound state exists.

However, since the preparation of a pure eigenstate for studying its spontaneous
emission may, admittedly, pose severe challenges from a practical point of view,
we turned to the investigation of the atomic system's response to a weak and 
incoherent driving field in \secref{sec:steadystatesigs}.
Starting from an external pump field with a single spatial Fourier component,
we explained the relevant excitation and relaxation mechanisms and identified 
the corresponding reduced level scheme.
Within this framework of a single-pump setup, we were able to extract 
signatures for the existence of a two-body bound state from the steady-sate intensity
and emission spectrum in the far field.

Still, this scheme failed to reveal detailed information on the characteristics 
of the involved scattering states.
As another approach to inferring information on the collective atomic eigenstates 
from the far field, we therefore extended the scheme to a special two-pump setup, 
where detectors were chosen such that they exclusively detect photons
from driven transitions.
With regard to the detected steady-state intensity and emission spectrum,
the Hilbert space accessible to the external fields now not only results in a distinct 
signature for the existence of a bound state.
Unlike for a single pump, we were also able to identify features that stem from 
the scattering states' momentum distribution and which are related to the 
decay channels which we termed \anfz{background fluorescence} and \anfz{direct fluorescence}.
However, due to the spectral overlap of scattering states with different relative wavenumbers
it is not possible to extract the momentum distribution of a single scattering state
and we were instead left with far-field observables resulting from the superposition
of different scattering states.
Moreover, we also contrasted the results for the two cases of non-interacting atoms
and strong atom--atom interactions, which exhibit very different far-field patterns.
In addition to that, we showed that in the framework of such a two-pump setup, 
an angle-dependent intensity correlation function represents another tool for 
analyzing the scattering states' properties for both interacting and non-interacting 
atoms.

Future studies might include the dynamics induced by one or more coherent driving 
fields, requiring a description beyond the rate equations utilized in this work.
In the context of coherent fields,
an analysis of how interference effects modify the far-field emission patterns 
discussed in this paper would be highly interesting.
This would also address whether these signatures
may reveal additional or complimentary information on the properties of the collective 
few-excitation wavefunction.
Besides that, going from weak driving fields to higher pump powers would in principle 
open up a way to probe the more complicated Hilbert space beyond two-excitations, which is 
still largely unexplored in detail.
Finally, our analysis provides a promising approach to the theoretical modeling and 
understanding of recent experiments in nuclear quantum optics on a microscopic level.

\appendix

\section{Integrating out the Photonic Degrees of Freedom}
\label{sec:appphdof}
The Heisenberg equations of motion for Hamiltonian~(\ref{eq:origH}) read
\begin{eqnarray}
 \label{eq:Heisa}
 \ii \partial_t a^{\phantom \dagger}_k & = & \epsilon_k a_k + \sum_{n} g^{*}_{nk} \sigma^{-}_n \mathcomma \\
 \label{eq:Heisb}
 \nonumber
 \ii \partial_t \sigma^{-}_n & = & - \omega_n \sigma^z_n \sigma^-_n - \sum_k g^{\phantom *}_{nk} \sigma^z_n a_k
      \\
    && ~~~ ~+~ \frac{1}{2} \sum_{m} V_{\abs{n-m}} \sigma^+_m \sigma^-_m \sigma^-_n \mathperiod
\end{eqnarray}
Transforming to a rotating frame
(\ie, $a^{\phantom \dagger}_k \rightarrow a^{\phantom \dagger}_k \ee^{-\ii \epsilon_k t}$, 
$\sigma^{-}_i \rightarrow \sigma^{-}_i \ee^{-\ii \omega_i t}$), 
formally integrating \Eqref{eq:Heisa}, tracing out the photonic degrees of freedom 
(we assume the reservoir to be initially in the vacuum state), 
and inserting the result back into \Eqref{eq:Heisb},
we arrive at the integro-differential equation for the atomic operators:
\begin{eqnarray}
 \nonumber
 \label{eq:integrodiffeq}
 \partial_t \sigma^-_n(t) &=& 
  \displaystyle \int \limits_0^t \mathrm{d} t^{\prime} \sum_{m} K_{nm}(t,t^{\prime}) \sigma^z_n(t) \sigma^-_m(t^{\prime})	\\ 
	 && ~~~ - \frac{\ii}{2} \sum_{m} V_{\abs{n-m}} \sigma^+_m(t) \sigma^-_m(t) \sigma^-_n(t) \mathperiod
\end{eqnarray}
Here, $K_{nm}(t,t^{\prime})$ denotes a memory kernel, which we assume
to describe atom--photon coupling to a featureless continuum of modes.
In other words, we apply a Markov approximation, 
which reduces the memory kernel to a set of complex rates, \ie,
\begin{equation}
K_{nm}(t-t^{\prime}) = \frac{\Gamma_{\abs{n-m}}}{2} \delta(t-t^{\prime}) 
\end{equation}
(the details of the rates depend on the reservoir, see Eqs.~(\ref{eq:fullrates})--(\ref{eq:fullrateslast})),
turning \Eqref{eq:integrodiffeq} into
\begin{equation}
 \label{eq:diffeqintermed}
 \partial_t \sigma^-_n = \frac{1}{2} \sum_{m} \Gamma_{\abs{n-m}} \sigma^z_n \sigma^-_m 
		      - \frac{\ii}{2} \sum_{m} V_{\abs{n-m}} \sigma^+_m \sigma^-_m \sigma^-_n \mathperiod
\end{equation}
Hamiltonian~(\ref{eq:Heff}) is the effective non-Hermitian Hamiltonian corresponding
to the equations of motion~(\ref{eq:diffeqintermed}).

\section{Eigenequations}
\label{sec:appeigeq}

\subsection{Single-Excitation Eigenstates}
\label{subsec:appeigeq1}
The eigenproblem with respect to Hamiltonian~(\ref{eq:Heff}) for 
a single-excitation state 
\begin{equation}
 \ket{\varphi}=\sum_n \varphi_n \sigma^+_n \ket{0}
\end{equation}
yields the difference equation
\begin{equation}
 \label{eq:1exeigeq}
 0 = - \frac{\ii}{2} \sum_m \Gamma_{\abs{n-m}} \varphi_m - E \varphi_n \mathperiod
\end{equation}
Note that the atom--atom interaction terms are absent in the single-excitation submanifold.

\subsection{Two-Excitation Eigenstates}
\label{subsec:appeigeq2}
The eigenproblem with respect to Hamiltonian~(\ref{eq:Heff}) for 
a two-excitation state
\begin{equation}
 \label{eq:twobodyansatz0}
 \ket{\Phi} = \sum_{n_1 n_2} \Phi_{n_1 n_2} \sigma^+_{n_1} \sigma^+_{n_2} \ket{0}
 \mathcomma
\end{equation}
which we rewrite
as a product of the center-of-mass motion 
(described by a plane wave with a center-of-mass wavenumber $K$) and a
relative wavefunction $\Psi^{(K \nu)}_{\abs{n_1-n_2}}$, \ie,
\begin{equation}
 \label{eq:twobodyansatz}
 \ket{K \nu} = \frac{1}{2\sqrt{M}} \sum_{n_1 n_2} 
	      \ee^{\ii \frac{Ka}{2} \left(n_1+n_2\right)} 
	      \cdot \Psi^{(K \nu)}_{\abs{n_1-n_2}} 
	      ~\sigma^+_{n_1} \sigma^+_{n_2} \ket{0}
	      \mathcomma
\end{equation}
reduces to an effective single-particle problem on 
a half-infinite lattice ($x > 0$, $\Psi^{(K \nu)}_0 = 0$):
\begin{eqnarray}
 \nonumber
 \label{eq:eigen2ex1dlatt}
 0 &=& - \ii \sum_{j \neq x} \Gamma_j \cos\left(\frac{Kaj}{2} \right) \left( \Psi^{(K \nu)}_{\abs{x-j}} + \Psi^{(K \nu)}_{\abs{x+j}} \right)
 \\ 
 && ~~~ ~+~ \left( V_{x} - E - \ii \Gamma_0 \right) \Psi^{(K \nu)}_{x} \mathperiod
\end{eqnarray}
Here, $K/2$ is from the first Brillouin zone
and $\nu$ is a quantum number still to be determined (see main text).
Note that for each center-of-mass wavenumber $K$ the effective particle \anfz{sees}
a different lattice (different hopping terms).
The two-body atom--atom interaction terms now play the role of a potential for the effective particle.
For the tight-binding Hamiltonian~(\ref{eq:Hefftb}), the difference equation simplifies to 
($x>0$)
\begin{eqnarray}
 \nonumber
 \label{eq:eigen2ex1dlatttb}
 0 &=& - \ii \Gamma_1 \cos\left(\frac{Ka}{2} \right) \left( \Psi^{(K \nu)}_{x-1} + \Psi^{(K \nu)}_{x+1} \right)
 \\ 
 && ~~~ ~+~ \left( U \delta_{x,1} - E - \ii \Gamma_0 \right) \Psi^{(K \nu)}_x \mathperiod
\end{eqnarray}

\section{Expansion of the Atomic Operator $\sigma^-_n$ in the Eigenbasis}
\label{sec:appexpineig}
We transform the operator $\sigma^-_n$ to the basis of the system's single- and two-excitation eigenstates 
by virtue of the expansion
\begin{eqnarray}
 && \sigma^{-}_n =
 \sum_k \ketsmall{0} \underbrace{\brasmall{0} \sigma^-_n \ketsmall{k} }_{=\frac{1}{\sqrt{M}} \ee^{\ii ka n}} \brasmall{k}
 \\ \nonumber
 && ~~~~~~ ~+~
 \sum_{k; K \nu} \ketsmall{k} \underbrace{\brasmall{k} \sigma^-_n \ketsmall{K \nu}}_{=\brasmall{k} \sum_{x_1 x_2} \Phi_{x_1 x_2} \sigma^-_n \sigma^+_{x_1} \sigma^+_{x_2} \ketsmall{0}} \brasmall{K \nu}
 \\ \nonumber
 && ~~~ =
 \frac{1}{\sqrt{M}} \sum_k \ee^{\ii ka n} \hat{S}_{0; k} 
 \\ \nonumber
 && ~~~~~~ ~+~ 
 2 \sum_{k; K \nu} \sum_{m=-N/2}^{N/2} 
  \underbrace{\Phi_{n m}}_{\frac{1}{2\sqrt{M}} \ee^{\ii \frac{Ka}{2} (n+m)} \Psi^{(K \nu)}_{n-m} } 
  \underbrace{\brasmall{k} \sigma^+_{m} \ketsmall{0}}_{~~ \frac{1}{\sqrt{M}} \ee^{-\ii ka m}}
  \hat{S}_{k; K \nu}
 \\ 
 && ~~~ =
 \frac{1}{\sqrt{M}} \sum_k \ee^{\ii ka n} \hat{S}_{0; k}
 \\ \nonumber
 && ~~~~~~ ~+~ \frac{1}{2\sqrt{M}} \sum_k \sum_{K \nu} 
		\ee^{\ii (K-k)a n} \eta^{(K \nu)}_{\frac{K}{2}-k; n} 
		\hat{S}_{k; K \nu} \mathcomma
\end{eqnarray}
where $\hat{S}_{0; k} = \ketsmall{0} \brasmall{k}$ and 
$\hat{S}_{k; K \nu} = \ketsmall{k} \brasmall{K \nu}$.
The quantum number $\nu$ runs over all scattering states ($\nu=p$)
and a possible bound state ($\nu=\mathrm{BS}$).

The quantity 
\begin{equation}
 \label{eq:etafacdef}
 \eta^{(K \nu)}_{q;n} = \frac{2}{\sqrt{M}} 
			\sum_{z=-N/2-n}^{N/2-n} 
			\ee^{\ii qa z}~
			\Psi^{(K \nu)}_{z}
\end{equation}
may be interpreted as the windowed Fourier lattice transform 
of the relative wavefunction (with respect to the relative coordinate).
According to \secref{subsec:ffobs}, the electric field operator in the eigenbasis
involves a sum over all atom positions, which is of the form
\begin{eqnarray}
 \nonumber
  && {\hat{\vv{E}}}^{(-)}(\vv{r},t) =
    \xi \vv{w}_0(\vv{r})
      \sum_{n k} \Bigg{(} \dots 
      \\ \nonumber
 && ~~~
      ~+~
      \sum_{K \nu}
      \frac{1}{2 \sqrt{M}} 
      \ee^{-\ii \left( K-k + \frac{\Delta^{K \nu}_k}{c} \sin\beta_{\mathrm{det}}  \right)a n}
      \left(\eta^{(K \nu)}_{\frac{K}{2}-k;n}\right)^*
      \\ \nonumber
 && ~~~~~~~~~~~~ ~\times~
      {\hat{S}}_{K \nu;k}(t-\frac{r}{c})
      \Bigg{)} 
 \mathperiod
\end{eqnarray}
Performing the sum over $n$ yields
\begin{eqnarray}
 \nonumber
  && {\hat{\vv{E}}}^{(-)}(\vv{r},t) =
    \xi \vv{w}(\vv{r}) 
      \sum_{k} 
  \Bigg{(} \dots
  \\
  \nonumber
      && ~~~~~ + ~
      \sqrt{M} \sum_{K \nu} \left( \bar{\eta}^{(K \nu)}_{\frac{K}{2}-k} \right)^*
      \delta_{k \bar{k}}
      {\hat{S}}_{K \nu;k}\left( t -\frac{r}{c}\right)
      \Bigg{)} \mathcomma
\end{eqnarray}
where we now have the \anfz{reduced} quantity $\bar{\eta}^{(K \nu)}_{\frac{K}{2}-k}$,
which is defined via
\begin{equation}
 \label{eq:etafacred}
 \frac{1}{2M} \sum_{n=-N/2}^{N/2} 
			    \ee^{\ii \kappa a n} \eta^{(K,\nu)}_{q; n} 
 = 
 \bar{\eta}^{(K,\nu)}_{q} \delta_{ \left[ \kappa \right]_{\frac{2\pi}{a}} , 0}
 \mathperiod
\end{equation}
We will discuss 
$\bar{\eta}^{(K \nu)}_{q;n}$ and $\bar{\eta}^{(K \nu)}_{q}$
in more detail in \appref{sec:eta}.

\section{Expectation Value of the Field--Field Auto Correlation}
\label{sec:appautocorr}
In order to arrive at an expression for 
${G}^{(1)}(\vv{r},t,t+\tau) = \expvalsmall{ \hat{G}^{(1)}(\vv{r},t,t+\tau) }$,
where
\begin{eqnarray}
 && \hat{G}^{(1)}(\vv{r},t,t+\tau) \equiv 
    {\hat{\vv{E}}}^{(-)}(\vv{r},t) ~ {\hat{\vv{E}}}^{(+)}(\vv{r},t+\tau) \\
 \nonumber
 && = ~~ \xi^2 \abs{\vv{w}(\vv{r})}^2 M
 \\ \nonumber
 && ~~~~~~ \times
  \Bigg{(}
      {\hat{S}}_{\bar{k};0}\left( t_{\mathrm{ret}} \right) {\hat{S}}_{0;\bar{k}}\left( t_{\mathrm{ret}} + \tau \right) 
      \\ \nonumber
      && \hspace{1.5cm} 
      + ~ \sum_{K \nu}
      \bar{\eta}^{(K \nu)}_{\frac{K}{2}-\bar{k}} 
      {\hat{S}}_{\bar{k}; 0}\left( t_{\mathrm{ret}} \right) {\hat{S}}_{K-\bar{k}; K \nu}\left( t_{\mathrm{ret}} + \tau \right)
      \\ \nonumber
      && \hspace{1.5cm} 
      + ~ \sum_{K \nu}
      \left( \bar{\eta}^{(K \nu)}_{\frac{K}{2}-\bar{k}} \right)^* 
      {\hat{S}}_{K \nu; K-\bar{k}}\left( t_{\mathrm{ret}} \right) {\hat{S}}_{0; \bar{k}}\left( t_{\mathrm{ret}} + \tau \right)
      \\ \nonumber
      && \hspace{1.5cm} 
      + ~ \sum_{K \nu} \sum_{K^\prime \nu^\prime}
      \left( \bar{\eta}^{(K \nu)}_{\frac{K}{2}-\bar{k}} \right)^* 
      \bar{\eta}^{(K^\prime \nu^\prime)}_{\frac{K^\prime}{2}-\bar{k}} 
      \\ \nonumber
      && \hspace{2.0cm} 
      \times
      {\hat{S}}_{K \nu; K-\bar{k}}\left( t_{\mathrm{ret}} \right) 
      {\hat{S}}_{K^\prime-\bar{k}; K^\prime \nu^\prime}\left( t_{\mathrm{ret}} + \tau \right)
  \Bigg{)}
  \mathcomma
\end{eqnarray}
we ultimately need to calculate (cross-)correlations such as
$\expvalsmall{ \hat{S}_{k;0}(t) \hat{S}_{0;k^\prime}(t+\tau) }$,
$\expvalsmall{ \hat{S}_{k;0}(t) \hat{S}_{k^\prime,K \nu}(t+\tau) }$,
$\expvalsmall{ \hat{S}_{K \nu;k}(t) \hat{S}_{0;k^\prime}(t+\tau) }$,
and
$\expvalsmall{ \hat{S}_{K \nu;k}(t) \hat{S}_{k^\prime,K^\prime \nu^\prime}(t+\tau) }$.
Generally, this can be achieved by exploiting the quantum regression theorem \cite{mandelwolf}, 
which we will use in the form 
(in the following, we have $t \nolinebreak \rightarrow \nolinebreak \infty$, denoting a steady state)
\begin{eqnarray}
 \label{eq:regtheo}
 \expvalsmall{\hat{A}(t) \hat{B}(t+\tau)}_{t \rightarrow \infty} 
    &=& \sum_j \mathcal{G}_j(\tau) \expvalsmall{\hat{A} \hat{B}_j}_{t \rightarrow \infty} \mathcomma
    \\ \nonumber
 \expvalsmall{\hat{B}(\tau)} &=& \sum_j \mathcal{G}_j(\tau) \expvalsmall{\hat{B}_j(0)}
 \mathperiod
\end{eqnarray}

Explicitly, we have
\begin{eqnarray}
 \expvalsmall{ \hat{S}_{k;0}(t) \hat{S}_{0;k^\prime}(t+\tau) }
 &=&
 \mathcal{G}_{0;k^\prime}(\tau) ~ \expvalsmall{ \hat{S}_{k;0} \hat{S}_{0;k^\prime} }
 \\ \nonumber
 &=&
 \mathcal{G}_{0;k^\prime}(\tau) ~ \varrho_{k^\prime; k}
 \mathcomma \\ \nonumber
 \expvalsmall{ \hat{S}_{0;k}(\tau) }
 &=&
 \mathcal{G}_{0;k}(\tau) ~ \expvalsmall{ \hat{S}_{0;k}(0) }
 \\ \nonumber
 &=&
 \mathcal{G}_{0;k}(\tau) ~ \varrho_{k;0}(0)
 \mathcomma \\
 \expvalsmall{ \hat{S}_{k;0}(t) \hat{S}_{k^\prime; K \nu}(t+\tau) }
 &=&
 \mathcal{G}_{k^\prime;K \nu}(\tau) ~ \expvalsmall{ \hat{S}_{k;0} \hat{S}_{k^\prime; K \nu} }
 \\ \nonumber
 &=& 0 \mathcomma
 \\
 \expvalsmall{ \hat{S}_{K \nu;k}(t) \hat{S}_{0;k^\prime}(t+\tau) }
 &=&
 \mathcal{G}_{0;k^\prime}(\tau) ~ \expvalsmall{ \hat{S}_{K \nu;k} \hat{S}_{0;k^\prime} }
 \\ \nonumber
 &=& 0 
 \mathcomma \\
 \expvalsmall{ \hat{S}_{K \nu;k}(t) \hat{S}_{k^\prime; K^\prime \nu^\prime}(t+\tau) }
 &=&
 \mathcal{G}_{k^\prime; K^\prime \nu^\prime}(\tau) ~ \expvalsmall{ \hat{S}_{K \nu;k} \hat{S}_{k^\prime; K^\prime \nu^\prime} }
 \\ \nonumber
 &=&
 \mathcal{G}_{k^\prime; K^\prime \nu^\prime}(\tau) ~ \delta_{k k^\prime} ~ 
 \varrho_{K^\prime \nu^\prime; K \nu}
 \mathcomma
 \\ 
 \expvalsmall{ \hat{S}_{k;K \nu}(\tau) }
 &=&
 \mathcal{G}_{k;K \nu}(\tau) ~ \expvalsmall{ \hat{S}_{k;K \nu}(0) }
 \\ \nonumber
 &=& \mathcal{G}_{k;K \nu}(\tau) ~ \varrho_{K \nu;k}(0)
 \mathperiod
\end{eqnarray}
We can extract the propagators from Eqs.~(\ref{eq:incpumpcohstart})--(\ref{eq:incpumpcohend}),
yielding
\begin{eqnarray}
 \mathcal{G}_{0;k}(\tau) &=& \ee^{ - \ii \Delta^{k}_0  \tau}~\ee^{-\frac{1}{2} \left( \tilde{\Gamma}_k + \sum_n \abs{ \mathcal{P}_n }^2 \right) \tau}  \mathcomma \\
 \mathcal{G}_{k;K \nu}(\tau) &=& \ee^{ - \ii \Delta^{K \nu}_k  \tau}~\ee^{-\frac{1}{2} \left( \Gamma^{K \nu}_\mathrm{tot} + \tilde{\Gamma}_k \right) \tau}
 \mathcomma
\end{eqnarray}
and therefore
\begin{eqnarray}
 && \expvalsmall{ \hat{S}_{k;0}(t) \hat{S}_{0;k^\prime}(t+\tau) }
 = \\ \nonumber
 && ~~~~~~~~~
 \ee^{ - \ii \Delta^{k}_0  \tau}~\ee^{-\frac{1}{2} \left( \tilde{\Gamma}_k + \sum_n \abs{ \mathcal{P}_n }^2 \right) \tau} ~ \varrho_{k^\prime;k} \mathcomma
 \\ 
 && \expvalsmall{ \hat{S}_{K \nu;k}(t) \hat{S}_{k^\prime; K^\prime \nu^\prime}(t+\tau) }
 = \\ \nonumber
 && ~~~~~~~~~
 \ee^{ - \ii \Delta^{K \nu}_k  \tau}~\ee^{-\frac{1}{2} \left( \Gamma^{K \nu}_\mathrm{tot} + \tilde{\Gamma}_k \right) \tau}
 ~ \delta_{k k^\prime}
 ~ \varrho_{K^\prime \nu^\prime; K \nu}
 \mathperiod
\end{eqnarray}
The simplified expressions (for $\omega_0 \gg U \gg \gamma_0 \gg \abssmall{ \mathcal{P}_n }^2$) 
for the propagators are
\begin{eqnarray}
 \mathcal{G}_{0;k}(\tau) 	 &\simeq& \ee^{ - \ii \omega_0  \tau}~\ee^{-\frac{1}{2} \gamma_0 \tau}  \mathcomma \\
 \mathcal{G}_{k;K p}(\tau) &\simeq& \ee^{ - \ii \omega_0 \tau}~\ee^{-\frac{3}{2} \gamma_0 \tau} \mathcomma \\
 \mathcal{G}_{k;K, \mathrm{BS}}(\tau) &\simeq& \ee^{ - \ii \left( \omega_0+U \right) \tau}~\ee^{-\frac{3}{2} \gamma_0 \tau}
 \mathperiod
\end{eqnarray}
Note that in the presence of an incoherent pump, we have
$\varrho_{k^\prime;k} \propto \delta_{k k^\prime}$ and 
$\varrho_{K^\prime \nu^\prime; K \nu} \propto \delta_{K K^\prime} \delta_{\nu \nu^\prime}$.

Hence,
\begin{eqnarray}
 && \frac{{G}^{(1)}(\vv{r},t,t+\tau)}{\xi^2 \abs{\vv{w}(\vv{r})}^2 M}  = 
 \\ \nonumber
 && ~~~~ 
  \left(
      \mathcal{G}_{0; \bar{k}}(\tau) ~ N_{\bar{k}}
      ~+~ \sum_{K \nu} 
      \abs{ \bar{\eta}^{(K \nu)}_{\frac{K}{2}-\bar{k}} }^2
      \mathcal{G}_{K-\bar{k};K \nu}(\tau) ~ N_{K \nu}
  \right)
  \mathperiod
\end{eqnarray}
Specifically, for the single-pump setup 
(see Eqs.~(\ref{eq:sssol1appr})--(\ref{eq:sssol2appr}) for the nonzero occupation numbers) 
\begin{eqnarray}
 && \frac{{G}^{(1)}(\vv{r},t,t+\tau)}{\xi^2 \abs{\vv{w}(\vv{r})}^2 M}  = 
 \\ \nonumber
 && ~~~~ 
  \left(
      \mathcal{G}_{0; \bar{k}}(\tau) ~ N_{\bar{k}}
      ~+~ \sum_{\nu} 
      \abs{ \bar{\eta}^{(2k_P, \nu)}_{k_P-\bar{k}} }^2
      \mathcal{G}_{2k_P-\bar{k};2k_P, \nu}(\tau) ~ N_{2k_P, \nu}
  \right)
  \mathcomma
\end{eqnarray}
whereas for the two-pump setup as introduced in \secref{subsec:twopump}
we have (see \appref{subsec:apptwopump} for the steady-state solution)
\begin{eqnarray}
 && \frac{{G}^{(1)}(\vv{r}_1,t,t+\tau)}{\xi^2 \abs{\vv{w}(\vv{r}_1)}^2 M}  = 
 \\ \nonumber
 && ~~~~
 \Bigg{(}
      \mathcal{G}_{0; k_1}(\tau) ~ N_{k_1}
      ~+~ \sum_{\nu} 
      \abs{ \bar{\eta}^{(2k_1, \nu)}_{0} }^2
      \mathcal{G}_{k_1;2k_1, \nu}(\tau) ~ N_{2k_1, \nu}
      \\ \nonumber
      && ~~~~~~
      ~+~ 
      \left( 1 - \delta_{k_1 k_2} \right)
      \sum_{\nu} 
      \abs{ \bar{\eta}^{(k_1+k_2, \nu)}_{\frac{1}{2}\left(k_1-k_2\right)} }^2
      \mathcal{G}_{k_2;k_1+k_2, \nu}(\tau) ~ N_{k_1+k_2, \nu}
 \Bigg{)}
 \mathperiod
\end{eqnarray}
The Fourier transform according to \Eqref{eq:specdef} then finally yields 
the emission spectrum.

\section{Calculation of $\bar{\eta}^{(K \nu)}_q$}
\label{sec:eta}
Let us start with a closer inspection of \Eqref{eq:etafacdef} (remember that $\Psi^{(K \nu)}_0=0$):
\begin{eqnarray}
 \label{eq:etaredximed}
 \eta^{(K \nu)}_{q;n} &=& \frac{2}{\sqrt{M}} 
			\sum_{z=-N/2-n}^{N/2-n} 
			\ee^{\ii qa z}
			\Psi^{(K \nu)}_{z} 
 \\ \nonumber
 &=&
 \frac{2}{\sqrt{M}} 
 \left( \sum_{z=-N/2-n}^0 \ee^{\ii qa z} \Psi_z + \sum_{z=0}^{N/2-n} \ee^{\ii qa z} \Psi^{(K \nu)}_z \right)
 \\ \nonumber
 &=&
 \frac{2}{\sqrt{M}} 
 \left( \sum_{z=0}^{N/2-n} \ee^{\ii qa z} \Psi_z + \sum_{z=0}^{N/2+n} \ee^{-\ii qa z} \Psi^{(K \nu)}_z \right)
 \\ \nonumber
 &=&
 \frac{2}{\sqrt{M}} 
 \Bigg{(} \theta\left(\frac{N}{2}-n-1\right) \sum_{z=1}^{N/2-n} \ee^{\ii qa z} \Psi^{(K \nu)}_z 
 \\ \nonumber
 && ~~~~~
 ~+~ \theta\left(n+\frac{N}{2}-1\right) \sum_{z=1}^{N/2+n} \ee^{-\ii qa z} \Psi^{(K \nu)}_z \Bigg{)}
 \\ \nonumber
 &=&
 2 
 \sum_{\zeta = \pm}
 \theta\left(\frac{N}{2} + \zeta n - 1 \right) \sum_{z=1}^{N/2 + \zeta n} \frac{\ee^{ - \ii \zeta qa z}}{\sqrt{M}} \Psi^{(K \nu)}_z 
 \mathperiod
\end{eqnarray}
Here, we have used the discrete version of the Heaviside step function according to
$\theta(n) = 1$ if $n \geq 0$ and $\theta(n) = 0$ if $n < 0$.
From \Eqref{eq:etafacred}, we see that the quantity $\bar{\eta}^{(K \nu)}_{q}$ 
can be determined via (setting $\kappa=0$ in \Eqref{eq:etafacred})
\begin{eqnarray}
 \label{eq:etaredimed}
 \bar{\eta}^{(K \nu)}_{q} &=& 
 \frac{1}{2M} \sum_{n=-N/2}^{N/2} \eta^{(K \nu)}_{q; n}
 \\ \nonumber
 &=& 
 \frac{1}{M} 
 \sum_{\zeta = \pm}
 \sum_{n=-N/2}^{N/2} 
 \theta\left(\frac{N}{2} + \zeta n - 1 \right) 
 \\ \nonumber
 && ~~~~~
 \times
 \sum_{z=1}^{N/2 + \zeta n} \frac{\ee^{ - \ii \zeta qa z}}{\sqrt{M}} \Psi^{(K \nu)}_z 
 \mathperiod
\end{eqnarray}

Before we proceed with the actual calculation, we need to be cautious regarding the wavefunction's boundary conditions.
Usually, the boundary conditions must not play a role for a lattice with many atoms ($M \gg 1$).
However, being pedantic, the eigenstates presented in \secref{subsec:eigenstates} 
actually do not satisfy hard-wall boundary conditions for a finite lattice 
(in the calculation, it is assumed that $M$ is still finite).
At the boundary, the wavefunction should actually vanish.
This is not the case here as the ansatz we employed (both for single- and two-excitation states) 
represents open boundary conditions, \ie,
in- and outgoing waves that are suited in the context of an infinite lattice.
To be precise,
Equations~(\ref{eq:etaredximed}) and~(\ref{eq:etaredimed}) explicitly depend on the boundary values 
(for instance, the term for $n=\zeta N/2$ in \Eqref{eq:etaredximed} requires the relative wavefunction 
at the relative coordinate $z=N$, which may stand for two excitations at the boundaries $x_1=N/2$ and $x_2=-N/2$).
Furthermore, the sum over $z$ depends on $n$ and represents a finite \anfz{window}. 
If we, however, combine the demand that \anfz{integrated} quantities such as \Eqref{eq:etaredimed} must be 
independent of the boundary conditions in the limit of a large (eventually infinite) lattice,
we come to the conclusion that the \anfz{window length} should not affect Eqs.~(\ref{eq:etaredximed}) and~(\ref{eq:etaredimed}).
Hence, we proceed with the expressions
\begin{eqnarray}
 \eta^{(K \nu)}_{q;n} &=& 
 2 \sum_{\zeta = \pm}
 \sum_{z=1}^{N/2} \frac{\ee^{ - \ii \zeta qa z}}{\sqrt{M}} \Psi^{(K \nu)}_z 
 \mathcomma
 \\
 \bar{\eta}^{(K \nu)}_{q} &=& 
 \frac{1}{M} 
 \sum_{\zeta = \pm}
 \sum_{n=-N/2}^{N/2} 
 \sum_{z=1}^{N/2} \frac{\ee^{ - \ii \zeta qa z}}{\sqrt{M}} \Psi^{(K \nu)}_z 
 \mathcomma
\end{eqnarray}
(where we additionally ignored the $\theta$-function for similar reasons).
Performing the sum over $n$, 
we are ultimately left with
\begin{equation}
 \label{eq:etafacredexpr2calc}
 \bar{\eta}^{(K \nu)}_{q} =
 \sum_{\zeta = \pm}
 \sum_{z=1}^{N/2} \frac{\ee^{ - \ii \zeta qa z}}{\sqrt{M}} \Psi^{(K \nu)}_z 
 \mathperiod
\end{equation}

\subsection{Scattering States ($\bar{\eta}^{(K p)}_q$)}
For scattering states, we arrive at
(the Kronecker symbols $\delta_{p , \pm q}$ should not be confused with the scattering 
phase shift in $\exp(\ii \delta_{Kp})$)
\begin{eqnarray}
 && \bar{\eta}^{(K p)}_{q} =
 \sum_{\zeta = \pm}
 \frac{1}{M} 
 \sum_{z=1}^{N/2}   
 \left( \ee^{ \ii \left( p - \zeta q \right)a z}  +  \ee^{\ii \delta_{Kp}} \ee^{ -\ii \left( p + \zeta q \right)a z}  \right)
 \\
 \nonumber
 && =
 \sum_{\zeta = \pm}
 \Bigg{[}
 \frac{1}{2} \left( \delta_{p, \zeta q} + \ee^{\ii \delta_{Kp}} \delta_{p,-\zeta q} \right)
 \\ \nonumber
 && ~~~~~~~ 
 + \frac{1}{M} \left( 1 - \delta_{p,\zeta q} \right) \left(  \frac{1 - \ee^{\ii \left( p-\zeta q \right)a\left( \frac{N}{2}+1\right)}}{1 - \ee^{\ii \left( p-\zeta q \right)a}}  - 1 \right)
 \\ \nonumber
 && ~~~~~~~ 
 + \frac{1}{M} \left( 1 - \delta_{p,-\zeta q} \right) \left(  \frac{1 - \ee^{-\ii \left( p+\zeta q \right)a\left( \frac{N}{2}+1\right)}}{1 - \ee^{-\ii \left( p+\zeta q \right)a}}  - 1 \right) \ee^{\ii \delta_{Kp}}
 \Bigg{]}
 \\ \nonumber
 && =
 \frac{1}{2} \left( 1+\ee^{\ii \delta_{Kp}} \right) \left( \delta_{pq} + \delta_{p,-q} \right)
 \\ \nonumber
 && ~~~
 + \frac{1}{M} \left( 1-\delta_{pq} \right) 
 \left(  
  \frac{ 1-\ee^{\ii \left(p-q\right)a \frac{N}{2} } }{\ee^{-\ii\left( p-q \right)a}-1}
  + \ee^{\ii \delta_{Kp}} \frac{ 1-\ee^{-\ii \left(p-q\right)a \frac{N}{2} } }{\ee^{\ii\left( p-q \right)a}-1}  
 \right)
 \\ \nonumber
 && ~~~
 + \frac{1}{M} \left( 1-\delta_{p,-q} \right) 
 \left(  
  \frac{ 1-\ee^{\ii \left(p+q\right)a \frac{N}{2} } }{\ee^{-\ii\left( p+q \right)a}-1}
  + \ee^{\ii \delta_{Kp}} \frac{ 1-\ee^{-\ii \left(p+q\right)a \frac{N}{2} } }{\ee^{\ii\left( p+q \right)a}-1}  
 \right)
 \mathperiod
\end{eqnarray}
During the derivation, we have used the formula for finite geometric sums,
and we also assumed $M,N \gg 1$.
Note the symmetry property $\bar{\eta}^{(K p)}_{q} = \bar{\eta}^{(K p)}_{-q}$.
For $q \geq 0$ and $p>0$, we may therefore write 
\begin{eqnarray}
 \label{eq:etascattred}
 \bar{\eta}^{(K, p>0)}_{q \geq 0} &=&
 \frac{1+\ee^{\ii \delta_{Kp}}}{2} \delta_{pq} 
 \\ \nonumber
 && ~
 +~
 \frac{2 \ee^{\ii \frac{\delta_{Kp}}{2}}}{M} \cdot
 \frac{\sin\left[ \frac{1}{2} \left( \delta_{Kp} - \left(p-q\right)a \right) \right]}{\sin\left[\frac{1}{2}\left(p-q\right)a\right]}
 \\ \nonumber
 && ~~~~~~~
 ~\times~
 \left( 1-\delta_{pq} \right) 
 \left( 1-\delta_{\left[\frac{\left(p-q\right)aN}{2\pi}\right]_{2} ,0} \right) 
 \mathperiod
\end{eqnarray}
During this derivation, we have exploited the fact that the difference of the wavenumbers 
$p$ and $q$ can be written as 
$p-q = m \cdot 2 \pi/Ma$, where $m$ is an integer.
Hence, $(p-q)Na/2 \overset{M \gg 1}{=} (p-q)Ma/2 =  m \pi$ 
and $\exp\left[ (p-q)N/2 \right] = (-1)^m$
so that we only have a contribution from the \anfz{background term} if $m$ is odd
(expressed through the second Krocker-$\delta$).

\subsection{Bound States ($\bar{\eta}^{(K, \mathrm{BS})}_q$)}
Proceeding along the same lines for the case of a bound state, we have the 
expression
\begin{eqnarray}
 \bar{\eta}^{(K,\mathrm{BS})}_{q} &=&
 \sum_{\zeta = \pm}
 \sum_{z=1}^{N/2} \frac{\ee^{ - \ii \zeta qa z}}{\sqrt{M}} \alpha_K^{z-1}
 \\ \nonumber
 &=&
 \frac{1}{\sqrt{M} \alpha_K}
 \sum_{\zeta = \pm}
 \sum_{z=1}^{N/2} \left( \ee^{ - \ii \zeta qa} \alpha_K \right)^z
 \\ \nonumber
 &=&
 \frac{1}{\sqrt{M} \alpha_K}
 \sum_{\zeta = \pm}
 \left( \frac{1 - \left( \ee^{ - \ii \zeta qa} \alpha_K \right)^{N/2+1} }{ 1 - \ee^{ - \ii \zeta qa} \alpha_K } - 1 \right)
 \\ \nonumber
 &=&
 \frac{2}{\sqrt{M}} \cdot \frac{\cos(qa) - \alpha_K}{1-2\alpha_K \cos(qa) + \alpha_K^2}
 \\ \nonumber
 &\overset{U \gg \gamma_0}{\simeq}&
 \frac{2 \cos(qa)}{\sqrt{M}}
 \mathperiod
\end{eqnarray}
In the derivation, we have exploited the fact that $\alpha_K^{N/2}$ vanishes 
since $N \gg 1$.
In the last line, we have additionally assumed $U \gg \gamma_0$, meaning
the bound state is tightly confined with respect to the relative coordinate.

\subsection{Sum over $\abs{\bar{\eta}^{(K \nu)}_q}^2$}
Consider the quantity
\begin{eqnarray}
 Z^{(K \nu)} &\equiv& \sum_{qa=\frac{K}{2}-\pi}^{\frac{K}{2}+\pi} \abs{\bar{\eta}^{(K \nu)}_q}^2
		  = \sum_{qa=-\pi}^{\pi} \abs{\bar{\eta}^{(K \nu)}_q}^2
 \\ \nonumber
 &=& 2 \sum_{qa=0}^{\pi} \abs{\bar{\eta}^{(K \nu)}_q}^2  - \abs{\bar{\eta}^{(K \nu)}_0}^2
 \\ \nonumber
 &\overset{M\gg1}{=}& 2 \sum_{qa=0}^{\pi} \abs{\bar{\eta}^{(K \nu)}_q}^2
 \mathperiod
\end{eqnarray}

For scattering states, we have
\begin{eqnarray}
 Z^{(K p)} &=& 2 \left( \sum_{qa=0, q \neq p}^{\pi} \abs{\bar{\eta}^{(K p)}_q}^2 + \abs{\bar{\eta}^{(K p)}_p}^2 \right)
 \\ \nonumber
 &=& 2 \Bigg{(}
 \cos^2 \left( \frac{\delta_{Kp}}{2} \right) 
 \\ \nonumber
 && ~~~
 ~+~
 \frac{4}{M^2} 
 \sum_{qa=0, q \neq p}^{\pi}
 \frac{\cos\left[\delta_{Kp} +(q-p)a \right]-1}{\cos\left[(p-q)a\right]-1}
 \\ \nonumber
 && ~~~~~~
 ~\times~ \left( 1-\delta_{\left[\frac{\left(p-q\right)aN}{2\pi}\right]_{2} ,0} \right) 
 \Bigg{)}
 \mathperiod
\end{eqnarray}
The lattice sum can be performed in the limit $M \gg 1$.
Defining $\delta \equiv 2\pi/aM \ll 1$ and rewriting $q=p+\delta \cdot n$, we
essentially have to calculate
\begin{equation}
 \frac{\delta^2}{\pi^2} \sideset{}{^\prime}\sum_{n}^{} \frac{\cos\left(\delta_{Kp} + \delta n \right)-1}{\cos\left(\delta n\right)-1}
 \mathcomma
\end{equation}
which can be expanded into a Taylor series in $\delta$ in which only terms of order $\delta^0$ remain 
in the limit $M \gg 1$,
yielding
\begin{equation}
 \frac{2}{\pi^2} \left( 1-\cos\delta_{Kp} \right) \sideset{}{^\prime}\sum_n \frac{1}{n^2}
 \mathperiod
\end{equation}
The sum runs over those odd values of $n$ such that all wavenumbers from 
$0 \rightarrow p-\delta$ as well as from $p + \delta \rightarrow \pi/a$ are covered.
In the limit $M\gg1$, $n$ runs over all odd values from $1$ to $\infty$ as well as from $-1$ to $-\infty$,
resulting in
\begin{equation}
 2 \underbrace{\sum_{n=0}^{\infty} \frac{1}{\left( 2n + 1\right)^2}}_{\frac{\pi^2}{8}} = \frac{\pi^2}{4}
 \mathperiod
\end{equation}
This eventually leads to 
\begin{equation}
 Z^{(K p)} = 2 \cdot \left( 
 \cos^2 \left( \frac{\delta_{Kp}}{2} \right) 
 +
 \frac{1}{2} \left( 1 - \cos\delta_{Kp} \right)
 \right)
 = 2
 \mathperiod
\end{equation}

Similarly, the calculation for a bound states is ($U \nolinebreak \gg \nolinebreak \gamma_0$)
\begin{eqnarray}
 Z^{(K, \mathrm{BS})} &=& \sum_{qa=-\pi}^{\pi} \abs{\bar{\eta}^{(K, \mathrm{BS})}_q}^2 
 \\ \nonumber
 &=& \frac{4}{M} \sum_{qa=-\pi}^{\pi} \cos^2\left(qa\right) \\ \nonumber
 &\rightarrow&
 4 \int \limits_{-\frac{\pi}{a}}^{\frac{\pi}{a}} \frac{\mathrm{d}q}{2 \pi} \cos^2\left(qa\right)
 = 2 
 \mathperiod
\end{eqnarray}

Hence, we can generally write $Z^{(K \nu)} = 2$.

\section{Equations of Motion with Incoherent Driving Field}
\label{sec:eqominc}
We can extend Eqs.~(\ref{eq:eqstart})--(\ref{eq:eqend}) to account for an 
external, incoherent driving field,
yielding the following equations of motion for the density
matrix elements:
\begin{eqnarray}
\label{eq:glA}
 \partial_t \varrho_{K \nu;K^\prime \nu^\prime} &=& 
 \left[ - \ii \Delta^{K \nu}_{K^\prime \nu^\prime} - \frac{1}{2} \left( \Gamma^{K \nu}_{\mathrm{tot}}  +  \Gamma^{K^\prime \nu^\prime}_{\mathrm{tot}} \right) \right] 
  ~ \varrho_{K \nu; K^\prime \nu^\prime}
 \\ \nonumber
 && 
 \hspace{-1.2cm} ~+~ \delta_{K K^\prime} \delta_{\nu \nu^\prime} \sum_{k} \sum_n
 \delta_{K, k+k_n} \abs{\mathcal{P}_n}^2 \abs{ \bar{\eta}^{(k+k_n,\nu)}_{\frac{1}{2} \left( k_n -k \right)} }^2 
 ~\varrho_{k;k}
 \mathcomma \\
 \label{eq:glB}
 \partial_t \varrho_{k;k^\prime} &=& 
 \left[ -\ii \Delta^k_{k^\prime} - \frac{1}{2} \left( \Gamma_{k} + \Gamma_{k^\prime} \right) \right] ~ \varrho_{k; k^\prime}
 \\ \nonumber
 && \hspace{-1.2cm}
  ~-~ \frac{1}{2} \sum_n \sum_{\nu}
  \abs{ \mathcal{P}_n }^2
  \left( 
    \abs{ \bar{\eta}^{(k+k_n,\nu)}_{\frac{1}{2} \left( k_n -k \right)} }^2 
    +
    \abs{ \bar{\eta}^{(k^\prime+k_n,\nu)}_{\frac{1}{2} \left( k_n -k^\prime \right)} }^2
  \right)
  ~\varrho_{k;k^\prime}
  \\ \nonumber
  && \hspace{-1.2cm}
  ~+~  \delta_{k, k^\prime} ~ \sum_n \abs{ \mathcal{P}_n }^2 \delta_{k, k_n} ~ \varrho_{0;0}
  \\ \nonumber
  && \hspace{-1.2cm}
  ~+~ \delta_{k, k^\prime} ~ \sum_{K^\prime \nu^\prime}
    \Gamma^{K^\prime \nu^\prime}_k \varrho_{K^\prime \nu^\prime; K^\prime \nu^\prime}
 \mathcomma \\
 \label{eq:glC}
 \partial_t \varrho_{0;0} &=&
 \sum_k \Gamma_k \varrho_{k;k} 
  ~-~ \sum_n \abs{ \mathcal{P}_n }^2 ~ \varrho_{0;0}
 \mathcomma
 \\
 \label{eq:glD}
 \partial_t \varrho_{K \nu;k} &=&
 \left[ -\ii \Delta^{K \nu}_k - \frac{1}{2} \left( \Gamma^{K \nu}_{\mathrm{tot}} + \Gamma_k \right) \right] \varrho_{K\nu; k} \\ \nonumber
 && \hspace{0.3cm}
 - ~ \frac{1}{2} \sum_n \sum_{\nu^\prime} ~
  \abs{ \mathcal{P}_n }^2 \abs{ \bar{\eta}^{(k+k_n,\nu^\prime)}_{\frac{1}{2} \left( k_n -k \right)} }^2
 ~\varrho_{K \nu;k}
 \mathcomma \\
 \label{eq:glE}
 \partial_t \varrho_{K \nu;0} &=&
 \left[ -\ii \mathrm{Re}(E^{(2)}_{K \nu}) - \frac{1}{2} \Gamma^{K \nu}_{\mathrm{tot}} \right] \varrho_{K\nu; 0}
 \\ \nonumber
 && \hspace{0.3cm}
 -~ \frac{1}{2} \sum_n \abs{ \mathcal{P}_n }^2 ~ \varrho_{K \nu; 0}
 \mathcomma \\
 \label{eq:glF}
 \partial_t \varrho_{k;0} &=&
 \left[ -\ii \Delta^k_0 -\frac{1}{2} \Gamma_k \right] \varrho_{k;0}
 \\ \nonumber
 && \hspace{0.3cm}
 -~ \frac{1}{2} \sum_n \sum_{\nu} 
 \abs{ \mathcal{P}_n }^2 \abs{ \bar{\eta}^{(k+k_n,\nu)}_{\frac{1}{2} \left( k_n -k \right)} }^2
 ~\varrho_{k;0}  \\ \nonumber
 && \hspace{0.3cm}
 -~ \frac{1}{2} \sum_n \abs{ \mathcal{P}_n }^2 ~\varrho_{k;0}
 \mathperiod
\end{eqnarray}
The coherences ($K \neq K^\prime$, $\nu \neq \nu^\prime$, $k \neq k^\prime$)
evolve in time according to
\begin{eqnarray}
 \label{eq:incpumpcohstart}
 \varrho_{K \nu; K^\prime \nu^\prime}(t)
 &=&
 \varrho_{K \nu; K^\prime \nu^\prime}(0)
 ~\ee^{ - \ii \Delta^{K \nu}_{K^\prime \nu^\prime}  t}
 ~\ee^{- \frac{1}{2} \left( \Gamma^{K \nu}_{\mathrm{tot}} + \Gamma^{K^\prime \nu^\prime}_{\mathrm{tot}} \right) t}
 \mathcomma \\
 \varrho_{k; k^\prime}(t)
 &=&
 \varrho_{k; k^\prime}(0)
 ~\ee^{ - \ii \Delta^{k}_{k^\prime}  t}
 ~\ee^{-\frac{1}{2} \left( \tilde{\Gamma}_k + \tilde{\Gamma}_{k^\prime} \right) t}
 \mathcomma \\
 \varrho_{K \nu; k}(t)
 &=&
 \varrho_{K \nu; k}(0)
 ~\ee^{ - \ii \Delta^{K \nu}_k  t}
 ~\ee^{-\frac{1}{2} \left( \Gamma^{K \nu}_\mathrm{tot} + \tilde{\Gamma}_k \right) t}
 \mathcomma \\
 \varrho_{K \nu; 0}(t)
 &=&
 \varrho_{K \nu; 0}(0)
 ~\ee^{ - \ii \mathrm{Re}(E^{(2)}_{K \nu})  t}
 ~\ee^{-\frac{1}{2} \left[ \Gamma^{K \nu}_\mathrm{tot} + \sum_n \abs{ \mathcal{P}_n }^2 \right] t}
 \mathcomma \\
 \label{eq:incpumpcohend}
 \varrho_{k; 0}(t)
 &=&
 \varrho_{k; 0}(0)
 ~\ee^{ - \ii \Delta^{k}_0  t}
 ~\ee^{-\frac{1}{2} \left( \tilde{\Gamma}_k + \sum_n \abs{ \mathcal{P}_n }^2 \right) t}
 \mathcomma
\end{eqnarray}
where
$\tilde{\Gamma}_k \equiv {\Gamma}_k  + \sum_n \sum_{\nu}  \abs{\mathcal{P}_n}^2 ~ \abssmall{\bar{\eta}^{(k+k_n,\nu)}_{\frac{1}{2} \left( k_n -k \right)}}^2$.
For $\omega_0 \gg \gamma_0 \gg \abs{\mathcal{P}_n}^2$ we are left with
\begin{eqnarray}
 \label{eq:cohsimpA}
 \varrho_{K \nu; K^\prime \nu^\prime}(t)
 &\simeq&
 \varrho_{K \nu; K^\prime \nu^\prime}(0)
 ~\ee^{- 2 \gamma_0 t}
 \mathcomma \\
 \label{eq:cohsimpB}
 \varrho_{k; k^\prime}(t)
 &\simeq&
 \varrho_{k; k^\prime}(0)
 ~\ee^{- \gamma_0 t}
 \mathcomma \\
 \label{eq:cohsimpC1}
 \varrho_{K p; k}(t)
 &\simeq&
 \varrho_{K p; k}(0)
 ~\ee^{ - \ii \omega_0 t}
 ~\ee^{-\frac{3}{2} \gamma_0 t}
 \mathcomma \\
 \label{eq:cohsimpC2}
 \varrho_{K, \mathrm{BS}; k}(t)
 &\simeq&
 \varrho_{K, \mathrm{BS}; k}(0)
 ~\ee^{ - \ii \left( \omega_0 + U \right)  t}
 ~\ee^{-\frac{3}{2} \gamma_0 t}
 \mathcomma \\
 \label{eq:cohsimpD1}
 \varrho_{K p; 0}(t)
 &\simeq&
 \varrho_{K p; 0}(0)
 ~\ee^{ - 2 \ii \omega_0 t}
 ~\ee^{- \gamma_0 t}
 \mathcomma \\
 \label{eq:cohsimpD2}
 \varrho_{K, \mathrm{BS}; 0}(t)
 &\simeq&
 \varrho_{K, \mathrm{BS}; 0}(0)
 ~\ee^{ - \ii \left( 2 \omega_0 + U \right)  t}
 ~\ee^{- \gamma_0 t}
 \mathcomma \\
 \label{eq:cohsimpE}
 \varrho_{k; 0}(t)
 &\simeq&
 \varrho_{k; 0}(0)
 ~\ee^{ - \ii \omega_0  t}
 ~\ee^{-\frac{1}{2} \gamma_0 t}
 \mathperiod
\end{eqnarray}

For the diagonal elements $N_m \equiv \varrho_{m;m}$, we arrive at a set of coupled
rate equations,
which can be compactly written as 
\begin{eqnarray}
 \label{eq:incpumpratesI}
 \left[ \partial_t + \Gamma^{K \nu}_{\mathrm{tot}} \right] N_{K \nu} 
 &=&
 \sum_n \abs{\mathcal{P}_n}^2 \abs{\bar{\eta}^{(K,\nu)}_{\frac{K}{2} - k_n}}^2 N_{K-k_n}
 \mathcomma
 \\ 
 \label{eq:incpumpratesII}
 \left[ \partial_t + \tilde{\Gamma}_k \right] N_{k} 
 &=&
 \sum_n \abs{\mathcal{P}_n}^2 \delta_{k k_n} \left( 1- \sum_k N_k - \sum_{K \nu} N_{K \nu} \right)
 \\ \nonumber
 && ~~~ ~+~ \sum_{K \nu} \Gamma^{K \nu}_k N_{K \nu}
 \mathcomma
\end{eqnarray}
where we have exploited the conservation of the total probability
\begin{equation}
 N_0 + \sum_{k} N_k + \sum_{K \nu} N_{K \nu} = 1
 \mathperiod
\end{equation}
Consequently, the steady state obeys
\begin{eqnarray}
 \label{eq:ssIapp}
 N_{K \nu} 
 &=&
 \sum_n \frac{\abs{\mathcal{P}_n}^2}{\Gamma^{K \nu}_{\mathrm{tot}}} \abs{\bar{\eta}^{(K,\nu)}_{\frac{K}{2} - k_n}}^2 N_{K-k_n}
 \mathcomma
 \\ 
 \label{eq:ssIIapp}
 N_{k} 
 &=&
 \sum_n \frac{\abs{\mathcal{P}_n}^2}{\tilde{\Gamma}_k} \delta_{k k_n} \left( 1- \sum_k N_k - \sum_{K \nu} N_{K \nu} \right)
 \\ \nonumber
 && ~~~ ~+~ \sum_{K \nu} \frac{\Gamma^{K \nu}_k}{\tilde{\Gamma}_k} N_{K \nu}
 \mathperiod
\end{eqnarray}

\subsection{Steady-State Solution: Single Pump}
\label{subsec:appsingpump}
The steady-state solution to Eqs.~(\ref{eq:ssIapp}) and~(\ref{eq:ssIIapp})
for the case of a driving field with a single Fourier component
(pump rate $\abssmall{P}^2$, \anfz{imprinted} wavenumber $k_P$) 
can be constructed as follows.
We insert \Eqref{eq:ssIapp} into \Eqref{eq:ssIIapp} and approximate
$N_0 \approx 1$ (weak pump) and arrive at
\begin{equation}
 \label{eq:steadstatimed}
 N_k \simeq \frac{\abs{\mathcal{P}}^2}{\tilde{\Gamma}_k} 
  \left( \delta_{k k_P} 
    + \sum_{K \nu} \frac{\Gamma^{K \nu}_k}{\Gamma^{K \nu}_{\mathrm{tot}}} \abs{\bar{\eta}^{(K \nu)}_{\frac{K}{2}-k_P}}^2 N_{K-k_P}  \right)
 \mathperiod
\end{equation}
Evaluating this equation for $k \neq k_P$
yields the occupation of a single excitation state $\ketsmall{k}$
that is not directly connected to the pump field and can only be populated
via spontaneous emission from a driven two-excitation state.
The dominant contributions in the sum on the right-hand side of 
\Eqref{eq:steadstatimed} is thus from the driven states with $K-k_P=k_P$.
The other states ($K-k_P \neq k_P$) are of even higher order in the 
pump rates and may be neglected.
This idea leads us to
\begin{eqnarray}
 \label{eq:imedofpump}
 N_{k \neq k_P}
 &\simeq&
 \frac{\abs{\mathcal{P}}^2}{\tilde{\Gamma}_k} 
 \sum_{\nu} \frac{\Gamma^{2 k_P, \nu}_k}{\Gamma^{2 k_P, \nu}_{\mathrm{tot}}} \abs{\bar{\eta}^{(2 k_P, \nu)}_{0}}^2 N_{k_P}
 \mathperiod
\end{eqnarray}
Conversely,
if we evaluate \Eqref{eq:steadstatimed} for $k=k_P$
we may assume 
\begin{equation}
 1 \gg \sum_{K \neq 2 k_P} \sum_{\nu} \frac{\Gamma^{K \nu}_k}{\Gamma^{K \nu}_{\mathrm{tot}}} \abs{\bar{\eta}^{(K \nu)}_{\frac{K}{2}-k_P}}^2 N_{K-k_P}
\end{equation}
so that 
\begin{equation}
 N_{k=k_P} \simeq  
 \frac{\abs{\mathcal{P}}^2}{\tilde{\Gamma}_{k_P}} 
  \left( 1
    +  \sum_{\nu} \frac{\Gamma^{2 k_P, \nu}_{k_P}}{\Gamma^{2 k_P, \nu}_{\mathrm{tot}}} \abs{\bar{\eta}^{(2 k_P, \nu)}_{0}}^2 N_{k_P}  \right)
 \mathperiod
\end{equation}
Up to second order in the pump rates, we can therefore write
(remember $1/(1-x) \approx 1+x,~x \ll 1$)
\begin{equation}
 N_{k=k_p} \simeq 
 \frac{\abs{\mathcal{P}}^2}{\tilde{\Gamma}_{k_P}}  
 + \left( \frac{\abs{\mathcal{P}}^2}{\tilde{\Gamma}_{k_P}} \right)^2 \sum_{\nu} \frac{\Gamma^{2 k_P, \nu}_{k_P}}{\Gamma^{2 k_P, \nu}_{\mathrm{tot}}} \abs{\bar{\eta}^{(2 k_P, \nu)}_{0}}^2 
 \mathperiod
\end{equation}
Inserted back into \Eqref{eq:imedofpump}, we arrive at
\begin{equation}
 N_{k \neq k_P}
 \simeq
 \frac{\abs{\mathcal{P}}^2}{\tilde{\Gamma}_k} 
 \frac{\abs{\mathcal{P}}^2}{\tilde{\Gamma}_{k_P}}
 \sum_{\nu} \frac{\Gamma^{2 k_P, \nu}_k}{\Gamma^{2 k_P, \nu}_{\mathrm{tot}}} \abs{\bar{\eta}^{(2 k_P, \nu)}_{0}}^2
\end{equation}
for the undriven single-excitation states.
Going back to \Eqref{eq:ssIapp} and using the previous results,
we see that the dominant terms on the right-hand side stem from 
$K=2 k_P$.
Any other $K$ value implies a scaling with the third power in
the pump rates, which we neglect.

For $\omega_0 \gg \gamma_0$, we can then finally write
\begin{eqnarray}
 \label{eq:sssol1appr}
 N_{k} &\simeq& 
  \Xi \delta_{k k_P} 
   + \Xi^2 \sum_{\nu} b^{(2 k_P, \nu)}_k \abs{\bar{\eta}^{(2 k_P, \nu)}_{0}}^2
   \\ \nonumber
   &=& 
  \Xi \delta_{k k_P} 
   + \frac{\Xi^2}{2} \sum_{\nu} \abs{\bar{\eta}^{(2 k_P, \nu)}_{k_P-k}}^2 \abs{\bar{\eta}^{(2 k_P, \nu)}_{0}}^2
  \mathcomma \\  
 \label{eq:sssol2appr}
 N_{K \nu} &\simeq& 
  \frac{\Xi^2}{2} 
  \abs{\bar{\eta}^{(2 k_P, \nu)}_{0}}^2
  \delta_{K, 2 k_P}
 \mathcomma
\end{eqnarray}
where $\Xi \equiv \abssmall{\mathcal{P}}^2 / \gamma_0$.

\subsection{Steady-State Solution: Two Pumps}
\label{subsec:apptwopump}
Using the same reasoning as in \appref{subsec:appsingpump},
the steady-state occupation numbers can be obtained for the
case of an external field with two spatial Fourier components.
The pump rates are, respectively, 
$\abssmall{\mathcal{P}_1}^2 \equiv \abssmall{\mathcal{P}}^2$ and 
$\abssmall{\mathcal{P}_2}^2 \equiv \epsilon^2 \abssmall{\mathcal{P}}^2$, 
and $k_1$ and $k_2$ denote the \anfz{imprinted} wavenumbers.
We only need the solution for wavenumbers 
$K=2 k_1, 2 k_2, k_1+k_2$ and $k=k_1,k_2$ (see main text).
For $k_1 \neq k_2$, the solution reads (again, $\Xi \equiv \abssmall{\mathcal{P}}^2 / \gamma_0$)
\begin{eqnarray}
 \label{eq:2pssbegin}
 N_{2k_1, \nu} &\simeq& \frac{\Xi^2}{2} \abs{\bar{\eta}^{(2 k_1, \nu)}_{0}}^2 
 \mathcomma \\
 N_{2k_2, \nu} &\simeq& \epsilon^4 \frac{\Xi^2}{2} \abs{\bar{\eta}^{(2 k_2, \nu)}_{0}}^2 
 \mathcomma \\
 N_{k_1+k_2, \nu} &\simeq& \epsilon^2 \Xi^2 \abs{\bar{\eta}^{(k_1+k_2, \nu)}_{\frac{1}{2}\left(k_1-k_2\right)}}^2 
 \mathcomma \\
 N_{k_1} &\simeq& \Xi 
		~+~ \frac{\Xi^2}{2} \sum_{\nu} \abs{\bar{\eta}^{(2 k_1, \nu)}_{0}}^4
		\\ \nonumber
		&& ~~~~~ ~+~ \epsilon^2 \Xi^2 \sum_{\nu} \abs{\bar{\eta}^{(k_1+k_2, \nu)}_{\frac{1}{2}\left(k_1-k_2\right)}}^4
				\\ \nonumber
		&& ~~~~~ ~+~ \epsilon^4 \frac{\Xi^2}{2} \sum_{\nu} \abs{\bar{\eta}^{(2 k_2, \nu)}_{k_2-k_1}}^2 \abs{\bar{\eta}^{(2 k_2, \nu)}_{0}}^2
 \mathcomma \\
 N_{k_2} &\simeq& \epsilon^2 \Xi 
		~+~ \epsilon^4 \frac{\Xi^2}{2} \sum_{\nu} \abs{\bar{\eta}^{(2 k_2, \nu)}_{0}}^4
		\\ \nonumber
		&& ~~~~~ ~+~ \epsilon^2 \Xi^2 \sum_{\nu} \abs{\bar{\eta}^{(k_1+k_2, \nu)}_{\frac{1}{2}\left(k_1-k_2\right)}}^4
				\\ \nonumber
		&& ~~~~~ ~+~ \frac{\Xi^2}{2} \sum_{\nu} \abs{\bar{\eta}^{(2 k_1, \nu)}_{k_2-k_1}}^2 \abs{\bar{\eta}^{(2 k_1, \nu)}_{0}}^2
 \mathperiod
\end{eqnarray}
For $k_1=k_2 \equiv k_P$, we use the single-pump results~(\ref{eq:sssol1appr})--(\ref{eq:sssol2appr}),
where we need to replace 
$\abs{\mathcal{P}}^2 \rightarrow \abs{\mathcal{P}}^2 \left( 1 + \epsilon^2 \right)$, yielding
\begin{eqnarray}
 N_{2 k_P, \nu} &\simeq& 
   \left( 1 + \epsilon^2 \right)^2 \frac{\Xi^2}{2} \abs{\bar{\eta}^{(2 k_P, \nu)}_{0}}^2
   \mathcomma \\
 \label{eq:2pssend}
 N_{k_P} &\simeq& 
  \left( 1 + \epsilon^2 \right) \Xi  
   + \left( 1 + \epsilon^2 \right)^2 \frac{\Xi^2}{2}
     \sum_{\nu} \abs{\bar{\eta}^{(2 k_P, \nu)}_{0}}^4
 \mathperiod
\end{eqnarray}

\section{Sums over Products of $\abs{\bar{\eta}^{(K \nu)}_q}^2$}
\label{sec:appetasums}
In this paper, we encounter sums of the form 
\begin{eqnarray}
\sum_\nu \abs{\bar{\eta}^{(K \nu)}_q}^4 &=& \abs{\bar{\eta}^{(K, \mathrm{BS})}_q}^4  +  \sum_p \abs{\bar{\eta}^{(K p)}_q}^4
 \\ \nonumber
 &\equiv& \begin{cases}
      Q^{(\mathrm{bg})}_q + Q^{(\mathrm{dir})}_q  & U=0 \\
      R^{(\mathrm{bg})}_q + R^{(\mathrm{dir})}_q + R^{(\mathrm{BS})}_q & U \gg \gamma_0
     \end{cases}
 \mathperiod
\end{eqnarray}
Exploiting the symmetry properties, we can further write
\begin{eqnarray}
 && \sum_p \abs{\bar{\eta}^{(K p)}_q}^4 
  = 
  2 \cdot \sum_{p>0} \abs{\bar{\eta}^{(K p)}_q}^4 + \abs{\bar{\eta}^{(K, p=0)}_q}^4
  \\ \nonumber
  && ~~~
  = 2 \cdot \left( \sum_{p>0, p \neq q} \abs{\bar{\eta}^{(K p)}_q}^4  + \abs{\bar{\eta}^{(K q)}_q}^4 \right) + \abs{\bar{\eta}^{(K, p=0)}_q}^4.
\end{eqnarray}
Here, the first part (the sum over $p>0$ but $p \neq q$) represents the contribution from the 
\anfz{background fluorescence} (\anfz{bg}), 
whereas the second term ($p=q$) stems from the emission via a
\anfz{direct channel} (\anfz{dir}).
The last term refers to a $p=0$-relative wavefunction and therefore vanishes 
for both $U=0$ and $U \gg \gamma_0$ (see also \secref{subsec:eigenstates}).

The relevant quantities to be calculated are
\begin{eqnarray}
 Q^{(\mathrm{bg})}_q  &\equiv& 2 \sum_{p>0, p \neq q} \abs{\bar{\eta}^{(K p)}_q}^4~~~\mathrm{for~U=0} \mathcomma \\
 R^{(\mathrm{bg})}_q  &\equiv& 2 \sum_{p>0, p \neq q} \abs{\bar{\eta}^{(K p)}_q}^4~~~\mathrm{for~}U \gg \gamma_0 \mathcomma \\
 R^{(\mathrm{dir})}_q &\equiv& 2 \abs{\bar{\eta}^{(K q)}_q}^4~~~\mathrm{for~}U \gg \gamma_0 \mathcomma \\
 R^{(\mathrm{BS})}_q  &\equiv& \abs{\bar{\eta}^{(K, \mathrm{BS})}_q}^4~~~\mathrm{for~}U \gg \gamma_0 \mathcomma \\
 Q^{(\mathrm{cross})}_q &\equiv& \sum_{p} \abs{\bar{\eta}^{(K p)}_0}^2 \abs{\bar{\eta}^{(K p)}_q}^2~~~\mathrm{for~U=0} \\ \nonumber
      &=& 2 \Bigg{(} \sum_{p>0, p \neq q} \abs{\bar{\eta}^{(K p)}_0}^2 \abs{\bar{\eta}^{(K p)}_q}^2 
      \\ \nonumber
      && ~~~~~~~~~~
	~+~ \underbrace{\abs{\bar{\eta}^{(K p)}_0}^2 \abs{\bar{\eta}^{(K q)}_q}^2}_{\overset{U=0}{=} 0} \Bigg{)} \\ \nonumber
      &=& 2 \sum_{p>0, p \neq q} \abs{\bar{\eta}^{(K p)}_0}^2 \abs{\bar{\eta}^{(K p)}_q}^2 
	\mathcomma \\
 R^{(\mathrm{cross})}_q &\equiv& \sum_{p} \abs{\bar{\eta}^{(K p)}_0}^2 \abs{\bar{\eta}^{(K p)}_q}^2
 ~~~\mathrm{for~}U \gg \gamma_0 \mathcomma \\
 R^{(\mathrm{cross~BS})}_q &\equiv& \abs{\bar{\eta}^{(K, \mathrm{BS})}_0}^2 \abs{\bar{\eta}^{(K, \mathrm{BS})}_q}^2
 ~~~\mathrm{for~}U \gg \gamma_0 
 \mathperiod
\end{eqnarray}
Inserting the expressions for the momentum distributions from \secref{subsubsec:dirbgetc},
we ultimately have to calculate ($M \gg 1$)
\begin{eqnarray}
 \label{eq:lattsumQ}
 Q^{(\mathrm{bg})}_q &=& 
 \frac{32}{M^4} 
 \sum_{p>0, p \neq q} 
 \frac{1}{\tan^4\left[\frac{1}{2}\left(p-q\right)a\right]}
 \\ \nonumber
 && \hspace{2cm} ~\times~ \left( 1-\delta_{\left[\frac{\left(p-q\right)Na}{2\pi}\right]_{2} ,0} \right) 
 \mathcomma \\
 R^{(\mathrm{bg})}_q &=& 
 \frac{32}{M^4} 
 \sum_{p>0, p \neq q} 
 \frac{\cos^4\left[\frac{1}{2}\left(p+q\right)a\right]}{\sin^4\left[\frac{1}{2}\left(p-q\right)a\right]} 
 \\ \nonumber
 && \hspace{2cm} ~\times~ 
 \left( 1-\delta_{\left[\frac{\left(p-q\right)Na}{2\pi}\right]_{2} ,0} \right) 
 \mathcomma \\
 R^{(\mathrm{dir})}_q &=& 2 \sin^4(qa)
 \mathcomma \\
 R^{(\mathrm{BS})}_q &=& 
 \frac{16}{M^2} \cos^4(qa)
 \mathcomma \\
 Q^{(\mathrm{cross})}_{q} &=& 
  \frac{32}{M^4} \sum_{p>0, p \neq q} 
  \frac{1}{\tan^2\left(\frac{pa}{2}\right)} \frac{1}{ \tan^2\left[\frac{1}{2}\left(p-q\right)a \right] }
  \\ \nonumber
  && \hspace{2cm} ~\times~ 
  \left( 1-\delta_{\left[\frac{\left(p-q\right)Na}{2\pi}\right]_{2} ,0} \right) 
 \mathcomma \\
 R^{(\mathrm{cross})}_{q} &=& 
 \frac{32}{M^4} \sum_{p>0, p \neq q} 
 \frac{1}{\tan^2 \left(\frac{pa}{2}\right)} 
 \frac{1+\cos\left[(p+q)a\right]}{1-\cos\left[(p-q)a\right]}
 \\ \nonumber
 && \hspace{2cm} ~\times~ 
 \left( 1-\delta_{\left[\frac{\left(p-q\right)Na}{2\pi}\right]_{2} ,0} \right) 
 \mathcomma \\
 R^{(\mathrm{cross~BS})}_q &=& \frac{16}{M^2} \cos^2(qa)
 \mathperiod
\end{eqnarray}

The remaining lattice sums can be performed for $M \gg 1$ as follows.
We explain the procedure for \Eqref{eq:lattsumQ}.
The other quantities can be obtained in a similar manner.
Defining $\delta \equiv 2\pi/aM \ll 1$ and rewriting $p=q+\delta \cdot n$ 
($n$ is an integer), we essentially need to calculate
\begin{equation}
 Q^{(\mathrm{bg})}_q = 
 \frac{2 \delta^4 a^4}{\pi^4} \sideset{}{^\prime}\sum_n  \frac{1}{\tan^4\left( \frac{\delta a}{2} \cdot n \right)}
 \mathcomma
\end{equation}
where the sum runs over those odd values of $n$ such that all wavenumbers from 
$0 \rightarrow q-\delta$ as well as from 
$q + \delta \rightarrow \pi/a$ are covered.
Since $M \gg 1$, this expression can be expanded into a Taylor series in $\delta$ (around $\delta=0$).
In the limit $M \gg 1$, only terms of order $\delta^0$ remain, yielding
\begin{eqnarray}
 Q^{(\mathrm{bg})}_q &=& 
 \frac{2 a^4}{\pi^4} \cdot \frac{16}{a^4} \cdot 
 \left( 2 - \delta_{q0} \right) \cdot
 \underbrace{\sum_{n=0}^{\infty}  \frac{1}{\left(2n+1\right)^4}}_{= \frac{\pi^4}{96}}
 \\ \nonumber
 &=&
 \frac{1}{3} \left( 2- \delta_{q0} \right)
 \mathperiod
\end{eqnarray}
If $q \neq 0$, then there are (for $M \rightarrow \infty$) 
infinitely many wavenumbers on either side of $q$
(\ie, from $0 \rightarrow q-\delta$ as well as from $q + \delta \rightarrow \pi/a$).
Hence, we have a factor of $2$ in front of the sum over all odd values.
In contrast to this, if $q=0$, there is only a single interval (and therefore we just have
the factor of $1$).

The other quantities can be calculated in a similar manner, yielding
\begin{eqnarray}
  R^{(\mathrm{bg})}_{q} &=& 
  \cos^4\left( qa \right) Q^{(bg)}_q
 \mathcomma \\
 Q^{(\mathrm{cross})}_{q} &=& \frac{1}{3} \delta_{q 0} = Q^{(bg)}_q \delta_{q0}
 \mathcomma \\
 R^{(\mathrm{cross})}_{q} &=& Q^{(bg)}_q \delta_{q0}
 \mathperiod
\end{eqnarray}
Note that $R^{(\mathrm{cross})}_{q=0}=Q^{(\mathrm{cross})}_{q=0} = R^{(\mathrm{bg})}_{q=0} = Q^{(\mathrm{bg})}_{q=0}=1/3$.


\begin{thebibliography}{99}
\bibitem{noteequiv}
This procedure is equivalent 
to setting up a Lindblad
equation with respect to the bare atomic basis in the first place and then diagonalizing the
coherent part.
\bibitem{fnphaseshift}
Note that by redefining the phase shift as 
$\delta^\prime_{Kp} = - \delta_{Kp}/2 + \pi/2$
our ansatz~(\ref{eq:ansatzscattstat})
for the relative wavefunction can be brought into the form 
$\Psi^{(K p)}_x \propto \sin\left( pa \abs{x} + \delta^\prime_{Kp} \right)$,
which is often used in the context of cold-atom scattering.
\bibitem{commentficek}
In the expressions for $g^{(2)}(0)$ in Refs.~\cite{wiegand,ficekPRA}, we 
neglect the \anfz{cooperativity parameters} and assume $(1+a)^2+b^2 \approx 1$
(assuming the inter-atomic dipole--dipole coupling to be small).
Furthermore, the angle-dependence which is encoded in the $\cos$-terms drops out
for the special geometry considered in \secref{subsec:twopump}.
\end{thebibliography}

\end{document}